\documentclass[11pt,a4paper]{article}
\usepackage{jheppub}
\usepackage{slashed}
\usepackage{multirow}
\usepackage{subcaption}
\usepackage{rotating}
\usepackage[table]{xcolor}
\usepackage{enumitem}
\usepackage{colortbl}
\usepackage{pdflscape}
\usepackage{float}
\usepackage{siunitx}
\usepackage[compat=1.1.0]{tikz-feynman}
\usepackage{lscape}
\usepackage{mathtools}
\usepackage{amsfonts}
\usepackage{bbold}
\usepackage{soul}
\usepackage{comment}
\usepackage{adjustbox}
\usepackage{lineno}
\usepackage{tocloft}

\newcommand{\mcO}{\ensuremath{\mathcal{O}}}
\newcommand{\mcM}{\ensuremath{\mathcal{M}}}
\definecolor{nicered}{rgb}{0.6,0.1,0.1}
\definecolor{nicegreen}{rgb}{0.1,0.5,0.1}
\definecolor{mediumcandyapplered}{rgb}{0.99, 0.12, 0.07}
\definecolor{red}{rgb}{1.0, 0, 0}

\newcommand{\squa}[3]{\ensuremath{{[#1|}_{#3}^{#2}}}
\newcommand{\squet}[3]{\ensuremath{|#1]_{#3}^{#2}}}
\newcommand{\tra}[3]{\ensuremath{\langle#1|_{#3}^{#2}}}
\newcommand{\tret}[3]{\ensuremath{{|#1\rangle}_{#3}^{#2}}}

\newcommand{\squaket}[2]{\ensuremath{[#1|#2]}}
\newcommand{\traket}[2]{\ensuremath{\langle#1|#2\rangle}}
\newcommand{\squam}[3]{\ensuremath{{[\mathbf{#1}|}_{#3}^{#2}}}
\newcommand{\squetm}[3]{\ensuremath{|\mathbf{#1}]_{#3}^{#2}}}
\newcommand{\tram}[3]{\ensuremath{\langle\mathbf{#1}|_{#3}^{#2}}}
\newcommand{\tretm}[3]{\ensuremath{{|\mathbf{#1}\rangle}_{#3}^{#2}}}

\renewcommand{\bar}{\overline}

\definecolor{LightCyan}{rgb}{0.88,1,1}
\definecolor{piggypink}{rgb}{0.99, 0.87, 0.9}
\definecolor{applegreen}{rgb}{0.55, 0.71, 0.0}
\definecolor{darkpastelgreen}{rgb}{0.01, 0.75, 0.24}
\definecolor{green-yellow}{rgb}{0.68, 1.0, 0.18}

\newcommand{\beq}{\begin{equation}}
	\newcommand{\eeq}{\end{equation}}
\newcommand{\bea}{\begin{eqnarray}}
	\newcommand{\eea}{\end{eqnarray}}

\DeclareSIUnit\barn{b}

\arxivnumber{2509.02680}

\title{\bf{Multi-Higgs Amplitudes Bootstrapped: \\ Dissecting SMEFT and HEFT}}

\author[a]{Ramona Gr\"ober,}
\author[a]{Alejo N. Rossia,}
\author[a]{Micha\l{} Ryczkowski}
\emailAdd{ramona.groeber@pd.infn.it, alejonahuel.rossia@unipd.it, michaljakub.ryczkowski@unipd.it}
\affiliation[a]{ Dipartimento di Fisica e Astronomia ``G. Galilei'', Universit\`a di Padova and Istituto Nazionale di Fisica Nucleare, Sezione di Padova, Via F. Marzolo 8, I-35131, Padova, Italy}

\abstract{The precise measurement of the Higgs boson properties requires a robust framework to parametrize possible deviations from Standard Model (SM) predictions in the most model-independent way possible. The Effective Field Theory (EFT) framework has become the most widely used since it offers a broad scope and a consistent path to increase the precision of the computations. Two prominent EFTs are the Standard Model Effective Field Theory (SMEFT) and the Higgs Effective Field Theory (HEFT). While similar in many aspects, their phenomenological differences are nowhere more pronounced than in multi-Higgs production. To precisely chart the separation between both EFTs, we study gluon-fusion double and triple Higgs production using bootstrapped on-shell amplitudes. This allows us to get the kinematic dependence of the gauge-invariant amplitude without field-redefinition ambiguities. As part of our study, we develop a technique that allows to build tree-level five-point on-shell amplitudes from lower-point on-shell amplitudes and bootstrapped contact terms. We then match the bootstrapped on-shell scattering amplitudes to the amplitudes computed in SMEFT (up to order $1/\Lambda^4$) and HEFT (at NNLO) and analyze the EFT order at which each kinematic structure appears. We also show how certain structures in $gg\to hhh$ appear only at dimension-12 in SMEFT or N$^3$LO in HEFT.}

\preprint{COMETA-2025-34}
\begin{document}

\maketitle
\flushbottom

\addtocontents{toc}{\protect\setcounter{tocdepth}{2}} 
\section{Introduction} 
\label{sec:introduction}
The precise measurement of the Higgs boson properties is one of the main goals of the LHC program. The Higgs interactions can be tested model-independently adopting an Effective Field Theory (EFT) approach. The two most popular EFTs for Higgs physics are the Standard Model Effective Field Theory (SMEFT) \cite{BUCHMULLER1986621, Grzadkowski:2010es} and the Higgs Effective Field Theory (HEFT) \cite{Feruglio:1992wf, Grinstein:2007iv, Buchalla:2012qq, Alonso:2012px, Brivio:2013pma}. They both assume the SM field content, but they differ in their gauge symmetries and, more precisely, in the Higgs boson embedding. In SMEFT, the Higgs comes together with the unphysical Goldstone bosons in a SU(2) doublet while in HEFT, the Higgs boson is a gauge singlet. Furthermore, these two EFTs typically differ in the power counting \cite{Gavela:2016bzc, Buchalla:2013eza, Brivioinprep}. 

But what is the right EFT to apply? 
This question has been addressed in a series of papers~\cite{Alonso:2015fsp, Alonso:2016oah,Falkowski:2019tft, Cohen:2020xca, Cohen:2021ucp, Alminawi:2023qtf}, that use analyticity and geometric arguments to identify cases in which one should use the more general HEFT instead of SMEFT. In particular, in \cite{Cohen:2020xca} two types of models were identified that realise HEFT instead of SMEFT: i.)~models where the new particles get most of their mass from electroweak symmetry breaking and ii.)~models that have additional sources of electroweak symmetry breaking even in the limit of $v \to 0 $, where $v$ denotes the SM vacuum expectation value. 
But does HEFT just show better convergence behavior in these cases, or can it give us something fundamentally different from SMEFT? 

In this paper, we address this question making use of on-shell amplitude techniques from the spinor-helicity formalism. Ultimately, with the help of general principles such as Lorentz invariance, unitarity and Bose or Fermi statistics one can bypass the Lagrangian by bootstrapping the physical quantities of interest, scattering amplitudes. These techniques, together with the spinor-helicity formalism, have proven to be very useful in building up bases of EFT operators~\cite{Shadmi:2018xan, Aoude:2019tzn,Durieux:2019eor, Durieux:2019siw, Durieux:2020gip, AccettulliHuber:2021uoa, Dong:2022jru}, computing the EFT RGE running~\cite{Caron-Huot:2016cwu,
EliasMiro:2020tdv,Baratella:2020lzz,Baratella:2020dvw,Bresciani:2023jsu,Bresciani:2024shu, Aebischer:2025zxg} and matching to UV models~\cite{DeAngelis:2023bmd,Chala:2024llp}. 
The complete set of four-point on-shell scattering amplitudes with energy growth up to order $E^4$ and their matching to HEFT and SMEFT operators was studied in~\cite{Liu:2023jbq}.

Here, we consider on-shell scattering amplitudes for multi-Higgs production up to five points, particularly relevant for the question of EFT convergence, and match them to their SMEFT and HEFT counterparts. Apart from providing an ultimate test of electroweak symmetry breaking, by allowing for measurement of the Higgs self-couplings, multi-Higgs production allows for study of the differences between SMEFT and HEFT. This originates from the fact that in HEFT multi-Higgs interactions arise at the same order as single-Higgs interactions, with the Higgs fields entering via a polynomial in $h/v$ with independent coefficients, the so-called flare function. This is why multi-Higgs production is an ideal showcase to explore the question of convergence between HEFT and SMEFT. 

For the purpose of disentangling HEFT and SMEFT, our analysis is restricted to tree-level diagrams. In the SM, multi-Higgs production arises for the first time at one loop, predominantly through top-quark mediation. In the limit where the top-quark mass is large compared to the external momenta, these loops reduce to effective Higgs–gluon couplings, which is the setup we adopt here. Of course, from a phenomenological perspective, the full top-mass dependence of the SM loops must be retained rather than approximated by an EFT. However, for the conceptual goal of our study — highlighting the structural differences between the two EFT frameworks — restricting ourselves to tree-level is fully justified.

In a similar spirit to this study, Ref.~\cite{Goldberg:2024eot} studied vector-pair production via on-shell methods and found that only a single structure arising in HEFT does not arise at dimension-8 in SMEFT. Other studies in the context of multi-Higgs production were limited to SMEFT and HEFT at the LO in the EFT expansion \cite{Gomez-Ambrosio:2022qsi, Gomez-Ambrosio:2022why, Delgado:2023ynh}, while in Ref.~\cite{Domenech:2025gmn} also a matching to the NLO HEFT Lagrangian was taken into account. Indeed, in contrast to these latter references, our study shows that the differences in HEFT and SMEFT are purely a question of convergence of the EFT expansion. 

Our paper is structured as follows: in section~\ref{sec:spinor_hel_eft}, we introduce the three theoretical frameworks to be used throughout: the spinor-helicity formalism, the SMEFT and HEFT. In section~\ref{sec:HH}, we apply them to the case of gluon-fusion double Higgs production. We review the derivation of the on-shell amplitudes for $gg\to hh$ in section~\ref{sec:HH_nonfact}
and~\ref{sec:HH_factorizable_contributions}. Then, we introduce our on-shell matching procedure and we apply it to match the $gg\to hh$ amplitudes to their SMEFT and HEFT counterparts in section~\ref{sec:matching_SMEFT_HEFT_3_4pt}. Section~\ref{sec:HHH} presents our results for triple Higgs production. We present our methodology to bootstrap the full five-point on-shell amplitude and apply it to the simple case of five Higgs bosons in section~\ref{sec:5h_amplitudes}. We then apply the same techniques to build the amplitudes of $gg\to hhh$ in section~\ref{sec:gghhh_onshell_bootstrap}. Finally, we match the on-shell $gg\to hhh$ amplitudes to the SMEFT and HEFT amplitudes in section~\ref{sec:gghhh_matching} and discuss how different structures appear at different EFT orders. Our conclusions and outlook are given in section~\ref{sec:conclusions}.
We provide the results and further details of our study in various appendices.

%%%%%%%%%%%%%%%%%%%%%%%%%%%%%%%
\section{Theoretical framework} 
\label{sec:spinor_hel_eft}
In this section, we review the theoretical framework used in our work and establish the notation used throughout the paper. 

\subsection{Spinor-helicity formalism}
\label{sec:spinor_hel_formalism}
Massless scattering amplitudes can be written in a very compact form by specifying the helicity of the initial and final particles and by making its Lorentz- and little-group structure manifest. Such simplification is easily achieved with the spinor-helicity formalism that allows to write the momenta and polarization vectors in terms of massless spinors with well-defined Lorentz- and little-group transformation properties. This formalism (see Ref.~\cite{Elvang:2013cua} for a review) is based on the observation that a lightlike momentum can be written in terms of two independent two-dimensional commuting spinors,
\begin{equation}
    p_{\alpha\dot\alpha}\equiv p_{\mu}\sigma_{\alpha\dot\alpha}^{\mu}\equiv \tret{p}{}{\alpha}\squa{p}{}{\dot\alpha}\,,\quad \bar{p}^{\dot\alpha\alpha}\equiv p_{\mu}\bar{\sigma}^{\mu\,\dot\alpha\alpha} \equiv \squet{p}{\dot\alpha}{}\tra{p}{\alpha}{}\,,
\end{equation}
where $\sigma^\mu=\left(\mathbb{1},\sigma^i\right)$, $\bar\sigma^\mu=\left(\mathbb{1},-\sigma^i\right)$, and $\sigma^i$ are the Pauli matrices. $\tra{p}{\alpha}{}$, $\tret{p}{}{\alpha}$, $\squa{p}{}{\dot\alpha}$, and $\squet{p}{\dot\alpha}{}$ are the chiral (angle) and anti-chiral (square) spinors that are solutions to the massless Dirac (Weyl) equation of motion. The angle and square spinors transform as the $(1/2,0)$ and $(0,1/2)$ representations of the $SL(2,\mathbb{C})$ Lorentz group and carry opposite little-group weights,
\begin{equation}
    \tret{p}{}{}\to e^{-i\xi}\tret{p}{}{}\,, \quad \squa{p}{}{}\to e^{i\xi} \squa{p}{}{}\,,
\end{equation}
where $\xi$ parametrizes a transformation of the massless little-group, U(1). Throughout this work, we follow the conventions from Ref.~\cite{Durieux:2019eor}, and for further details, see App.~\ref{app:spinor_helicity_details}.

The helicities of the external particles in an on-shell scattering amplitudes determine the weight of the little-group transformation of the amplitude. This gives powerful selection rules for the spinor-helicity structures allowed in a given on-shell massless scattering amplitude. Furthermore, these selection rules, together with locality and special three-point kinematics, fully fix the structure of the massless (complex momenta) three-point on-shell amplitudes~\cite{Elvang:2013cua}.

Several works explored how to generalize the previous formalism to the massive case. We follow the little-group covariant extension from Ref.~\cite{Arkani-Hamed:2017jhn}. The massive little group is SO(3)$\simeq$SU(2)$/\mathbb{Z}_2$ and a massive momenta can be expressed as an SU(2)-invariant combination of two pairs of spinors,
\begin{equation}
    \mathbf{p}_{\alpha\dot\alpha}=\tretm{p}{I}{\alpha}\squam{p}{}{I,\dot\alpha}\,,\quad \bar{\mathbf{p}}^{\dot\alpha\alpha}=-\squetm{p}{I,\dot\alpha}{}\tram{p}{\alpha}{I}\,,
\end{equation}
where we use bold face to denote a massive momentum and its corresponding spinors, and $I=1,2$ is an SU(2) index. The choice of $\mathbf{p}^{I}$ is arbitrary and can be used to select different polarization axes or to smooth the transition towards massless helicity amplitudes in the high-energy limit. Since they are SU(2) doublets, these massive spinors can be combined in symmetric tensors to reproduce the polarization tensors of massive particles of any spin. 
The massive little-group indices of an on-shell scattering amplitude must be fully symmetrized. However, during the rest of this work, all our external states will be either massless spin-1 or massive spin-0 states, so our amplitudes will have no little-group indices. For further details, see App.~\ref{app:spinor_helicity_details}.

A general $n-$point on-shell scattering amplitude can be written as,
\begin{equation}
    \mcM_n = \mathcal{T} \sum_{i} c_{i} \langle...\rangle^{N_i} \left[...\right]^{M_i},
\label{eq:gral_onshell_amplitude}
\end{equation}
where $\langle...\rangle$ and $\left[ ...\right]$ are contracted helicity spinor structures and the sum is over all the ones allowed by the symmetries of the amplitude and momentum conservation. $\mathcal{T}$ is a tensor that contains the gauge-group invariant structure of the amplitude, which in the cases of our interest will be either $1$ or $\delta^{ab}$. The coefficients $c_{i}$ are free coefficients that might depend on Lorentz-invariant quantities such as the generalized Mandelstam variables,
\begin{equation}
    s_{ij}=(p_i+p_j)^2 \, .
\end{equation}
When computing the on-shell scattering amplitudes, we adopt the all-incoming convention.

In this work, we consider only tree-level scattering amplitudes. Under the assumption of locality, these amplitudes have a simple analytical structure characterized by their poles, given by the possible internal propagators\footnote{Hence, poles can only appear for $n\ge4$.}. These poles must be simple and their residues are determined by a product of two lower-point on-shell amplitudes. Thus, the amplitude can be split into factorizable and non-factorizable pieces,
\begin{equation}
    \mcM_{n} = \mcM_{n}^{\text{NF}} + \mcM_{n}^{\text{F}} \, , 
\end{equation}
where the non-factorizable piece $\mcM_{n}^{\text{NF}}$ can be built by bootstrapping as in Eq.~\eqref{eq:gral_onshell_amplitude}. The factorizable piece, $\mcM_{n}^{\text{F}}$, can be built out of products of lower-point amplitudes, in what is called ``gluing'', or via recursion relations~\cite{Britto:2004nc,Britto:2005fq,Risager:2005vk,Arkani-Hamed:2008bsc,Cohen:2010mi,Kampf:2012fn,Elvang:2013cua,Cheung:2014dqa,Cheung:2015cba,Cheung:2015ota,Durieux:2019eor,AccettulliHuber:2021uoa}. The use of recursion relations in the context of EFTs is limited by uncomputable boundary terms~\cite{Cheung:2015cba,AccettulliHuber:2021uoa}. Thus, we will use the gluing technique to construct the factorizable part of four- and five-point amplitudes. The factorizable piece should include all the possible factorization channels allowed by the non-vanishing lower-point amplitudes. The general formula to construct the factorizable part of a four-point amplitude is then\footnote{As written here, this formula is valid for either massless or massive and spinless intermediate states, which is enough for our purposes. For its generalization to massive intermediate particles with spin, see Ref.~\cite{Durieux:2019eor}.},
\begin{equation}
   \mcM_{4}^{\text{F}}(\psi_1,\psi_2,\psi_3,\psi_4) = - \sum \frac{\mcM_3(\psi_i,\psi_j,\psi_V)\mcM_3(\psi_k,\psi_\ell,\tilde\psi_V)}{s_{ij} - m_V^2} \, , 
\end{equation}
where $i$, $j$, $k$, $\ell\in\lbrace1,2,3,4\rbrace$, $\psi_V$ is the particle of mass $m_V$ being exchanged and the sum runs over all possible factorization channels. We introduce $\tilde\psi_V$ to denote a particle like $\psi_V$ but with opposite helicity and momentum since each three-point amplitude requires all particles to be incoming. This prescription allows the full amplitude to reproduce the right poles and residues. We will see examples of its usage in section~\ref{sec:HH_factorizable_contributions}, and discuss how to extend it to the case of five-point amplitudes in section~\ref{sec:5h_amplitudes}.

Since all the singularities of the amplitude are contained in the factorizable piece, the coefficients $c_{i}$ in Eq.~\eqref{eq:gral_onshell_amplitude} are analytical functions of the Mandelstam variables, hence admit a Taylor expansion in the relevant variables,
\begin{equation}
    c_{i} = c_{i}^{(00)}+c_{i}^{(10)}\ s_{12} + c_{i}^{(01)}\ s_{13} + c_{i}^{(20)}\ s_{12}^2 +c_{i}^{(02)}\ s_{13}^2+c_{i}^{(11)}\ s_{12}\ s_{13}\ + ... \, ,
\end{equation}
in the case of a four-point amplitude.
The overall energy dimension of $c_i$ can be inferred from the energy dimension of the amplitude and the spinor structures that it multiplies. Higher powers in the Taylor expansion of $c_{i}$ as a function of the Mandelstam variables will be suppressed by some energy scale for dimensional reasons. Since a high-enough power of the Mandelstam variables will induce a unitarity-violating behavior in the amplitude, their coefficients should be suppressed by a certain scale. 
Considering only a finite amount of terms, a possible choice of the suppressing scale is the minimal one at which unitarity is lost.  
These observations establish a clear path to connect the on-shell scattering amplitudes with EFTs: each unitarity-violating term in the expansion of the $c_{i}$'s would be related to a certain combination of WCs.

To conclude this section, let us remark that when bootstrapping the amplitude, we do not assume any gauge symmetries\footnote{$\mathcal{T}$ could be seen as a tensor in ``flavor'' indices that can later be identified as, e.g., color indices.}. The algebraic structures related to gauge symmetries can be obtained by requiring consistent factorization of four-point tree-level amplitudes into three-point amplitudes and unitarity in the SM limit~\cite{Benincasa:2007xk,Shadmi:2018xan,Aoude:2019tzn,Durieux:2019eor,AccettulliHuber:2021uoa}. Additionally, the constraints on hypercharges from gauge anomaly cancellation arise from requiring both locality and unitarity in four-point one-loop amplitudes~\cite{Huang:2013vha,Chen:2014eva,AccettulliHuber:2021uoa}. This feature represents a clear advantage to compare EFTs that have the same physical degrees of freedom but differ in their gauge symmetry, such as HEFT and SMEFT.
Finally, we note that our amplitude construction corresponds to a HEFT-like setup from the viewpoint of particle content and symmetry realization, as the Higgs field is treated as a singlet, independent of the Goldstone bosons associated with electroweak symmetry breaking.

\subsection{SMEFT}

The Standard Model Effective Field Theory provides a model-independent framework to parametrize the effects of heavy and decoupling beyond-the-Standard-Model (BSM) physics through higher-dimensional operators, suppressed by powers of a heavy new physics scale $\Lambda$. The SMEFT expansion is organized via a power counting in $1/\Lambda$ with a Lagrangian of the form,
\begin{equation}\mathcal{L}_{\mathrm{SMEFT}}=\mathcal{L}_{\mathrm{SM}}+
\sum_i
\frac{C_i}{\Lambda^{d_i-4}}\mathcal{O}_i \, ,
  \label{eq:SMEFT:Lagr:General}
\end{equation}
where $\mathcal{L}_{\mathrm{SM}}$ denotes the SM Lagrangian, $\mathcal{O}_i$ are higher-dimensional
operators of dimension $d_i$, parametrized by dimensionless Wilson coefficients $C_i$ and invariant under the SM gauge group, see~\cite{Brivio:2017vri,Isidori:2023pyp,Aebischer:2025qhh} for reviews.

In recent years, SMEFT has become one of the primary tools in the search for BSM physics. The lack of direct evidence for new particles at collider experiments has motivated a shift of interest towards precision measurements as a probe of new physics, where SMEFT offers a systematic and well-defined approach.

Although highly successful and widely used, SMEFT suffers from a number of limitations. For example, it assumes the Higgs boson as a part of the $SU(2)_L$ doublet and that the electroweak symmetry breaking is linearly realized. As such, SMEFT properly describes only heavy and decoupling BSM scenarios. Examples of models for which the more general HEFT is necessary rather than SMEFT are discussed, for instance, in Ref.~\cite{Banta:2021dek}.

In the context of this work, we focus our attention on the effects of dim-6 and dim-8 SMEFT operators that can affect Higgs-gluon interactions and alter processes of interest, including single, double and triple Higgs production via gluon-fusion. For dim-6 operators, we employ the \textit{Warsaw} basis~\cite{Grzadkowski:2010es}, while for dim-8 operators we follow the Murphy basis~\cite{Murphy:2020rsh}.\footnote{Ref.~\cite{Li:2020gnx} introduced an alternative dim-8 basis, which is equivalent for our work since it only differs from the Murphy basis in the flavor sector.}

Amplitude calculations were carried out using \texttt{SmeftFRv3}~\cite{Dedes:2023zws}, which provides the complete set of Feynman rules for SMEFT including the full dim-6 basis and all bosonic dim-8 operators, interfaced with \texttt{FeynArts}~\cite{Hahn:2000kx} and \texttt{FeynCalc}~\cite{Shtabovenko:2020gxv} for symbolic evaluation. Moreover, for the generation of Feynman rules for dim-10 and -12 operators~\cite{Harlander:2023psl} used in section~\ref{sec:gghhh_matching} we used a private and modified version of \texttt{SmeftFRv3}. The complete list of SMEFT operators used in this work together with relevant conventions is provided in the App.~\ref{appendix:SMEFT}.

\subsection{HEFT}
Higgs Effective Field Theory, sometimes also called electroweak chiral Lagrangian, is a mix of SMEFT and the chiral Lagrangian as employed for QCD.  With respect to the SMEFT, it relaxes the assumption that the Higgs boson transforms as a SU(2) doublet under the SM gauge symmetry. Instead, the physical Higgs boson transforms in a singlet, while the Goldstone bosons come separately from the Higgs boson in a Goldstone matrix transforming as a bi-doublet under a global $SU(2)_L \times SU(2)_R$ symmetry,
\begin{equation}
    {\bf U} = \exp \left( \frac{i\,\pi^a \sigma^a}{v}\right)\,,
\end{equation} 
where $\pi^a$ are the Goldstone bosons and $\sigma^a$ the Pauli matrices. 
Its covariant derivative is given by,
\begin{align}
    D_{\mu} {\bf U} = \partial_{\mu} {\bf U} + \frac{i g}{2} W_{\mu}^I\, \sigma^I {\bf U} - \frac{i g'}{2} B_{\mu}\, {\bf U} \sigma^3\,.
\end{align}
For convenience, it is useful to define the objects,
\begin{align}
{\bf V}_{\mu} &= (D_{\mu} {\bf U}) {\bf U}^{\dagger}\,,
    &
 {\bf T} &= {\bf U }\sigma^3{\bf U}^\dagger   \,,
\end{align}
which both transform in the adjoint representation of $SU(2)_L$. ${\bf V}_\mu$ is a singlet under $SU(2)_R$, and ${\bf T}$ breaks $SU(2)_R$ and hence constitutes a spurion injecting an explicit breaking of the custodial symmetry.  
The LO Lagrangian in the EFT expansion -- differently from SMEFT -- does not correspond to the SM but allows for generic deviations given by,
\begin{equation}
    \begin{aligned}
\mathcal{L}_{\text{LO}}^{\text{HEFT}} &= 
- \frac{1}{4}G_{\mu\nu}^\alpha G^{\alpha\,\mu\nu}
- \frac{g_s^2}{16\pi^2} \theta_s \,G_{\mu\nu}^\alpha \widetilde{G}^{\alpha\,\mu\nu} 
- \frac{1}{4}  W^a_{\mu\nu} W^{a\mu\nu}
- \frac{1}{4} B_{\mu\nu} B^{\mu\nu}
\\
&+ \frac{1}{2} \partial_\mu h \partial^\mu h - V(h)
- \frac{v^2}{4} \text{Tr}\left( \mathbf{V}_\mu \mathbf{V}^\mu \right) \mathcal{F}_C(h) \\
&+ i \bar{Q}_L \slashed{D} Q_L + i \bar{Q}_R \slashed{D} Q_R
+ i \bar{L}_L \slashed{D} L_L + i \bar{L}_R \slashed{D} L_R \\
&- \frac{v}{\sqrt{2}} \left( \bar{Q}_L \mathbf{U} \mathcal{Y}_Q(h) Q_R + \text{h.c.} \right)
- \frac{v}{\sqrt{2}} \left( \bar{L}_L \mathbf{U} \mathcal{Y}_L(h) L_R + \text{h.c.} \right),
    \end{aligned}
    \label{HEFT:LO:Lagrangian}
\end{equation}
with the Higgs potential (relevant for the processes studied in this work) of the following form,
\begin{equation}
    \begin{aligned}
            V(h) = -\frac{1}{4} \lambda v^4 + \lambda v^2 h^2 
             + (1+\Delta \lambda_3)\lambda v h^3 + (1+\Delta \lambda_4) \frac{1}{4}\lambda h^4 + \kappa_{\lambda_5} \frac{h^5}{v} + \ldots\,. \label{eq:pot}
    \end{aligned}
\end{equation}
The functions $\mathcal{F}_C(h)$, $\mathcal{Y}_{Q}(h)$ and $\mathcal{Y}_{L}(h)$ are so-called flare functions and are 
infinite power series in $h/v$ with independent coefficients. Higgs interactions arise all at the same order due to its singlet nature. In principle, the LO HEFT Lagrangian can contain custodial symmetry violating operators, which are though commonly accounted for in the NLO basis mirroring the fact that their coefficients must be suppressed. We will not consider them here due to the nature of the processes we consider.
For the full list of operators employed in this work at NLO, NNLO and N$^3$LO see App.~\ref{appendix:HEFT}.

In HEFT, the order at which a given operator appears depends on the underlying assumptions about the UV completion and the adopted counting scheme. Using naïve dimensional analysis (NDA) one can identify the minimal order at which an operator may arise~\cite{Gavela:2016bzc}, while chiral-dimension power counting~\cite{Buchalla:2013eza,Brivioinprep} also assigns weights to the SM couplings. As a consequence, the chiral dimension of a given operator can vary depending on whether SM couplings are included in, or factored out of, the Wilson coefficients -- making the precise counting ultimately UV dependent.
In the following, we apply the amplitude-based construction to remain as general as possible and avoid fixing a specific HEFT counting convention. When matching our results to HEFT, we adopt the notation 
$N_{HEFT}^s=N_{HEFT}+N_{g_s}$
 ~\cite{Brivioinprep}, which combines the chiral-dimension order (set by derivatives, loops and external legs, as well as by the power of SM electroweak couplings) with the power of the strong coupling
$N_{g_s}$.
We note that while the amplitude construction shows the same symmetries as the HEFT, this does not hold true for the power counting, since the power counting is no longer global in the on-shell amplitude approach. This is one of the advantages of the amplitude approach: it allows us to discuss operator structures and helicity configurations independently of the specific chiral-dimension assignment, providing a transparent link between EFT operators and on-shell observables.
For the HEFT amplitude computations we employ a private \texttt{FeynRules}~\cite{Alloul:2013bka} implementation of the model, which will be made public in the near future.

 \section{Double Higgs production}
 \label{sec:HH}
 
We first consider double Higgs production as a pedagogical introduction and to verify our techniques. In sections~\ref{sec:HH_nonfact},~\ref{sec:HH_factorizable_contributions}, and parts of~\ref{sec:matching_SMEFT_HEFT_3_4pt} we review the results obtained in~\cite{Shadmi:2018xan}.
While in the SM double Higgs production via gluon-fusion arises at the one-loop level, for the sake of studying the differences between the two EFTs, we concentrate on tree-level contributions not present in the SM. 
Similarly in the SM, such structures would arise when considering the limit in which the top-quark mass is much larger than the external momenta of the considered process. 

\subsection{Non-factorizable contributions}
\label{sec:HH_nonfact}
 The non-factorizable part can be obtained by bootstrapping the Lorentz structures according to the external momenta, helicities and their little-group weights. Let $p_{1},\,p_{2}$ be the gluon momenta and $p_{3},\, p_{4}$ the Higgs momenta. Then there exist two independent helicity-spinor structures after accounting for momentum conservation,
\begin{equation}
    \left[1|2\right],\quad \left[ 1| \mathbf{3} | 2 \rangle \right.\,,
\end{equation}
where $\mathbf{3}$ corresponds to $p_3$, the massive momentum of one of the Higgs bosons. All other structures can be obtained via complex conjugation.
In general, the $gg\to hh$ scattering amplitude can be written as~\cite{Shadmi:2018xan},
\begin{equation}
\begin{aligned}
\mathcal{M}\left(g^{a,\,+}\left(p_1\right);g^{b,\,+}\left(p_2\right);h\left(p_3\right);h\left(p_4\right) \right)_{\text{NF}} = & i \,\delta^{ab} c_{gghh}^{++} \, \left[1|2\right]^2 \,, \\
\mathcal{M}\left(g^{a,\,+}\left(p_1\right);g^{b,\,-}\left(p_2\right);h\left(p_3\right);h\left(p_4\right) \right)_{\text{NF}} = & i \delta^{ab} c_{gghh}^{+-} \, \left[ 1| \mathbf{3} - \mathbf{4} | 2 \rangle \right.^2 \, ,\\
\end{aligned}
\label{eq:Mgghh}
\end{equation}
where the expression for the $--00$ ($-+00$) helicity configuration can be obtained from $++00$ ($+-00$) by hermitian conjugation. We have used momentum conservation to make the $+-00$ amplitude explicitly Bose symmetric under $3\leftrightarrow 4$. 

The coefficients $c_{gghh}^{++}$ and $c_{gghh}^{+-}$ are analytical functions of the Mandelstam variables  $s_{ij}=(p_i+p_j)^2$
that, by dimensional analysis, enter as a dimensionless ratios with the EFT cutoff $\bar\Lambda$. 
They can be expanded in powers of $s_{ij}$ as,
\begin{equation}
    c_{gghh}^{++} =\frac{c_{gghh}^{++,(00)}}{\bar{\Lambda}^2} + \frac{c_{gghh}^{++,(10)}}{\bar{\Lambda}^4} \, s_{12}  + \mathcal{O}\left(1/\bar{\Lambda}^6\right),
    \label{eq:amp_gghh_++}
\end{equation}
\begin{equation}
    c_{gghh}^{+-} =\frac{c_{gghh}^{+-,(00)}}{\bar{\Lambda}^4} + \frac{c_{gghh}^{+-,(10)}}{\bar{\Lambda}^6} \, s_{12} + \mathcal{O}\left(1/\bar{\Lambda}^8\right),
    \label{eq:amp_gghh_+-}
\end{equation}
where the coefficients $c_{gghh}^{+\pm,(ij)}$ are dimensionless. The Mandelstam variables $s_{13}$ and $s_{23}$ appear only from second order onwards due to the $3 \leftrightarrow 4$ symmetry and their correlation with $s_{12}$~\cite{Shadmi:2018xan}.
Notice that if one computes these functions in the SM in the heavy-top limit, they will correspond to the limit value of boxes of fermions, and the scale $\Lambda$ can be chosen as the top mass, in particular if the top is the only fermion considered as massive. Indeed, after imposing Lorentz invariance and Ward identities one can write the amplitude in terms of two form factors \cite{Plehn:1996wb} that arise in analogy to our considerations at different order in the expansion in $1/m_t^2$, i.e. the part corresponding to the $+-$ helicity amplitude arises an order higher in $1/m_t^2$ than the $++$ helicity configuration, see for instance \cite{Degrassi:2016vss}. The latter approach has been discussed in the context of EFT in \cite{Chang:2022crb}. 

\subsection{Factorizable contributions}
\label{sec:HH_factorizable_contributions}

There are three different factorizable contributions to the $gg\to hh$ amplitude: the $s$-channel which involves gluing a $gg\to h$ and a $hh\to h$ amplitude via a virtual $h$, and the $t/u$-channels that involve gluing two $gg\to h$ amplitudes via a propagating gluon. In the SM, the $s$-channel appears at one-loop order and the $t/u$-channels at two-loop order, while in the SMEFT, both could appear at tree level at dimension-6 and dimension-8, respectively.\footnote{Under the assumption of a weakly-interacting theory, the operator giving rise to the effective $gg\to h$ coupling arises
at one-loop order \cite{Arzt:1994gp, Grojean:2024tcw}. }

\subsubsection{$s$-channel}
 The $gg\to h$ on-shell amplitude can be parametrized as
\begin{equation*}
\begin{aligned}
\begin{tikzpicture}[baseline=(v1.base)]
\begin{feynman}[inline=v1.base]            \vertex (v1) at (0,0) [dot] {};
            \vertex (g1) at (-1,-0.75) {\( g^{h_2}(p_2) \)};
            \vertex (g2) at (-1,0.75) {\( g^{h_1}(p_1) \)};
            \vertex (h1) at (1,0) {\( h(p_3) \)};
            \diagram* {
                (g1) -- [gluon] (v1),
                (g2) -- [gluon] (v1),
                (v1) -- [scalar] (h1),
            };
        \end{feynman}
    \end{tikzpicture} = 
        \mcM\left( g_{1}^{a,h_1};g_{2}^{b,h_2};h \right) = i\, \delta^{ab} \left[1|2\right]^n g_{-\ell}(s_{12},\Lambda)\, ,
        \end{aligned}
\end{equation*}
where $n=2h_1=2h_2$ and $-\ell=1-n$ is the energy dimension of the function $g$.
The restrictive three-point kinematics tells us that $s_{12}=m_h^2$ and hence $g_{-\ell}$ is a constant that can only depend on couplings and masses.
Furthermore, little-group covariance makes the $+-0$ and $-+0$ configurations vanish on-shell, 
\begin{equation}
    \mcM\left( g_{1}^{a,+};g_{2}^{b,-};h \right) = \mcM\left( g_{1}^{a,-};g_{2}^{b,+};h \right) = 0 \, .
    \label{eq:ggh:opp_helicity}
\end{equation}
Thus, we only need to consider the $++0$ helicity configuration ($--0$ is its complex conjugate),
\begin{equation}
    \mcM\left( g_{1}^{a,+};g_{2}^{b,+};h \right) = \delta^{ab} \left[1|2\right]^{2} g_{-1}(m_h^2,\Lambda^2)=\delta^{ab} \left[1|2\right]^{2} c_{ggh} \, ,
    \label{eq:ggh:helicity}
\end{equation}
where in the last step we renamed $g_{-1}(m_h^2,\Lambda^2)=c_{ggh}$. The three-scalar on-shell amplitude is just a constant, without any kinematic dependence,
\begin{equation*}
\begin{tikzpicture}[baseline=(h3.base)]
        \begin{feynman}[inline=h3.base]
            \vertex (v1) at (0,0) [dot] {};
            \vertex (h2) at (-1,-0.75) {\( h(p_2) \)};
            \vertex (h1) at (-1,0.75) {\( h(p_1) \)};
            \vertex (h3) at (1,0) {\( h(p_3) \)};
            \diagram* {
                (h1) -- [scalar] (v1),
                (h2) -- [scalar] (v1),
                (v1) -- [scalar] (h3),
            };
        \end{feynman}
    \end{tikzpicture} = \mcM\left( h;h;h \right) = i\,c_{3h} \, .
\end{equation*}
Gluing the two three-point amplitudes is straightforward and results in the desired $s$-channel amplitude~\cite{Shadmi:2018xan},
\begin{equation}
\begin{aligned}
   \begin{tikzpicture}[baseline=(v2.base)]
        \begin{feynman}[inline=(v2.base)]
            \vertex (v1) at (0,0) [dot] {};
            \vertex (v2) at (1,0) [dot] {};
            \vertex (g1) at (-1,-1) {\( g^{a_1}_{\mu_1}(p_2) \)};
            \vertex (g2) at (-1,1) {\( g^{a_2}_{\mu_2}(p_1) \)};
            \vertex (h1) at (2,1) {\( h(p_3) \)};
            \vertex (h2) at (2,-1) {\( h(p_4) \)};
            
            \diagram* {
                (g1) -- [gluon] (v1),
                (g2) -- [gluon] (v1),
                (v1) -- [scalar] (v2),
                (v2) -- [scalar] (h1),
                (v2) -- [scalar] (h2),
            };
        \end{feynman}
    \end{tikzpicture}
   &=  \mcM\left( g_{1}^{a,+};g_{2}^{b,+};h_3;h_4 \right)_{s\text{--ch.}}=-\delta^{ab} \frac{c_{3h} \, c_{ggh} }{s_{12}-m_h^2}\squaket{1}{2}^2 \, .
   \end{aligned}
    \label{eq:amp_gghh_++_s}
\end{equation}
Notice that despite the virtual Higgs in the $s$-channel being off-shell, $c_{3h}$ and $c_{ggh}$ depend only on $m_h^2$ since we keep the glued-in three-point amplitudes on-shell. 
This expression can be very easily generalized to the case of having a scalar in the $s$-channel that is different from the final state scalars by replacing the dimensionless couplings and the mass of the intermediate state without modifying the general kinematic structure. 

\subsubsection{t/u-channels}

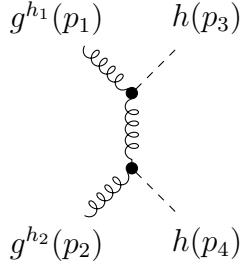
\begin{figure}[t!]
\begin{equation*}
\begin{tikzpicture}
        \begin{feynman}
            \vertex (v1) at (0,0.5) [dot] {};
            \vertex (v2) at (0,-0.5) [dot] {};
            \vertex (g1) at (-1,-1.5) {\( g^{h_2}(p_2) \)};
            \vertex (g2) at (-1,1.5) {\( g^{h_1}(p_1) \)};
            \vertex (h1) at (1,1.5) {\( h(p_3) \)};
            \vertex (h2) at (1,-1.5) {\( h(p_4) \)};
            
            \diagram* {
                (g1) -- [gluon] (v2),
                (g2) -- [gluon] (v1),
                (v1) -- [gluon] (v2),
                (v1) -- [scalar] (h1),
                (v2) -- [scalar] (h2),
            };
        \end{feynman}
    \end{tikzpicture}
\end{equation*}
\caption{$t/u$-channel factorizable contribution to $gg
\to hh$. \label{fig:gghhtu}}
\end{figure}

These channels require gluing two $gg\to h$ amplitudes via a virtual propagating gluon. An example diagram can be found in Fig.~\ref{fig:gghhtu}. Since an incoming particle is equivalent to an outgoing anti-particle with opposite momentum and helicity, there is no $++00$ helicity configuration for the $t/u$-channel diagrams.

Instead, for the $+-00$ configuration, we can combine a $++0$ and a $--0$ on-shell $gg\to h$ amplitude to get,
\begin{equation}
    \begin{aligned}
        \mcM\left(g_{1}^{a,+}; g_{2}^{b,-}; h_{3}; h_{4}\right)_{t+u\text{--ch.}} = & -\delta^{ab} \frac{|c_{ggh}|^2}{4} {\squa{1}{}{}\mathbf{3}-\mathbf{4}\tret{2}{}{}}^2 \left( \frac{1}{s_{13}} + \frac{1}{s_{23}} \right), \\
       = & -\delta^{ab} \frac{|c_{ggh}|^2}{4} {\squa{1}{}{}\mathbf{3}-\mathbf{4}\tret{2}{}{}}^2  \frac{2 m_h^2 - s_{12}}{s_{13}s_{23}} \, .
    \end{aligned}
    \label{eq:amp_gghh_+-_t_u}
\end{equation}
Notice that we have taken the absolute value of $c_{ggh}$ since it is a complex number to allow for CP-odd effects.

\subsection{Matching to SMEFT and HEFT}
\label{sec:matching_SMEFT_HEFT_3_4pt}

Now we turn to matching our amplitudes to SMEFT and HEFT. In Table~\ref{tab:amp_gghh_SMEFTHEFT} we summarize at which orders of the EFT expansion individual contributions arise,\footnote{We note that if we had adopted the power counting $N_{HEFT}$ that explicitly subtracts powers of $g_s$ from Ref.~\cite{Brivioinprep}, all the HEFT structures with gluons would effectively arise an order lower.} purely from naive power counting. 
The column indicating the dimension refers to the mass dimension in the amplitude approach and corresponds to what is called in \cite{Liu:2023jbq} ``HEFT dimension'' while we use as HEFT minimal order the counting in chiral dimension. Our first objective is to understand if matching the on-shell bootstrapped and EFT amplitudes yields the same result as this naive analysis.
\renewcommand{\arraystretch}{1.25}
\begin{table}[h]
    \centering
    \begin{tabular}{c|c|c|c|c|c|c}
    \multirow{2}{*}{Amplitude} & \multirow{2}{*}{Helicity} & \multirow{2}{*}{Spinor structure} & \multirow{2}{*}{Coeff.} & \multirow{2}{*}{Dimension} & \multicolumn{2}{c}{Minimal order}  \\
    & & & & & SMEFT & HEFT \\
    \hline
    \multicolumn{7}{c}{ Three-point}\\
    \hline
    $gg\to h$     & $++$ & $\squaket{1}{2}^2$ & $c_{ggh}$ & $-1\,(1/\bar\Lambda)$ & $6\,(v/\Lambda^2)$ & $\text{NLO}^*$ \\
    $hh\to h$     & - & - & $c_{hhh}$ & $1\,(\bar\Lambda) $ & $4$ & LO \\
    \hline
    \multicolumn{7}{c}{ Four-point}\\
    \hline
    $hh\to hh$    & - & - & $c_{4h}$ & $0$ & $4$ & LO \\
    \multirow{2}{*}{$gg\to hh$}    & $++$ & $\squaket{1}{2}^2$ & $c_{gghh}^{++}$ & $-2\,(1/\bar\Lambda^2)$ & $6\,(1/\Lambda^2)$ & $ \text{NLO}^*$ \\
        & $+-$ & ${\squa{1}{}{}\mathbf{3}-\mathbf{4}\tret{2}{}{}}^2$ & $c_{gghh}^{+-}$ & $-4\,(1/\bar\Lambda^4)$ & $8\,(1/\Lambda^4)$ & $\text{NNLO}^*$ \\
    \hline
    \end{tabular}
    \caption{Summary of on-shell coefficients and dimensions for $gg\to hh$, $hh\to hh$ and the related three-point amplitudes. $^*$ marks where we applied the power counting assuming that one order of $\alpha_s$ is factored out, i.e. that our HEFT operators are written as $\frac{\alpha_s}{\pi}\partial^m h^n G_{\mu\nu}G^{\mu\nu}$. }
    \label{tab:amp_gghh_SMEFTHEFT}
\end{table}

We perform rational numerical matching using the package \texttt{mosca}~\cite{LopezMiras:2025gar}, in particular its random phase-space point generator based on~\cite{DeAngelis:2022qco}. Our rational matching technique consists of expressing both the EFT and the on-shell amplitude as functions of the four-momentum components of the external particles and equating them. Then, we evaluate them in $N$ random on-shell phase-space points generated with \texttt{mosca}, where $N$ is the number of free on-shell coefficients to be matched. This yields a linear system of equations for the on-shell coefficients as a function of the WCs and other Lagrangian parameters, which we solve with standard techniques to obtain the matching results. We verified that different sets of random phase-space points yield the same matching results. This technique allows us to obtain exact results, which we verified analytically for three- and four-point amplitudes, while being easily extendable to higher-point amplitudes.
The full matching results can be found in App.~\ref{app:full_matching_res}. 

In case of the three-point amplitudes, the minimal order at which the amplitudes arise are equivalent in SMEFT and HEFT, see Table~\ref{tab:amp_gghh_SMEFTHEFT} or Eqs.~\eqref{eq:match_ons_smeft_ggh_Gf} and \eqref{eq:match_ons_smeft_ggh_HEFT} for $c_{ggh}$ and Eqs.~\eqref{eq:match_ons_smeft_hhh_Gf} and \eqref{eq:match_ons_smeft_hhh_HEFT} for $c_{3h}$.
 It is worth noting an interesting difference between SMEFT and HEFT. In HEFT the $gg\to h$ vertex receives contributions from two NLO operators, $\mathcal{P}_{GH}$ and $\mathcal{P}_{\tilde{G}H}$. Instead, in SMEFT we not only have dim-6 contributions from $Q_{\varphi G}$ and $Q_{\varphi \tilde{G}}$, but also dim-$6^2$ contributions from normalization of the kinetic terms and dim-8 contributions from operators $Q_{G^2 \varphi^4}^{(1)}$ and $Q_{G^2 \varphi^4}^{(2)}$.

Turning now our attention to the four-point results, $c_{gghh}^{++,(00)}$ receives leading contributions from dim-6 SMEFT operators (Eq.~\eqref{eq:match_Result_gghh_++00_00_rational_Gf}) and NLO HEFT operators (Eq.~\eqref{eq:match_Result_gghh_+-00_00_HEFTNNLO}). In the case of SMEFT, we identify $\bar\Lambda=\Lambda$ which automatically removes the scale from the leading contributions to each on-shell coefficient and generates a suppression factor of $m_h^2/\Lambda^2$ or $1/(G_F\,\Lambda^2)$ for dim-6 squared and dim-8 contributions. The identification of $\bar\Lambda$ with HEFT parameters is not straightforward, and we defer the discussion until after obtaining all the matching results.

A common feature in the results for both EFTs is the double insertion contributions to $c_{gghh}^{++,(00)}$. These can not be fully explained from field-redefinition contributions to the four-point EFT vertex. Instead, they arise, at least partially, from the non-factorizable on-shell amplitude absorbing parts of the propagator-mediated EFT amplitude that cannot be absorbed by the factorizable on-shell amplitude. We show this mechanism explicitly in App.~\ref{app:full_matching_importance}. In particular, this is crucial to reproduce the physical effect of the momentum-dependent three-Higgs vertex in SMEFT related to the factor $C_{\varphi\Box} - \frac{C_{\varphi D}}{4}$ in the on-shell amplitude.
From the EFT perspective, the presence of these momentum-dependent vertices is field-redefinition dependent. However, the bootstrapped on-shell amplitudes remains invariant under these field redefinitions, which is one of the advantages of this formalism.

Closer inspection of the matching results for $c_{gghh}^{++,(00)}$ unveils intriguing similarities and differences between the SMEFT and HEFT, particularly with respect to the power counting. First, we can clearly see the correspondence between contributions arising from single insertions of SMEFT dim-8 derivative operators $Q_{G^2 \varphi^2 D^2}^{(1,2,3)}$ and HEFT NNLO operators $\mathcal{P}_{GHD}^{(1,2)}, \, \mathcal{P}_{\tilde{G}HD}$. Then, on the side of differences, we can note that non-derivative single insertion contribution in this channel corresponds to two NLO operators, $\mathcal{P}_{GH}$ and $\mathcal{P}_{\tilde{G}H}$. In SMEFT, the corresponding effects arise from both two dim-6 operators, $Q_{\varphi G}$ and $Q_{\varphi \tilde{G}}$, and two dim-8 $Q_{G^2 \varphi^4}^{(1)}$ and $Q_{G^2 \varphi^4}^{(2)}$ operators. This correspondence can be made explicit as,
\begin{equation}
    - \sqrt{2}\, g_S^2 \,\bar\Lambda^2 \, G_F\left( C_{GH}^{(2)} +i\,C_{\widetilde{G}H}^{(2)} \right) \rightarrow 2 \left( C_{\varphi G} + i C_{\varphi \tilde{G}} \right) + 3 \frac{\sqrt{2}}{G_F\, \Lambda^2} \left( C_{G^2\varphi^4}^{(1)} + i C_{G^2\varphi^4}^{(2)}  \right).
\end{equation}
The above mapping is also manifest at the Lagrangian level: after matching the SMEFT mass-basis Lagrangian onto HEFT, one recovers precisely the same translation. The same effect could already be observed in the case of the $c_{ggh}$ coefficient discussed earlier in this section.

Moreover, the double insertion term, which in HEFT contains a purely Higgs-gluon operator contribution, in SMEFT contains additional contributions from the WCs corresponding to the canonical normalization of the Higgs field,
\begin{equation}
    \frac{g_S^4}{\sqrt{2}}\,G_F \bar\Lambda^2\, (C_{GH}^{(1)}+ i\, C_{\widetilde{G}H}^{(1)} )^2 \ \rightarrow  \frac{\sqrt{2}}{G_F \Lambda^2} \left( C_{\varphi G} + i C_{\varphi \tilde{G}}\right) \left( 4 C_{\varphi \Box} - C_{\varphi D} + 6 C_{\varphi G} + 4 i C_{\varphi \tilde{G}} \right).
\end{equation}

The next order in the $s_{ij}$ expansion shows a non-vanishing contribution only to the term linear in $s_{12}$, namely $c_{gghh}^{++,(10)}$. In SMEFT, this structure originates from dim-8 operators, while in HEFT it first arises at NNLO. Also, unlike in the case of the $c_{gghh}^{++,(00)}$ coefficient, there is no ``mixing'' between different orders in the EFT expansion across the two frameworks.

Finally, we shift focus towards the opposite-helicity amplitude, $+-00$. The matching results for $c_{gghh}^{+-,(00)}$ reveal leading contributions from dim-8 SMEFT operators (Eq.~\eqref{eq:match_Result_gghh_+-00_00_rational_Gf}) and NNLO HEFT operators (Eq.~\eqref{eq:match_Result_gghh_+-00_00_HEFTNNLO}), with all higher-order coefficients in the $s_{12},\,s_{13}$ expansion vanishing. In order to generate coefficients corresponding to these higher-order, one has to go beyond dim-8 SMEFT and NNLO HEFT.

In summary, all the on-shell amplitude structures appear at the expected orders in the EFT expansion, consistent with Table~\ref{tab:amp_gghh_SMEFTHEFT}. While more Wilson coefficients contribute in SMEFT than in HEFT, the underlying physics is the same, as both frameworks generate the same set of on-shell coefficients. Nevertheless, this difference can become important when considering correlations with other processes, where the larger number of SMEFT operators may lead to distinct patterns of constraints and interplay between them.

\section{Triple Higgs production}
\label{sec:HHH}

Here we turn to triple Higgs production. Since a number of subtleties arise in this case, we first illustrate them in terms of simpler example, namely the five scalar amplitude.

\subsection{Five-point amplitudes: the five-Higgs case}
\label{sec:5h_amplitudes}

The computation of five-point amplitudes from on-shell bootstrap requires an extension of the methods used so far, in particular the tree-level gluing procedure. Hence, we first study the five-Higgs amplitude since it lacks spinor structures but contains a general pole-and-residue structure. For $n-$Higgses, the non-factorizable part of the amplitude can be simply written as,
\begin{equation}
\mcM_{\text{NF}}\left(nh\right)=
\begin{tikzpicture}[baseline=(b1.base)]
\begin{feynman}[inline=b1.base]
\vertex (i1) {$h_{1}$};
\vertex[below=1cm of i1] (i2) {$h_{2}$};
\vertex[below right=0.5cm and 0.75cm of i1, blob, minimum height=0.35cm,minimum width=0.35cm] (b1) {};
\vertex[below right=0.5cm and 0.75cm of b1] (h5) { $h_n$ };
\vertex[right=0.75cm of b1] (h4) { $\cdot$ };
\vertex[above=0.1cm of h4] (h6) { $\cdot$ };
\vertex[below=0.1cm of h4] (h7) { $\cdot$ };
\vertex[above right=0.5cm and 0.75cm of b1] (h3) { $h_3$ };
\diagram* {
(i1) -- [scalar] (b1),
(i2) -- [scalar] (b1),
(b1) -- [scalar] (h5),
(b1) -- [scalar] (h4),
(b1) -- [scalar] (h6),
(b1) -- [scalar] (h7),
(b1) -- [scalar] (h3)
};
\end{feynman}
\end{tikzpicture}
=i\,c_{nh}
\end{equation}
where $c_{nh}$ is an analytical function of the generalized Mandelstam variables in the $n-$point amplitude.

A naive extension of the gluing procedure used before in which we would consider all of the possible complete on-shell three- and four-point amplitudes would lead to erroneous residues. More precisely, the five-point amplitude built from the factorizable piece of the glued-in four-point amplitude times a three-point one and the five-point amplitude built exclusively from three three-point amplitudes would contribute to the same residue. To avoid this issue, we propose extending the gluing technique to the five-point amplitude case by gluing only non-factorizable on-shell amplitudes. Hence, the factorizable five-point amplitude will be built out of non-factorizable four- and three-point on-shell amplitudes glued with one or two propagators along all the possible factorization channels. This technique reproduces the one used in the four-point amplitude case, since three-point on-shell amplitudes do not possess factorizable pieces.

In general, our five-point gluing prescription is,
\begin{equation}
    \mcM_5 = \mcM_{5,\text{NF}} - \sum_{j} \frac{\mcM_{4,\text{NF}}\ \mcM_{3}}{s_{j}-m_{j}^2} + \sum_{\substack{j,k\\ k>j}} \frac{\mcM_{3}\ \bar\mcM_{3}\ \hat\mcM_{3} }{(s_{j}-m_{j}^2)(s_{k}-m_{k}^2)}\, ,
\label{eq:5pt_gluing_prescription}
\end{equation}
where the sums run over all possible factorization channels (poles) of the amplitude under study, $\mcM_{4,\text{NF}}$ is a non-factorizable four-point on-shell amplitude, and  $\mcM_3$, $\bar\mcM_3$, and $\hat\mcM_3$ are three possibly different three-point on-shell amplitudes. The sum in the last term prevents double counting of equivalent diagrams. In the following, we show how this prescription reproduces the right residues by looking at some of the factorization channels of the five-Higgses amplitude.

\begin{figure}
    \centering
    \begin{tikzpicture}
\begin{feynman}
\vertex (i1) {$h_i$};
\vertex[below=2cm of i1] (i2) {$h_j$};
\vertex[below right=1cm and 1cm of i1, dot] (b1) {};
\vertex[right=1cm of b1, dot] (b2) {};
\vertex[below right=1cm and 1.25cm of b2] (h5) {$h_k$};
\vertex[right=1.25cm of b2] (h4) {$h_\ell$};
\vertex[above right=1cm and 1.25cm of b2] (h3) {$h_m$};
\diagram* {
  (i1) -- [scalar] (b1),
  (i2) -- [scalar] (b1),
  (b1) -- [scalar, edge label'=\(s_{ij}\)] (b2),
  (b2) -- [scalar] (h5),
  (b2) -- [scalar] (h4),
  (b2) -- [scalar] (h3)
};
\end{feynman}
\end{tikzpicture} \hspace{2cm}
\begin{tikzpicture}
\begin{feynman}
\vertex (i1) {$h_i$};
\vertex[below=2cm of i1] (i2) {$h_j$};
\vertex[below right=1cm and 1cm of i1, dot] (b1) {};
\vertex[right=1cm of b1, dot] (b2) {};
\vertex[below right=0.75cm and 1.5cm of b2] (h4) {$h_k$};
\vertex[above right=0.5cm and 1.25cm of b2, dot] (h1) {};
\vertex[above right=0.5cm and 0.75cm of h1] (h2) {$h_\ell$};
\vertex[below right=0.5cm and 0.75cm of h1] (h3) {$h_m$};
\diagram* {
  (i1) -- [scalar] (b1),
  (i2) -- [scalar] (b1),
  (b1) -- [scalar, edge label'=\(s_{ij}\)] (b2),
  (b2) -- [scalar, edge label=\(s_{\ell m}\)] (h1),
  (h1) -- [scalar] (h2),
  (h1) -- [scalar] (h3),
  (b2) -- [scalar] (h4)
};
\end{feynman}
\end{tikzpicture}
    \caption{Fundamental topologies of the factorizable contributions to the $5-$Higgses amplitude. All diagrams in App.~\ref{app:all_os_amps} can be obtained from here by adequate assignment of $i$, $j$, $k$, $\ell$, $m$, and rotations.}
    \label{fig:5h_onshell_diags}
\end{figure}
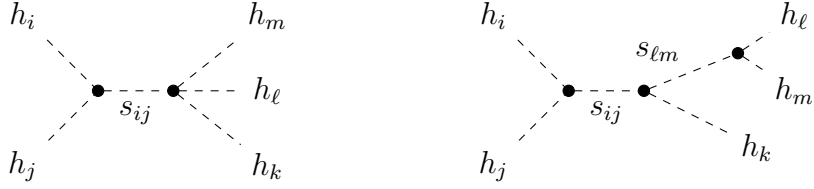

Consider the $s_{12}$ factorization channel and all the diagrams that will contribute to the corresponding residue,
\begin{equation*}
    \mcM\left(5 h\right)_{s_{12}}=
    \begin{tikzpicture}[baseline=(b1.base)]
\begin{feynman}[inline=b1.base]
\vertex (i1) {$1$};
\vertex[below=1cm of i1] (i2) {$2$};
\vertex[below right=0.5cm and 0.5cm of i1, dot] (b1) {};
\vertex[right=0.75cm of b1, dot] (b2) {};
\vertex[below right=0.5cm and 0.75cm of b2] (h1) { $5$ };
\vertex[right=0.75cm of b2] (h2) { $4$ };
\vertex[above right=0.5cm and 0.75cm of b2] (h3) { $3$ };
\diagram* {
(i1) -- [scalar] (b1),
(i2) -- [scalar] (b1),
(b1) -- [scalar, edge label' = \tiny\(s_{12}\)] (b2),
(b2) -- [scalar] (h1),
(b2) -- [scalar] (h2),
(b2) -- [scalar] (h3)
};
\end{feynman}
\end{tikzpicture} + 
\begin{tikzpicture}[baseline=(b1.base)]
\begin{feynman}[inline=h3.base]
\vertex (i1) {};
\vertex[below=1cm of i1] (i2) {};
\vertex[below right=0.5cm and 0.5cm of i1, dot] (b1) {};
\vertex[right=0.75cm of b1, dot] (b2) {};
\vertex[below right=0.33cm and 1cm of b2] (h4) {$5$};
\vertex[above right=0.25cm and 0.75cm of b2, dot] (h1) {};
\vertex[above right=0.25cm and 0.5cm of h1] (h2) {$3$};
\vertex[below right=0.25cm and 0.5cm of h1] (h3) {$4$};
\diagram* {
(i1) -- [scalar] (b1),
(i2) -- [scalar] (b1),
(b1) -- [scalar, edge label' = \tiny\(s_{12}\)] (b2),
(b2) -- [scalar, , edge label = \tiny\(s_{34}\)] (h1),
(h1) -- [scalar] (h2),
(h1) -- [scalar] (h3),
(b2) -- [scalar] (h4)};
\end{feynman}
\end{tikzpicture} +
\begin{tikzpicture}[baseline=(b1.base)]
\begin{feynman}[inline=h3.base]
\vertex (i1) {};
\vertex[below=1cm of i1] (i2) {};
\vertex[below right=0.5cm and 0.5cm of i1, dot] (b1) {};
\vertex[right=0.75cm of b1, dot] (b2) {};
\vertex[below right=0.33cm and 1cm of b2] (h4) {$4$};
\vertex[above right=0.25cm and 0.75cm of b2, dot] (h1) {};
\vertex[above right=0.25cm and 0.5cm of h1] (h2) {$3$};
\vertex[below right=0.25cm and 0.5cm of h1] (h3) {$5$};
\diagram* {
(i1) -- [scalar] (b1),
(i2) -- [scalar] (b1),
(b1) -- [scalar, edge label' = \tiny\(s_{12}\)] (b2),
(b2) -- [scalar, edge label = \tiny\(s_{35}\)] (h1),
(h1) -- [scalar] (h2),
(h1) -- [scalar] (h3),
(b2) -- [scalar] (h4)};
\end{feynman}
\end{tikzpicture} +
\begin{tikzpicture}[baseline=(b1.base)]
\begin{feynman}[inline=h3.base]
\vertex (i1) {};
\vertex[below=1cm of i1] (i2) {};
\vertex[below right=0.5cm and 0.5cm of i1, dot] (b1) {};
\vertex[right=0.75cm of b1, dot] (b2) {};
\vertex[below right=0.33cm and 1cm of b2] (h4) {$3$};
\vertex[above right=0.25cm and 0.75cm of b2, dot] (h1) {};
\vertex[above right=0.25cm and 0.5cm of h1] (h2) {$4$};
\vertex[below right=0.25cm and 0.5cm of h1] (h3) {$5$};
\diagram* {
(i1) -- [scalar] (b1),
(i2) -- [scalar] (b1),
(b1) -- [scalar, edge label' = \tiny\(s_{12}\)] (b2),
(b2) -- [scalar, edge label = \tiny\(s_{45}\)] (h1),
(h1) -- [scalar] (h2),
(h1) -- [scalar] (h3),
(b2) -- [scalar] (h4)};
\end{feynman}
\end{tikzpicture} \, 
\end{equation*}
and with our prescription we can build the relevant amplitude as,
\begin{equation}
    \begin{aligned}
           \mcM\left(5 h\right)_{s_{12}}= & -\frac{\mcM\left( 3h \right) \mcM\left(\scriptstyle{\tilde{h}_{1+2};h_3;h_4;h_5}\right)_{\text{NF}} }{s_{12}-m_h^2} + \sum_{k\in\lbrace 34,35,45\rbrace}\frac{\left(\mcM\left(3h\right)\right)^3}{s_{12}-m_h^2}\frac{1}{s_{k}-m_h^2}\\
           = & \frac{c_{3h}\ c_{4h}\left(s_{34},s_{35}\right) }{s_{12}-m_h^2}-i\ \sum_{k\in\lbrace 34,35,45\rbrace}\frac{c_{3h}^3}{s_{12}-m_h^2}\frac{1}{s_{k}-m_h^2} \, ,
    \label{eq:amp_5h_s12_facto_ch}
    \end{aligned}
\end{equation}
where $\tilde{h}_{1+2}$ is the virtual Higgs boson with momentum $p_1+p_2$ and we have used the $s_{ij}$ independence of the three-point on-shell amplitude from the identity of the involved Higgs bosons. The residue at $s_{12}=m_h^2$ is,
\begin{equation}
    \begin{aligned}
           \underset{s_{12}=m_h^2}{\text{Res}}\left(\mcM\left( 5 h \right)_{s_{12}}\right)= & - \mcM\left( 3 h\right) \mcM\left(\scriptstyle{\tilde{h}_{1+2};h_3;h_4;h_5}\right)_{\text{NF}} \Big|_{s_{12}=m_h^2} + \sum_{k\in\lbrace 34,35,45\rbrace}\frac{ \left(\mcM\left(3h \right)\right)^3 }{s_{k}-m_h^2}\Big|_{s_{12}=m_h^2} \\
           = & - \mcM\left(3h \right) \left(\mcM\left(\scriptstyle{\tilde{h}_{1+2};h_3;h_4;h_5}\right)_{\text{NF}} -  \sum_{k\in\lbrace 34,35,45\rbrace}\frac{\left(\mcM\left(3h \right)\right)^2}{s_{k}-m_h^2}\right)\Big|_{s_{12}=m_h^2}\\
           = & - \mcM\left(3h \right) \mcM\left(\scriptstyle{\tilde{h}_{1+2};h_3;h_4;h_5}\right) \, ,
    \end{aligned}
\end{equation}
where the term between parenthesis in the second line reproduces the full four-point on-shell amplitude $\mcM(\tilde{h}_{1+2};h_3;h_4;h_5)$ and hence we obtain the expected result: the product of a three-point and a four-point amplitude. Our gluing prescription ensures that Eq.~\eqref{eq:amp_5h_s12_facto_ch} is the only part of the whole five-Higgs amplitude that has a pole in $s_{12}=m_h^2$ and hence there are no further contributions to the residue. This can be extended to all factorization channels, depicted in Fig.~\ref{fig:5h_onshell_diags}, and to obtain the full amplitude reported in App.~\ref{app:all_os_amps}.

As a further cross check of the gluing prescription, we can check the collinear limit. Although two on-shell Higgs bosons cannot become exactly collinear in physical
kinematics due to their non-vanishing mass, it is nevertheless useful to consider this
formal massless (or high-energy) collinear limit as an additional
consistency check of the gluing construction in a regime where the universal
collinear factorisation properties of the S-matrix are well understood~\cite{Dixon:1996wi, Dixon:1993xd}.
In this limit, any tree-level amplitude must factorise as,
\begin{equation}
\mathcal{M}_5(\ldots,p_3,p_4,p_5)
\xrightarrow[k_\perp \to 0]{3 \parallel 4}
\mathrm{Split}_{h\to hh}(z)\;
\mathcal{M}_4(\ldots,P,p_5)\, ,
\label{eq:collinear-fact}
\end{equation}
with $P=(p_3+p_4)$ and the Sudakov parameterisation of the momenta,
\begin{align}
p_3^{\mu}= & z P^{\mu}+ k_{\perp}^{\mu}-\frac{k_{\perp}^2}{z}\frac{n^{\mu}}{(2 p_3\cdot n)}\, ,\\
p_4^{\mu}=&(1-z) P^{\mu}-  k_{\perp}^{\mu}-\frac{k_{\perp}^2}{1-z}\frac{n^{\mu}}{(2 p_3\cdot n)}\, ,
\end{align}
and $k_{\perp}\cdot P= k_{\perp}\cdot n=0$ and $n^2=1$.
This condition is trivially fulfilled with, 
\[
\mathrm{Split}_{h\to hh}(z)
=
\frac{c_{3h}}{(p_3+p_4)^2}.
\]
In analogy, also the double collinear limit of gluing the three three-point Higgs interactions can be shown. 

The on-shell amplitude has a remarkable feature: in terms with two propagators, all kinematic dependence is contained in the propagators because $c_{3h}$ is constant. In terms with one propagator, the kinematic dependence is dictated by the propagator and the expansion of $c_{4h}$ in powers of the Mandelstam variables. The EFT amplitudes, from either SMEFT or HEFT, might seem to depart from this behavior due to momentum-dependent three- and four-Higgs vertices. However, it is possible to show that the terms in the EFT amplitude that would not be reproduced by the on-shell amplitude cancel among different diagrams or can be rewritten such that the on-shell bootstrapped and EFT amplitudes have exactly the same dependence on the kinematics (see App.~\ref{app:cancellations_5_higgses}  for details). Furthermore, we checked analytically that our on-shell bootstrapped amplitude reproduces the poles and residues of the EFT amplitude once the three- and four-Higgs matching results are applied.
Thus, our five-Higgs on-shell bootstrapped amplitude can be matched exactly to the amplitudes obtained from SMEFT or HEFT to obtain the relation between $c_{5h}$ and the WCs of the chosen EFT. The matching technique and results will be discussed in section~\ref{sec:gghhh_matching}.

In this section, we did not need to worry about gauge (color) indices. Their addition to the gluing procedure is trivial: if the intermediate propagating particle carries a gauge index, we multiply the propagator by the adequate Kronecker delta and contract the index with the corresponding ones in the subamplitudes. An example of this will be seen at the end of the next section.

Finally, our gluing approach is inspired and closely related to the bootstrapping approach in~\cite{AccettulliHuber:2021uoa}, which we will call ``AHDA method'' from now on. While the AHDA method applies to any higher-point tree-level amplitude in generic EFTs, our method is proposed only for five-point amplitudes and its extension to higher-point amplitudes is left for future work. Our method does not assume any underlying theory, only a given particle content, and uses as building blocks (trivalent graphs in the language of the AHDA method) all the non-factorizable non-vanishing lower-point on-shell amplitudes allowed by that particle content. Our gluing approach does not have free coefficients to match afterwards, hence skipping the last two steps of the AHDA method. Finally, we use the gluing only for the factorizable part and always add the most general non-factorizable amplitude since we are not thinking of a particular basis of EFT operators to a given order when building the amplitude. On the other hand, our method is not directly suitable for cases with minimal couplings such as all-gluon amplitudes, while the AHDA method is. We leave these cases for future studies.

\subsection{On-shell amplitudes for triple Higgs production}
\label{sec:gghhh_onshell_bootstrap}

As before, we begin by discussing the non-factorizable pieces that can be obtained by bootstrapping. In the case of the $++000$ helicity configuration, we find two spinor-helicity structures that are independent in the sense that can not be related by multiplying them with generalized Mandelstam variables. We checked explicitly that these two structures are enough to generate the \textit{kinematical basis} up to dimension-8 inclusive~\cite{DeAngelis:2022qco}, while we leave for future work the exploration of higher-dimensional bases, where there could be new structures, including even the Levi-Civita tensor.

Thus, the general on-shell amplitude can be written as,
\begin{align}
\begin{tikzpicture}[baseline=(b1.base)]
\begin{feynman}[inline=(b1.base)]
\vertex (i1) {$g^{a,+}_1$};
\vertex[below=2cm of i1] (i2) {$g^{b,+}_2$};
\vertex[below right=1cm and 1cm of i1, blob] (b1) {};
\vertex[below right=1cm and 1cm of b1] (h3) {$h_5$};
\vertex[right=1cm of b1] (h4) {$h_4$};
\vertex[above right=1cm and 1cm of b1] (h5) {$h_3$};
\diagram* {
(i1) -- [gluon] (b1),
(i2) -- [gluon] (b1),
(b1) -- [scalar] (h3),
(b1) -- [scalar] (h4),
(b1) -- [scalar] (h5)};
\end{feynman}
\end{tikzpicture}  & =\mcM\left(g_{1}^{a,+}; g_{2}^{b,+}; h_{3}; h_{4}; h_{5}\right)_{\text{NF}} = i\,\delta^{ab}\, c_{gghhh}^{++,\, (1)} \squaket{1}{2}^2    \label{amp:gghhh:OS:++} \\
        & + i\,\delta^{ab}\, c_{gghhh}^{++,\, (2)}\left( {\squa{1}{}{}\mathbf{3}\mathbf{4}\squet{2}{}{}}  {\squa{1}{}{} \mathbf{4} \mathbf{3} \squet{2}{}{}} +
   {\squa{1}{}{}\mathbf{3}\mathbf{5}\squet{2}{}{}}  {\squa{1}{}{} \mathbf{5} \mathbf{3} \squet{2}{}{}} +
   {\squa{1}{}{}\mathbf{4}\mathbf{5}\squet{2}{}{}}  {\squa{1}{}{} \mathbf{5} \mathbf{4} \squet{2}{}{}}\right), \nonumber
    \end{align}
where $c_{gghhh}^{++,\, (1),(2)}$ are analytical functions of the generalized Mandelstam variables\footnote{We checked the independence of the two structures with the algorithm in~\cite{DeAngelis:2022qco} as implemented in the package~\texttt{MassiveGraphs}, available on \href{https://github.com/StefanoDeAngelis/MassiveEFT-Operators}{GitHub}. We are very grateful to Stefano De Angelis for his help with this check.}. 

The spinor-helicity structure $({\squa{1}{}{}\mathbf{3}\mathbf{4}\squet{2}{}{}}{\squa{1}{}{}\mathbf{4}\mathbf{3}\squet{2}{}{}}+...)$ is a hallmark of the five-point amplitude since momentum conservation in the four-point case would have reduced it to the first structure. There is a similar spinor structure, $\sum_{\substack{i,j=3\\ i\neq j}}^{5}{\squa{1}{}{}\mathbf{i}\mathbf{j}\squet{2}{}{}}^2$, to which is linearly related\footnote{The relation is $ \sum_{\substack{i,j=3\\ i\neq j}}^{5}{\squa{1}{}{}\mathbf{i}\mathbf{j}\squet{2}{}{}}^2 = -2 ( {\squa{1}{}{}\mathbf{3}\mathbf{4}\squet{2}{}{}}  {\squa{1}{}{} \mathbf{4} \mathbf{3} \squet{2}{}{}} +
   {\squa{1}{}{}\mathbf{3}\mathbf{5}\squet{2}{}{}}  {\squa{1}{}{} \mathbf{5} \mathbf{3} \squet{2}{}{}} +
   {\squa{1}{}{}\mathbf{4}\mathbf{5}\squet{2}{}{}}  {\squa{1}{}{} \mathbf{5} \mathbf{4} \squet{2}{}{}} ) + 4 \, ( (p_3 \cdot p_4)^2 + (p_3\cdot p_5)^2 + (p_4\cdot p_5)^2 ) {\squaket{1}{2}}^2.$}. 
Either of these structure can be used to form the basis of the amplitude, but we chose the one that facilitated the matching to higher-dimensional SMEFT and HEFT operators. 

When considering this amplitude beyond dimension-8 in energy, one should generalize the expressions of the second term in Eq.~\eqref{amp:gghhh:OS:++} since symmetrizing separately the spinor helicity structure and the Mandelstam polynomial is too restrictive. Instead, one should write $c_{gghhh}^{++,\, (2,1)}{\squa{1}{}{}\mathbf{3}\mathbf{4}\squet{2}{}{}}  {\squa{1}{}{} \mathbf{4} \mathbf{3} \squet{2}{}{}} +
c_{gghhh}^{++,\, (2,2)}   {\squa{1}{}{}\mathbf{3}\mathbf{5}\squet{2}{}{}}  {\squa{1}{}{} \mathbf{5} \mathbf{3} \squet{2}{}{}} +
c_{gghhh}^{++,\, (2,3)}   {\squa{1}{}{}\mathbf{4}\mathbf{5}\squet{2}{}{}}  {\squa{1}{}{} \mathbf{5} \mathbf{4} \squet{2}{}{}}$, expand the Mandelstam polynomials $c_{gghhh}^{++,\,(2,i)}$ to the desired order and then symmetrize. This will correlate the $c_{gghhh}^{++,\,(2,i)}$s which therefore should not be considered as three independent coefficients, but one.

In $n$-point amplitudes, with $n>5$, further structures could appear by inserting additional independent momenta to obtain ${\squa{1}{}{}\mathbf{3}\mathbf{4}...\mathbf{k}\squet{2}{}{}}^2$ for $k=5,...,n-1$.

The case of opposite-helicity gluons, $+-000$, yields,
\begin{equation}
    \begin{aligned}
\begin{tikzpicture}[baseline=(b1.base)]
\begin{feynman}[inline=(b1.base)]
\vertex (i1) {$g^{a,+}_1$};
\vertex[below=2cm of i1] (i2) {$g^{b,-}_2$};
\vertex[below right=1cm and 1cm of i1, blob] (b1) {};
\vertex[below right=1cm and 1cm of b1] (h3) {$h_5$};
\vertex[right=1cm of b1] (h4) {$h_4$};
\vertex[above right=1cm and 1cm of b1] (h5) {$h_3$};
\diagram* {
(i1) -- [gluon] (b1),
(i2) -- [gluon] (b1),
(b1) -- [scalar] (h3),
(b1) -- [scalar] (h4),
(b1) -- [scalar] (h5)};
\end{feynman}
\end{tikzpicture}       &=\mcM\left(g_{1}^{a,+}; g_{2}^{b,-}; h_{3}; h_{4}; h_{5} \right)_{\text{NF}} \\
        &=  
        i\, \delta^{ab}\, \left( c_{gghhh}^ {+-,(1)} \left(\squa{1}{}{}\mathbf{3}\tret{2}{}{}\right)^2 + c_{gghhh}^ {+-,(2)}  \left(\squa{1}{}{}\mathbf{4}\tret{2}{}{}\right)^2 + c_{gghhh}^ {+-,(3)}  \squa{1}{}{}\mathbf{3}\tret{2}{}{} \squa{1}{}{}\mathbf{4}\tret{2}{}{} \right),
    \end{aligned}
    \label{amp:gghhh:OS:+-}
\end{equation}
where the three coefficients $c_{gghhh}^ {+-,(i)}$ can be expanded in powers of the generalized Mandelstam variables. At each order, the Mandelstam polynomials $c_{gghhh}^ {+-,(i)}$ will be correlated among them to satisfy Bose symmetry. We have checked explicitly that these three structures are capable of generating the kinematical basis for the amplitude up to dimension-8 in energy, i.e. second order in the Mandelstam expansion.

For our purposes, however, it will be enough to consider only the zeroth order in the Mandelstam expansion. To this order, Bose symmetry forces the amplitude into the simpler form,
\begin{equation}
    \mcM\left(g_{1}^{a,+}; g_{2}^{b,-}; h_{3}; h_{4}; h_{5}\right)_{\text{NF}} =i\,\delta^{ab}\, c_{gghhh}^{+-}  \left(\left(\squa{1}{}{}\mathbf{3}\tret{2}{}{}\right)^2 + \left(\squa{1}{}{}\mathbf{4}\tret{2}{}{}\right)^2 + \squa{1}{}{}\mathbf{3}\tret{2}{}{} \squa{1}{}{}\mathbf{4}\tret{2}{}{}  \right) +\mcO(E^6).
\end{equation}
This expression can be made explicitly Bose symmetric via momentum conservation,
\begin{equation}
    \mcM\left(g_{1}^{a,+}; g_{2}^{b,-}; h_{3}; h_{4}; h_{5}\right)_{\text{NF}} =i\,\delta^{ab}\,  c_{gghhh}^{+-}  \left({\squa{1}{}{}\mathbf{3}\tret{2}{}{}}^2 + {\squa{1}{}{}\mathbf{4}\tret{2}{}{}}^2 + {\squa{1}{}{}\mathbf{5}\tret{2}{}{}}^2  \right)+\mcO(E^6),
     \label{amp:gghhh:OS:+-_simple}
\end{equation}
where we have absorbed a factor of $1/2$ inside the free coefficient.
For simplicity, we will work with the latter expression from now on.
Finally, the spinor-helicity structure in Eq.~\eqref{amp:gghhh:OS:+-_simple} can naively be generalized to $\sum_{i=3}^{n} \left(\squa{1}{}{}\mathbf{i}\tret{2}{}{}\right)^2$ for the case of a $n$-point $gg\to h^{n-2}$ amplitude. However, further structures could appear in such amplitudes and we leave their investigation for future work.

 It is instructive to compare our findings with the result of bootstrapping the Lorentz structure of the amplitude before contracting with the polarization vectors. By using axial gauge, momentum conservation and on-shell relations, and up to Levi-Civita tensor contributions, the Lorentz structure of the amplitude before Bose symmetrization should be,
\begin{equation}
    \mcM_{gghhh}^{\mu\nu} = a_1\, \eta^{\mu\nu} + a_2\, p_3^\mu\, p_3^\nu + a_3\, p_3^\mu\, p_4^\nu,
\end{equation}
where the coefficients $a_i$ are Lorentz invariant and depend on the generalized Mandelstam variables. Hence, we find three linearly independent structures, which agrees with our on-shell result since before symmetrizing we have the spinor structures $\squaket{1}{2}^2$, ${\squa{1}{}{} \mathbf{3} \mathbf{4}\squet{2}{}{}}^2$, and ${\squa{1}{}{}\mathbf{3}\tret{2}{}{}}^2$. We leave the general proof that these spinor structures generate the basis at all energy orders, up to Levi-Civita terms, for future work.

%%%%%%%%%%%%%%%%%%%%%%%%%%%%%%%%%%%%%%%%%%%%%%%%%%%%%%%%%%%%%%%%%%%%%%%%%%%%%%%%%%%
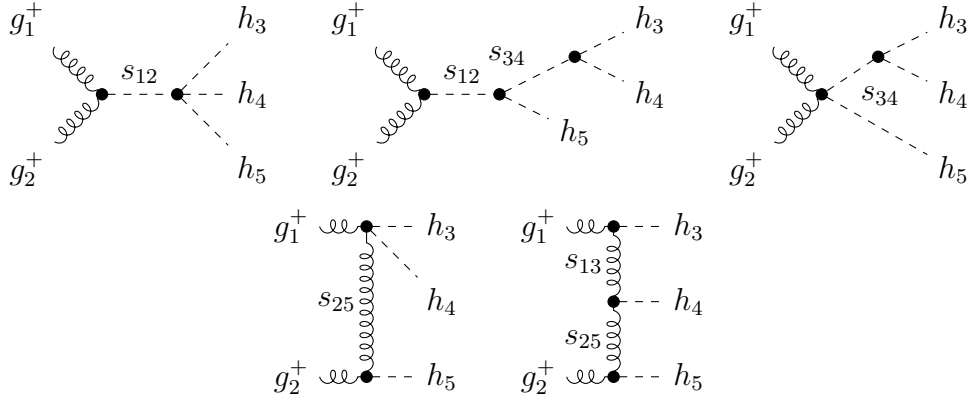
\begin{figure}[h!]
    \centering
\begin{tikzpicture}
\begin{feynman}
            \vertex (v1) at (0,0) [dot] {};
            \vertex (v2) at (1.0,0.0) [dot] {};
            \vertex (g1) at (-1.0,-1.0) {\( g^{+}_2 \)};
            \vertex (g2) at (-1.0,1.0) {\( g^{+}_1 \)};
            \vertex (h1) at (2.0,1.0) {\( h_3 \)};
            \vertex (h2) at (2.0,0) {\( h_4 \)};
            \vertex (h3) at (2.0,-1.0) {\( h_5 \)};
            \diagram* {
                (g1) -- [gluon] (v1),
                (g2) -- [gluon] (v1),
                (v1) -- [scalar, edge label=\(s_{12}\)] (v2),
                (v2) -- [scalar] (h1),
                (v2) -- [scalar] (h2),
                (v2) -- [scalar] (h3),
            };
\end{feynman}
\end{tikzpicture}
\hspace{3mm}
\begin{tikzpicture}
\begin{feynman}
\vertex (i1) {$g^{+}_{1}$};
\vertex[below=2cm of i1] (i2) {$g^{+}_{2}$};
\vertex[below right=1cm and 1cm of i1, dot] (b1) {};
\vertex[right=1cm of b1, dot] (b2) {};
\vertex[below right=0.5cm and 1cm of b2] (h4) {$h_5$};
\vertex[above right=0.5cm and 1cm of b2, dot] (h1) {};
\vertex[above right=0.5cm and 1cm of h1] (h2) {$h_3$};
\vertex[below right=0.5cm and 1cm of h1] (h3) {$h_4$};
\diagram*{
(i1) -- [gluon] (b1),
(i2) -- [gluon] (b1),
(b1) -- [scalar, edge label=\(s_{12}\)] (b2),
(b2) -- [scalar, edge label=\(s_{34}\)] (h1),
(h1) -- [scalar] (h2),
(h1) -- [scalar] (h3),
(b2) -- [scalar] (h4) };
\end{feynman}
\end{tikzpicture}
\hspace{3mm}
\begin{tikzpicture}
\begin{feynman}
\vertex (i1) {$g^{+}_1$};
\vertex[below=2cm of i1] (i2) {$g^{+}_{2}$};
\vertex[below right=1cm and 1cm of i1, dot] (b1) {};
\vertex[above right=0.5cm and 0.75cm of b1, dot] (b2) {};
\vertex[right=2.75cm of i2] (h3) {$h_5$};
\vertex[above=1cm of h3] (h4) {$h_4$};
\vertex[above=1cm of h4] (h5) {$h_3$};
\diagram* {
(i1) -- [gluon] (b1),
(i2) -- [gluon] (b1),
(b2) -- [scalar, edge label=\(s_{34}\)] (b1),
(b1) -- [scalar] (h3),
(b2) -- [scalar] (h4),
(b2) -- [scalar] (h5)};
\end{feynman}
\end{tikzpicture}
\\
\begin{tikzpicture}
\begin{feynman}
\vertex (i1) {$g^{+}_1$};
\vertex[below=2cm of i1] (i2) {$g^{+}_2$};
\vertex[right=1cm  of i1, dot] (b1) {};
\vertex[right=1cm  of i2, dot] (b2) {};
\vertex[right=1cm of b2] (h3) {$h_5$};
\vertex[below right=1cm and 1 cm of b1] (h4) {$h_4$};
\vertex[right=1cm of b1] (h5) {$h_3$};
\diagram*{
(i1) -- [gluon] (b1),
(i2) -- [gluon] (b2),
(b2) -- [gluon, edge label=\(s_{25}\)] (b1),
(b2) -- [scalar] (h3),
(b1) -- [scalar] (h4),
(b1) -- [scalar] (h5)
};
\end{feynman}
\end{tikzpicture}
\hspace{3mm}
\begin{tikzpicture}
\begin{feynman}
\vertex (i1) {$g^{+}_1$};
\vertex[below=2cm of i1] (i2) {$g^{+}_2$};
\vertex[right=1cm of i1, dot] (b1) {};
\vertex[right=1cm of i2, dot] (b2) {};
\vertex[above=1cm of b2, dot] (b3) {};
\vertex[right=1cm of b2] (h3) {$h_5$};
\vertex[right=1cm of b3] (h4) {$h_4$};
\vertex[right=1cm of b1] (h5) {$h_3$};
\diagram* {
(i1) -- [gluon] (b1),
(i2) -- [gluon] (b2),
(b2) -- [gluon, edge label=\(s_{25}\)] (b3),
(b3) -- [gluon, edge label=\(s_{13}\)] (b1),
(b2) -- [scalar] (h3),
(b3) -- [scalar] (h4),
(b1) -- [scalar] (h5)};
\end{feynman}
\end{tikzpicture}
    \caption{Factorization channels of $\mcM\left(g_{1}^{a,+}; g_{2}^{b,+}; h_{3}; h_{4}; h_{5}\right)$, up to permutations of final state particles. Each dot is a non-factorizable on-shell amplitude.}
    \label{fig:gghhh_facto_topologies_++}
\end{figure}
%%%%%%%%%%%%%%%%%%%%%%%%%%%%%%%%%%%%%%%%%%%%%%%%%%%%%%%%%%%%%%%%%%%%%%%%%%%%%%%%%
%%%%%%%%%%%%%%%%%%%%%%%%%%%%%%%%%%%%%%%%%%%%%%%%%%%%%%%%%%%%%%%%%%%%%%%%%%%%%%%%%%%

\begin{figure}
    \centering
\begin{tikzpicture}
\begin{feynman}
\vertex (i1) {$g^{+}_1$};
\vertex[below=2cm of i1] (i2) {$g^{-}_{2}$};
\vertex[below right=1cm and 1cm of i1, dot] (b1) {};
\vertex[above right=0.5cm and 0.75cm of b1, dot] (b2) {};
\vertex[right=2.75cm of i2] (h3) {$h_5$};
\vertex[above=1cm of h3] (h4) {$h_4$};
\vertex[above=1cm of h4] (h5) {$h_3$};
\diagram* {
(i1) -- [gluon] (b1),
(i2) -- [gluon] (b1),
(b2) -- [scalar, edge label=\(s_{34}\)] (b1),
(b1) -- [scalar] (h3),
(b2) -- [scalar] (h4),
(b2) -- [scalar] (h5)};
\end{feynman}
\end{tikzpicture}
\hspace{3mm}
\begin{tikzpicture}
\begin{feynman}
\vertex (i1) {$g^{+}_1$};
\vertex[below=2cm of i1] (i2) {$g^{-}_2$};
\vertex[right=1cm  of i1, dot] (b1) {};
\vertex[right=1cm  of i2, dot] (b2) {};
\vertex[right=1cm of b2] (h3) {$h_5$};
\vertex[below right=1cm and 1 cm of b1] (h4) {$h_4$};
\vertex[right=1cm of b1] (h5) {$h_3$};
\diagram*{
(i1) -- [gluon] (b1),
(i2) -- [gluon] (b2),
(b2) -- [gluon, edge label=\(s_{25}\)] (b1),
(b2) -- [scalar] (h3),
(b1) -- [scalar] (h4),
(b1) -- [scalar] (h5)
};
\end{feynman}
\end{tikzpicture}
\hspace{3mm}
\begin{tikzpicture}
\begin{feynman}
\vertex (i1) {$g^{+}_1$};
\vertex[below=2cm of i1] (i2) {$g^{-}_2$};
\vertex[right=1cm  of i1, dot] (b1) {};
\vertex[right=1cm  of i2, dot] (b2) {};
\vertex[right=1cm of b1, dot] (b3) {};
\vertex[right=2cm of b2] (h3) {$h_5$};
\vertex[below right=1cm and 2 cm of b1] (h4) {$h_4$};
\vertex[right=1cm of b3] (h5) {$h_3$};
\diagram*{
(i1) -- [gluon] (b1),
(i2) -- [gluon] (b2),
(b2) -- [gluon, edge label=\(s_{25}\)] (b1),
(b1) -- [scalar, edge label=\(s_{34}\)] (b3),
(b2) -- [scalar] (h3),
(b3) -- [scalar] (h4),
(b3) -- [scalar] (h5)
};
\end{feynman}
\end{tikzpicture}
\caption{Factorization channels of $\mcM\left(g_{1}^{a,+}; g_{2}^{b,-}; h_{3}; h_{4}; h_{5}\right)$, up to permutations of final state particles. Each dot is a non-factorizable on-shell amplitude. }
\label{fig:gghhh_facto_topologies_+-}
\end{figure}
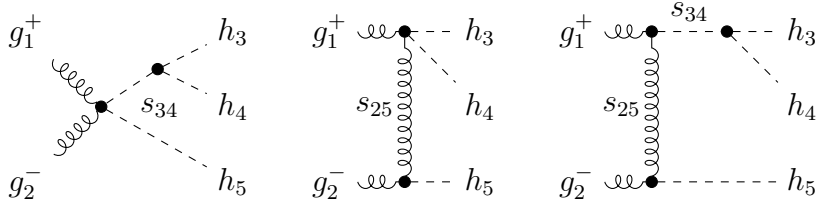

The factorizable pieces of these amplitudes are built with the gluing method described in Eq.~\eqref{eq:5pt_gluing_prescription}. For the $++000$ helicity configuration, we identify the factorization channels presented in Fig.~\ref{fig:gghhh_facto_topologies_++}, up to permutations of final state Higgs bosons, and where each black dot represents a non-factorizable three- or four-point amplitude.
In the case of $+-000$, there are only three factorization channels, depicted in Fig.~\ref{fig:gghhh_facto_topologies_+-}. The reduced number of topologies can be traced back to the vanishing of $\mcM\left(g_1^{a,+};g_2^{b,-};h\right)$.

We provide the full expression for all the factorizable amplitudes in terms of lower-point amplitudes in App.~\ref{app:all_os_amps}. Here, we explain their construction via two examples. First, consider the $s_{12}$ factorization channel of the $++000$ helicity configuration, top-left diagram in Fig.~\ref{fig:gghhh_facto_topologies_++}. This contribution is,
\begin{equation}
    \mcM \left(g_{1}^{a,+}; g_{2}^{b,+}; h_{3}; h_{4}; h_{5}\right)_{\text{F}1} = -\frac{\mcM\left( g_{1}^{+};g_{2}^{+};\tilde{h}_{1+2} \right) \mcM\left(\tilde{h}_{1+2};h_3;h_4;h_5\right)}{s_{12}-m_h^2} \, ,
\end{equation}
where $\tilde h_{1+2}$ is the virtual Higgs with momentum $k=p_1 +p_2$. Since $\tilde h_{1+2}$ must be considered as incoming in each of the two subamplitudes, it enters in $\mcM(\tilde{h}_{1+2};h_3;h_4;h_5)$ and $\mcM( g_{1}^{a,+};g_{2}^{b,+};\tilde{h}_{1+2} )$ with momentum $k$ or $-k$, respectively. In this case, the spinor structures and momentum dependencies allow us to completely neglect $k$ when evaluating the previous expression,
\begin{equation}
    \mcM \left(g_{1}^{a,+}; g_{2}^{b,+}; h_{3}; h_{4}; h_{5}\right)_{\text{F}1} = \delta^{ab}\, {\squaket{1}{2}}^2\ \frac{ c_{ggh} \ c_{4h}\left(s_{34},s_{35}\right)  }{s_{12}-m_h^2} \,.
\end{equation}
The construction of the topology with two Higgs propagators entering $++000$, top-center diagrams in Fig.~\ref{fig:gghhh_facto_topologies_++}, follows immediately,
\begin{equation}
\begin{aligned}
    \mcM \left(g_{1}^{a,+}; g_{2}^{b,+}; h_{3}; h_{4}; h_{5}\right)_{\text{F}2} = & \frac{\mcM\left( g_{1}^{a,+};g_{2}^{b,+};\tilde{h}_{1+2} \right) \mcM\left( \tilde{h}_{1+2},h_5,\tilde{h}_{3+4} \right)\mcM\left( \tilde{h}_{3+4},h_3,h_4 \right)}{(s_{12}-m_h^2)(s_{34}-m_h^2)} \\ & + \text{ perm.},\\
    = & -i\ \delta^{ab}\, {\squaket{1}{2}}^2\ \frac{ c_{ggh}^{++} \ (c_{hhh})^2  }{(s_{12}-m_h^2)\ (s_{34}-m_h^2)} + \text{ perm.} \,,
\end{aligned}
\end{equation}
where perm. stands for the result of exchanging the final-state Higgses.

The $s_{25}$ factorization channel of the $+-000$ amplitude can be written as,
\begin{equation}
    \mcM \left(g_{1}^{a,+}; g_{2}^{b,-}; h_{3}; h_{4}; h_{5}\right)_{\text{F}4} = -\frac{ \sum_{c,d=1}^{8}\mcM\left( g_{1}^{a,+};\tilde{g}_{2+5}^{c,+}; h_3; h_4 \right) \delta^{cd} \mcM\left(g_{2}^{b,-};\tilde{g}_{2+5}^{d,-}; h_5\right)}{s_{25}} + \text{ perm.} \, ,
\end{equation}
where $\tilde{g}_{2+5}$ is the virtual gluon that enters the subamplitudes with opposite momentum and helicity. This can be evaluated to,
\begin{equation}
    \begin{aligned}
        \mcM \left(g_{1}^{a,+}; g_{2}^{b,-}; h_{3}; h_{4}; h_{5}\right)_{\text{F}4} = & \delta^{ab}\ {\squaket{1}{-(2+5)}}^2 \ {\traket{2+5}{2}}^2\ \frac{ c_{ggh}^{--0}\ c_{gghh}^{++00}\left(s_{13},s_{14}\right) }{s_{25}} + \text{ perm.}, \\
        = & -\delta^{ab}\  {\squa{1}{}{} 2+\mathbf{5} \tret{2}{}{}}^2 \frac{(c_{ggh}^{++})^*\ c_{gghh}^{++00}\left(s_{13},s_{14}\right) }{s_{25}} + \text{ perm.}\,.
    \end{aligned}
\end{equation}
In the last line we used $-(c_{ggh}^{++})^* = c_{ggh}^{--0}$ and $\squet{-k}{}{}\tra{k}{}{}=-\bar k$. Notice that the spinor products $\squaket{1}{-(2+5)}$ and $\traket{2+5}{2}$ initially assumed $s_{25}=0$ since they arise from on-shell amplitudes. However, once combined and simplified we obtain the structure $\squa{1}{}{} 2+\mathbf{5} \tret{2}{}{}$ that can be easily evaluated in the off-pole region, $s_{25}\neq 0$, by evaluating $\squa{1}{}{} \gamma^{\mu} \tret{2}{}{}$ and contracting it with $(p_2+p_5)_\mu$. In this case, a further simplification is allowed: $ \squa{1}{}{} 2+\mathbf{5} \tret{2}{}{} = \squa{1}{}{} \mathbf{5} \tret{2}{}{}$, since $p_2$ is an on-shell massless momenta.
The final expression allows the full bootstrapped on-shell amplitude to reproduce the behavior of the EFT amplitudes both on- and off-pole. The remaining topologies can be computed in the same way. Finally, we have also explicitly verified that all the $gghhh$ amplitudes constructed here remain finite in the collinear limit, as anticipated (see also section~\ref{sec:5h_amplitudes}).

\subsection{Matching to SMEFT and HEFT for triple Higgs production}
\label{sec:gghhh_matching}

Now we turn to matching our five-point amplitudes to SMEFT and HEFT. In Table~\ref{tab:amp_gghhh_SMEFTHEFT} we summarize at which orders of the EFT expansion individual contributions are expected to arise from naive dimensional analysis. We perform the matching numerically as explained in section~\ref{sec:matching_SMEFT_HEFT_3_4pt}. In the case of the five-Higgses amplitude, the absence of spinor structures facilitates performing analytical matching, which we used to validate our numerical matching procedure.

\renewcommand{\arraystretch}{1.25}
\begin{table}[h]
    \begin{adjustbox}{width=\textwidth,max width=\textwidth,max totalheight=\textheight,keepaspectratio}
    \centering
    \begin{tabular}{c|c|c|c|c|c|c}
    Amplitude & Helicity & Spinor structure & Coeff. & Dimension & Minimal SMEFT order & Minimal HEFT order  \\
    \hline
    \multicolumn{7}{c}{ Five-point}\\
    \hline
    $hh\to hhh$   & - & - & $c_{5h}$ & $0$ & $6\,(v/\Lambda^2)$ & $\text{LO}$ \\
    \hline
    \multirow{3}{*}{$gg\to hhh$}   & $++$ & $\squaket{1}{2}^2$ & $c_{gghhh}^{++,(1)}$ & $-3\,(1/\bar\Lambda^3)$ & $8\,(v/\Lambda^4)$ & $\text{NLO}^*$ \\
      & $++$ & ${\squa{1}{}{}\mathbf{3}\mathbf{4}\squet{2}{}{}}^2$ & $c_{gghhh}^{++,(2)}$ & $-7\,(1/\bar\Lambda^7)$ & $12\,(v/\Lambda^8)$ & $  \text{N$^3$LO}^*$ \\
       & $+-$ & $ {\squa{1}{}{}\mathbf{3}\tret{2}{}{}}^2$ & $c_{gghhh}^{+-}$ & $-5\,(1/\bar\Lambda^5)$ & $10\,(v/\Lambda^6)$ & $\text{NNLO}^*$ \\
    \end{tabular}
    \end{adjustbox}
    \caption{Summary of on-shell coefficients and dimensions for $gg\to hhh$, $hh\to hhh$ amplitudes. $^*$ where we applied the power counting assuming that one order of $\alpha_s$ is factored out, i.e. that our HEFT operators are written as $\frac{\alpha_s}{\pi}\partial^m h^n G_{\mu\nu}G^{\mu\nu}$.
    For the sake of brevity, we represent the spinor structure associated to $c_{gghhh}^{++,(2)}$  by only ${\squa{1}{}{}\mathbf{3}\mathbf{4}\squet{2}{}{}}^2$, see main text for details.
    }
    \label{tab:amp_gghhh_SMEFTHEFT}
\end{table}

We begin by discussing the results for the five-Higgs amplitude.
We choose to expand $c_{5h}$ in powers of $s_{12}$, $s_{13}$, $s_{14}$, $s_{23}$ and $s_{24}$ with the following normalization,
\begin{equation}
    c_{5h}=\frac{1}{\bar\Lambda}\sum_{i,j,k,l,m=0} \frac{c_{5h}^{(ijklm)}}{\bar\Lambda^{\,2\left(i+j+k+l+m\right)}} s_{12}^{i}\ s_{13}^{j}\ s_{14}^{k}\ s_{23}^{l}\ s_{24}^{m} \, .
    \label{eq:expansion_Mandelstams_5h}
\end{equation}
We will consider only terms up to quadratic order in Mandelstam variables, i.e. order $1/\bar\Lambda^{5}$. 

The full results of the matching of three-, four- and five-point Higgs amplitudes can be found in App.~\ref{app:full_matching_res}. Here, we comment on some of their salient features. First of all, in the case of the three-point amplitude (see App~\ref{appendix:hhh}), SMEFT corrections are generated at the dim-6 and dim-8 order and originate from the Higgs potential modifications and operators affecting canonical normalization of the kinetic terms, as seen in Eq.~\eqref{eq:match_ons_smeft_hhh_Gf}. In contrast, the only modification in HEFT can be found at the LO, with the Higgs trilinear coupling modifier $\Delta\lambda_3$, see Eq.~\eqref{eq:match_ons_smeft_hhh_HEFT}. Then, SMEFT generates contributions to the four-point Higgs vertex that are constant (dim-6 and dim-8), linear, and quadratic (dim-8) in the Mandelstam variables (see Eq.~\eqref{eq:match_Result_hhhh_rational_SMEFT_Gf}), whereas in HEFT these are generated at LO and NLO, respectively (Eq.~\eqref{eq:match_Result_hhhh_rational_HEFT}). Here, we can note two interesting features. In the case of linear terms, both coefficients are equal, which reflects the underlying Bose symmetry. Also, in the limit of $m_h\rightarrow 0$ they vanish, in line with Ref.~\cite{Shadmi:2018xan}. Similarly, on-shell coefficients of quadratic terms are again equal, although non-vanishing in the massless limit.

Finally, in the case of five-point coefficient $c_{5h}^{(ijklm)}$,  SMEFT at dim-8 only generates the constant and linear in Mandelstam variables on-shell coefficients, as expected from the expansion order (see Eq.~\eqref{eq:match_Result_hhhhh_rational_SMEFT_Gf}). Moreover, linear terms do not get contributions from operators that generate the five-Higgs vertex, but from operators that generate a four-Higgses vertex with momentum dependence. The contribution of these operators is only partially accounted for via the four-point coefficients $c_{4h}^{(ij)}$ and the remaining contribution is absorbed by the linear five-point coefficients via the mechanism shown in App.~\ref{app:full_matching_importance}.
Instead, NNLO HEFT generates all the coefficients up to quadratic order (see Eq.~\eqref{eq:match_Result_hhhhh_rational_HEFT}). The uniform presence of $G_F$ in all the terms suggests the choice $\bar\Lambda=G_F^{-1/2}$ as the HEFT cutoff scale. The combination of the WCs $C_{DH}^{(1)}\ {C'}_{DH}^{(1)}$ appears in the HEFT contribution to all the on-shell coefficients.
Thus, NNLO HEFT generates a richer kinematic structure in the amplitude than dim-8 SMEFT, both by generating all the linear on-shell coefficients as well as the quadratic ones. We expect SMEFT to generate those at dim-10 and we leave its verification for future work.

Turning our attention towards the $gg\to hhh$ on-shell amplitudes presented in the previous section, which we match to the amplitudes computed in dim-8 SMEFT and in NNLO HEFT with the numerical procedure used before. The on-shell coefficients  $c_{gghhh}^{+\pm,(i)}$ are expanded in powers of the Mandelstam variables as in Eq.~\eqref{eq:expansion_Mandelstams_5h}, up to quadratic order, with appropriate adjustments of the overall scale power. The precise expansion used for each coefficient can be found in appendix~\ref{app:5pt_os_amps_gghhh}.
The full matching results for $gg\to hhh$ are  presented in App.~\ref{appendix:gghhh} and we analyze their main features in the following.

Closer examination of the $++$ helicity configuration reveals that, as expected from Table~\ref{tab:amp_gghhh_SMEFTHEFT}, SMEFT at dim-8 generates only the first, constant on-shell coefficient $c_{gghhh}^{++,(1)}$, receiving its contribution solely from the four-point $gg\to hh$ vertices with momentum dependence. 
In HEFT, NLO and NNLO contributions generate constant and linear contributions to the $c_{gghhh}^{++,(1)}$ on-shell coefficient and are combination of operators contributing to ``genuine'' five-point $gg\to hhh$ contact interaction, and from four-point $gg\to hh$ vertices with momentum dependence. 

In the case of $+-$ configuration, the only non-zero contribution to the $c_{gghhh}^{+-}$ on-shell coefficient is to its constant piece, and arises from NNLO HEFT. We discuss it in more detail in the next section.

\subsubsection{Effects of higher order EFT operators}
\label{sec:gghhh_matching_higher_dim_ops}
Up to now, we focused solely on the effects of at most dim-8 (SMEFT) and NNLO (HEFT) operators on the studied processes. However, as one can see from Table~\ref{tab:amp_gghhh_SMEFTHEFT}, non-trivial effects may arise when considering even higher-order terms in the EFT expansion. 

Starting with the $++$-channel of the $gg\to hhh$ amplitude, the subleading structure of the $c_{gghhh}^{++,(2)}$ coefficient in Eq.~\eqref{amp:gghhh:OS:++} can be generated by a dim-12 operator in SMEFT and by a N$^3$LO operators in HEFT. 
This observation illustrates the diagnostic power of the amplitude approach: we identify a helicity structure that, within SMEFT or HEFT, would only arise from operators appearing at very high order in the respective power counting. If experimental data were to show traces of such structures, it would indicate that the SMEFT expansion is insufficient and that the assumed HEFT power counting -- e.g. based on factoring out or counting powers of $g_s$ -- may not be appropriate. In that case, a less suppressed HEFT operator hierarchy, or even the breakdown of the EFT expansion altogether, would be implied.
However, given the difficulty of observing triple Higgs production at the LHC, this might remain a measurement for future high-energy colliders.

To be more specific, we identified one dim.-12 SMEFT operator,
\begin{equation}
    Q_{G^2\varphi^4 D^4 }^{(1)}=(\varphi^{\dagger} \varphi)\left(D^\mu D^\nu \varphi^{\dagger} D^\alpha D^\beta \varphi\right) G_{\mu \alpha}^A G_{\nu \beta}^{A} \, ,
    \label{eq:dim_12_SMEFT_Op}
\end{equation} 
and one N$^3$LO HEFT,
\begin{equation}
    \mathcal{P}(h)_{GHD^4}^{(1)}=-\frac{1}{4} g_s^2 (\partial^\mu \partial^\nu h \partial_\alpha \partial_\beta h) G_{\mu \alpha}^A G^{A \nu \beta}\mathcal{F}^{(1)}_{GHD^4} \, ,\quad
    \mathcal{F}_{GHD^4}^{(1)} = \sum_{n=0}^{\infty} C_{GHD^4(1)}^{(n)}\left( \frac{h}{v} \right)^n,
    \label{eq:dim_12_SMEFT_Op}
\end{equation} 
operator that generates such a structure. Any operator of this type with Lorentz indices contracted between the derivatives and the field-strength tensors would serve the same purpose. By contrast, contracting the indices within the derivatives and within the field-strength tensors separately leads instead to the structure associated with the $c_{gghhh}^{++,(1)}$ Wilson coefficient.
While in double Higgs production such operator contributes to $++$ and $+-$ channels generating higher powers of terms in $s_{ij}$, in the case of triple Higgs production, it generates a whole new structure.
Using our matching techniques, we find that $c_{gghhh}^{++,(2)}$ is generated by dim-12 SMEFT and N$^3$LO HEFT at the lowest order with coefficient,

\begin{equation}
-i \, c_{gghhh}^{++,(2) (00000)} =
\begin{cases}
-\dfrac{1}{\Lambda} \dfrac{1}{2^{9/4}\sqrt{G_F}} \; C^{(1)}_{G^2\varphi^4 D^4}, & \text{SMEFT}\, ,\\[1.2em]
\phantom{-}  \bar{\Lambda} \, \dfrac{g_s^2 \sqrt{G_F} }{2^{9/4}} \; C^{(1)}_{GHD^4(1)}, & \text{HEFT}\, .
\end{cases}
\label{eq:gghhh:++:dim12:N$^3$LO}
\end{equation}
These operators additionally generate a contribution to $c_{gghhh}^{++,(1) (00000)}$.
Note that no lower-order structures in SMEFT and HEFT will generate $c_{gghhh}^{++,(2)}$ coefficient as a result of gluing.

In the case of the $+-$ channel, the on-shell structure of the $c_{gghhh}^{+-}$ is already generated by NNLO HEFT operator. In SMEFT, this structure can be generated at the lowest order by a dim-10 operator,
\begin{equation}
    Q_{G^2\varphi^4 D^2 }^{(1)}=(\varphi^{\dagger} \varphi)\left(D^\mu \varphi^{\dagger} D^\nu \varphi\right) G_{\alpha \mu}^A G_\nu^{A \alpha} \, .
    \label{eq:dim_10_SMEFT_Op}
\end{equation} 
The result of matching these operators into $c_{gghhh}^{+-}$ is,
\begin{equation}
-i \, c_{gghhh}^{+-,(00000)} =
\begin{cases}
-\dfrac{1}{\Lambda} \dfrac{1}{2^{9/4}\sqrt{G_F}} \; C^{(1)}_{G^2\varphi^4 D^2}, & \text{SMEFT} \, ,\\[1.2em]
\phantom{-}  \bar{\Lambda} \, \dfrac{g_s^2 \sqrt{G_F} }{2^{9/4}} \; C^{(1)}_{GHD(2)}, & \text{HEFT} \, ,
\end{cases}
\label{eq:gghhh:+-:dim10:NNLO}
\end{equation}
where in the case of HEFT we included the full set of diagrams since this operator also generates the vertex $gg\to hh$ with $+-$ helicities.

In summary, while the study of the three- and four-point Higgs and Higgs–gluon amplitudes already proved to be highly instructive in revealing differences between SMEFT and HEFT, it was mainly tied to their distinct power counting schemes. The five-point amplitude turned out to be even more interesting. Here, the naive pattern observed at lower multiplicities, where SMEFT/HEFT contributions appear at respective EFT orders dim-4/LO, dim-6/NLO, and dim-8/NNLO, is broken. The additional Higgs boson in the amplitude leads to an increase of the SMEFT order in our order-mapping between SMEFT and HEFT by two: dim-6/LO, dim-8/NLO, dim-10/NNLO, and dim-12/N$^3$LO. Moreover, the $gg\to hhh$ amplitude in the $++$ helicity channel exhibits a new on-shell structure that had not been seen in the literature and is first generated at dim-12 in SMEFT and at N$^3$LO in HEFT. 

\section{Conclusions}
\label{sec:conclusions}
In this work, we have analyzed the tree-level scattering amplitudes for $gg\to hh$ and $gg\to hhh$, obtained both from a general on-shell bootstrap construction and from explicit computations in two commonly used EFT frameworks, SMEFT and HEFT. Our study included the most general on-shell amplitude structures, SMEFT contributions up to order $1/\Lambda^4$ in the EFT expansion -- including double insertions of dim-6 operators as well as the complete set of relevant dim-8 operators -- and the full NLO and NNLO contributions in HEFT.
We matched the on-shell scattering amplitudes to the ones obtained from SMEFT/HEFT to study how different kinematic dependencies appear in these different EFTs. This matching was performed with rational numerical matching techniques and validated analytically where appropriate.

Since the building of on-shell amplitudes is a recursive process, we first studied the lower-point $gg\to h$, $hh\to h$, $gg\to hh$ and $hh\to hh$ amplitudes.
We computed the full matching between the on-shell coefficients and the SMEFT and HEFT Wilson coefficients for the first time. 
Already at the level of three- and four-point amplitudes, characteristic differences between SMEFT and HEFT emerged. 
We observed a shift in the power counting between the $gg\to h$ and $gg\to hh$: contributions that appear at NLO in HEFT correspond to the combined effects of dim-6 and dim-8 operators in SMEFT. Then, for the 
$hh\to h$ case, SMEFT corrections arise at dim-6 and dim-8, while in HEFT they correspond to the LO trilinear coupling modifier $\Delta\lambda_3$. Finally, for the $hh\to hh$ amplitude, SMEFT corrections at dim-6 and dim-8 correspond respectively to LO and NLO contributions in HEFT.

When moving on to the five-point amplitudes, we first extended the gluing technique to build the factorizable part of the on-shell five-point amplitudes from on-shell three- and four-point amplitudes. Our technique is closely related to previous methods~\cite{AccettulliHuber:2021uoa}, with the difference that it does not require matching free coefficients after building the factorizable amplitude ansatz. The extension of our method to all possible five- and higher-point on-shell amplitudes is left for future work.

We first verified our method by studying the simpler five-point Higgs amplitude.
Its matching to the SMEFT and HEFT frameworks revealed interesting differences between the EFTs.
Dim-8 SMEFT generates only constant and linear terms in the Mandelstam variables of the five-point on-shell coefficient, whereas NNLO HEFT generates terms up to a quadratic order.

The gluon-fusion triple Higgs production amplitude, $gg\to hhh$, 
revealed further differences in convergence between HEFT and SMEFT. While SMEFT up to dim-8 could generate only the constant contribution to the lowest-order $++$ on-shell coefficient, HEFT at NNLO could also generate the linear terms in Mandelstam variables. Regarding the $+-$ helicity amplitude, the only non-zero contribution originates from HEFT at NNLO, with vanishing dim-8 contribution. For this channel, we identified a dim-10 SMEFT operator that can lead to such a structure. Finally, in the case of the $++$ helicity amplitude, we found a new kinematic structure generated at dim-12 in SMEFT and at $\text{N}^3$LO in HEFT, and we matched it explicitly to representative operators in both EFTs.

As an important conclusion from our work we find that HEFT and SMEFT are not fundamentally different in multi-Higgs production, but just show different \textit{convergence behavior}. This was evident as we could reproduce all of our coefficients corresponding to on-shell amplitude structures in both SMEFT and HEFT.

This work can be continued in several directions. The study of the high-energy limit of a more complete set of amplitudes could uncover relations between amplitudes usually attributed to gauge symmetry for WCs at all orders in the energy expansion. The comparison between SMEFT and HEFT can be extended to the entire set of five-point amplitudes allowed by the particle content in these EFTs, including massive vector bosons, as well as to higher-multiplicity amplitudes. Phenomenological studies based on our on-shell amplitudes could establish observables that maximize the sensitivity to dim-10 and dim-12 SMEFT and N$^3$LO HEFT operators. However, a thorough phenomenological analysis will require the inclusion of the SM loop-induced diagrams, which in turn demands techniques such as generalized unitarity. Although multi-Higgs production suffers from small cross sections, pursuing this effort is well motivated by the ongoing experimental efforts to measure double \cite{ATLAS:2025hhd, ATLAS:2024ish, CMS:2024awa} and even triple \cite{ATLAS:2024xcs, CMS-PAS-HIG-24-015}  Higgs production.

\section*{Acknowledgements}
We are deeply grateful to L.~Bresciani, I.~Brivio, S.~De Angelis, G.~Durieux, K.~Schmid  and Y.~Shadmi for useful discussions at different stages of this project.
We thank especially S.~De Angelis for his comments on this manuscript.
This project received funding from the University of Padua under the 2023 STARS Grants@UniPD Programme (Acronym and title of the project: HiggsPairs - Precise Theoretical Predictions for Higgs pair production at the LHC) and from the INFN Iniziativa Specifica APINE. This work was also partially supported by the Italian MUR Departments of Excellence grant 2023-2027 ``Quantum Frontiers''. We acknowledge support from the COMETA COST Action CA22130. ANR is grateful to the CERN Theory Department for its hospitality and support during the completion of this project.

\appendix

\section{Spinor-helicity conventions}
\label{app:spinor_helicity_details}
In this appendix, we collect our conventions for the spinor-helicity formalism. In general, we follow the conventions of~\cite{Durieux:2019eor}. Our choice for the Minkowski metric is,
\begin{equation}
    \eta_{\mu\nu}=\eta^{\mu\nu}=\text{diag}\left(1,-1,-1,-1\right).
\end{equation}
We choose the chiral representation of Dirac matrices,
\begin{equation}
\gamma^{\mu}=\left(\begin{array}{cc}
    0 & \left(\sigma^{\mu}\right)_{\alpha\dot{\beta}}  \\
    \left(\bar\sigma^{\mu}\right)^{\dot\alpha\beta} & 0  
\end{array}\right), \quad    \gamma^{5}=i\gamma^{0}\gamma^{1}\gamma^{2}\gamma^{3}=\left(\begin{array}{cc}
    -1 & 0  \\
    0 & 1  
\end{array}\right).
\end{equation}
The angle and square-spinors are related to the 2-component Weyl spinors by,
\begin{equation}    
\begin{array}{cc}
     u_+(p) = |p], & u_{-}(p) = |p\rangle \, ,  \\
     \bar{u}_{+}(p) = [p|, & \bar{u}_{-}(p) = \langle p | \, , 
\end{array}
\end{equation}
and we define the massive 4-component Dirac spinors as,
\begin{equation}
    u^{I}(p)=\left(\begin{array}{c}
        \lambda_{\alpha}^{I} \\
          \tilde\lambda^{I\dot\alpha}
    \end{array}\right)=\left(\begin{array}{c}
        \tret{p}{I}{\alpha} \\
          \squet{p}{I\dot\alpha}{}
    \end{array}\right),\quad v^{I}(p)=\left(\begin{array}{c}
        \lambda_{\alpha}^{I} \\
          -\tilde\lambda^{I\dot\alpha}
    \end{array}\right)=\left(\begin{array}{c}
        \tret{p}{I}{\alpha} \\
          -\squet{p}{I\dot\alpha}{}
    \end{array}\right).
\end{equation}
These satisfy the massive Dirac equation,
\begin{equation}
    \left(\slashed{p}-m\right)u^{I}(p)=0 \, ,\quad \left(\slashed{p}+m\right)v^{I}(p)=0 \, .
\end{equation}
The chiral components are in one-to-one correspondence to the angle and square spinors,
\begin{equation}
    u_{L}(p)=P_L\, u(p)=\left(\begin{array}{c}
        \tret{p}{I}{\alpha} \\
          0
    \end{array}\right), \quad u_{R}(p)=P_R\, u(p)=\left(\begin{array}{c}
        0 \\
          \squet{p}{I\dot\alpha}{}
    \end{array}\right).
\end{equation}
And in the massless limit, we have,
\begin{equation}
    \slashed{p}\,u(p)=\left(\begin{array}{cc}
    0 & p_{\mu}\left(\sigma^{\mu}\right)_{\alpha\dot{\beta}}  \\
    p_{\mu}\left(\bar\sigma^{\mu}\right)^{\dot\alpha\beta} & 0  
\end{array}\right)\left(\begin{array}{c}
        \tret{p}{}{\beta} \\
          \squet{p}{\dot\beta}{}
    \end{array}\right)=\left(\begin{array}{c}
p_{\mu}\left(\sigma^{\mu}\right)_{\alpha\dot\beta}\squet{p}{\dot\beta}{}\\
        p_{\mu}\left(\bar\sigma^{\mu}\right)^{\dot\alpha\beta} \tret{p}{}{\beta}
    \end{array}\right)=0 \,.
\end{equation}
The spinor products can be computed with ease from the previous identities as,
\begin{equation}
    \bar{u}_{L}(p)u_{R}(q)=\squaket{p}{q} \, ,\quad \bar{u}_{R}(p)u_{L}(q)=\traket{p}{q} \, .
\end{equation}
And for Lorentz vectors, it holds that,
\begin{equation}
 \begin{aligned}
    \squa{p}{}{}\gamma^\mu\tret{q}{}{}=&\bar{u}_L(p)\gamma^\mu u_L (q)\,,\\
    \tra{p}{}{}\gamma^\mu\squet{q}{}{}=&\bar{u}_R(p) \gamma^\mu u_R(q)\,.
 \end{aligned}
\end{equation}
The result of the previous identities in terms of the momenta depends on the chosen polarization axis. However, there is one fundamental Lorentz-invariant identity for spinor products,
\begin{equation}
    \langle p|q \rangle\left[ p|q \right]=(p+q)^2.
\end{equation}
This allows to easily convert between square and angle brackets.

The polarization vectors for massless spin-1 bosons can be expressed in this formalism as,
\begin{equation}
    \begin{aligned}
        \epsilon_{+}^{\mu}(p)=&\frac{1}{\sqrt{2}}\frac{\tra{\xi}{}{}\sigma^{\mu}\squet{p}{}{}}{\traket{p}{\xi}}\, ,\\
        \epsilon_{-}^{\mu}(p)=&\frac{1}{\sqrt{2}}\frac{\tra{p}{}{}\sigma^{\mu}\squet{\xi}{}{}}{\squaket{p}{\xi}}\, ,
    \end{aligned}
    \label{eq:pol_vec_spinor_hel}
\end{equation}
or equivalently,
\begin{equation}
    \begin{aligned}
        \epsilon_{+}^{\alpha\dot{\alpha}}(p)=&\sqrt{2}\frac{\squet{p}{\dot{\alpha}}{}\tra{\xi}{\alpha}{}}{\traket{p}{\xi}} \, ,\\
        \epsilon_{-}^{\alpha\dot{\alpha}}(p)=&\sqrt{2}\frac{\squet{\xi}{\dot{\alpha}}{}\tra{p}{\alpha}{}}{\squaket{p}{\xi}} \, ,
    \end{aligned}
    \label{eq:pol_vec_spinor_hel_spinor_notation}
\end{equation}
where $\squet{\xi}{}{}$ and $\tret{\xi}{}{}$ are arbitrary reference spinors that should not be collinear to $p$.

To evaluate explicitly the spinor products in terms of momenta, we choose the polarization axis to be the $y$-axis and use the following basis of Pauli matrices,
\begin{equation}
       \sigma^0=\mathbf{1}_{2\times 2},\quad \sigma^{1}=\left( \begin{array}{cc}
           0 & 1 \\
           1 & 0 
       \end{array} \right),\quad \sigma^{2}=\left( \begin{array}{cc}
           1 & 0 \\
           0 & -1 
      \end{array} \right), \quad \sigma^{3}=\left( \begin{array}{cc}
          0 & i \\
          -i & 0 
      \end{array} \right).
\end{equation} 
Thus, for two lightlike momenta $A^\mu$ and $B^\mu$, we obtain,
\begin{equation}
       \squaket{A}{B}=\eta_A\eta_B \frac{ B^+\,A^{-}_{T} - A^{+} \,B^{-}_T }{\sqrt{\eta_A^2\eta_B^2 A^{+}B^{+}}} \, ,
   \end{equation}
   and
   \begin{equation}
       \traket{A}{B}=-\left(\squaket{A}{B}\right)^{*} =  \eta_A \eta_B \frac{A^{+} B_{T}^{+} - B^{+} A_{T}^{+}  }{\sqrt{\eta_A^2\eta_B^2 A^{+}B^{+}}} \, ,
   \end{equation}
where,
\begin{equation}
 \begin{aligned}
      p^{+}=p^{0}+p^{2}, \quad & p^{-}_{T}=p^{1}-i\,p^{3}, \\
      p^{-}=p^{0}-p^{2}, \quad & p^{+}_{T}=p^{1}+i\,p^{3}, \\
 \end{aligned}
 \end{equation}
 and,
 \begin{equation}
    \eta_{p} =\begin{cases}
    1 & \quad p^0>0\\
    -i & \quad p^0<0
    \end{cases} \, . 
\end{equation}
The phase of the right-handed spinor was fixed by choosing $u_R = i\sigma^{3} u_{L}^{*}$. This also allows us to compute the polarization vector for a general momentum $p^\mu=\left(p^0,p^1,p^2,p^3\right)$,
\begin{equation}
    \epsilon_{\pm}^{\mu}(p)= \frac{1}{\sqrt{2}} \left( \pm\, \mathbf{f}_{\pm}(p), \pm\,1 ,\mp\,\mathbf{f}_{\pm}(p), -i \right),
\label{eq:pol_vector_gen_momenta}
\end{equation}
with
\begin{equation}
    \mathbf{f}_{\pm}(p)=\frac{p^1 \mp i\,p^3}{p^0 + p^2} \, .
\end{equation}

\section{SMEFT and HEFT bases}
\subsection{SMEFT}
\label{appendix:SMEFT}

In this work, we use the SMEFT Warsaw basis~\cite{Grzadkowski:2010es} for dim-6 operators, basis of~\cite{Murphy:2020rsh} for dim-8, and make use of~\cite{Harlander:2023psl} for dim-10 and -12 operators. Tables~\ref{tab:Dim6:SMEFT}--\ref{tab:Dim10:Dim12:SMEFT} collect all relevant operators.

\begin{table}[h!]
    \centering
    \begin{tabular}{|c|c|}
    \hline \multicolumn{2}{|c|}{$X^2 \varphi^2$} \\
    \hline$Q_{\varphi G}$ & $(\varphi^{\dagger} \varphi) G_{\mu \nu}^A G^{A \mu \nu}$ \\
    $Q_{\varphi \widetilde{G}}$ & $(\varphi^{\dagger} \varphi) \widetilde{G}_{\mu \nu}^A G^{A \mu \nu}$ \\
    \hline
    \end{tabular}
    \begin{tabular}{|c|c|}
    \hline \multicolumn{2}{|c|}{$\varphi^6$ and $\varphi^4 D^2$} \\
    \hline$Q_{\varphi}$ & $\left(\varphi^{\dagger} \varphi\right)^3$ \\
    $Q_{\varphi \square}$ & $\left(\varphi^{\dagger} \varphi\right) \square\left(\varphi^{\dagger} \varphi\right)$ \\
    $Q_{\varphi D}$ & $\left(\varphi^{\dagger} D^\mu \varphi\right)^*\left(\varphi^{\dagger} D_\mu \varphi\right)$ \\
\hline
\end{tabular}
    \caption{Subset of dimension-6 SMEFT operators used in this work, basis of Ref.~\cite{Grzadkowski:2010es}.}
    \label{tab:Dim6:SMEFT}
\end{table}

\begin{table}[h!]
\centering
\begin{tabular}{|c|c|}
\hline \multicolumn{2}{|c|}{$X^2 \varphi^4$} \\
\hline$Q_{G^2 \varphi^4}^{(1)}$ & $\left(\varphi^{\dagger} \varphi\right)^2 G_{\mu \nu}^A G^{A \mu \nu}$ \\
$Q_{G^2 \varphi^4}^{(2)}$ & $\left(\varphi^{\dagger} \varphi\right)^2 \widetilde{G}_{\mu \nu}^A G^{A \mu \nu}$ \\
\hline
\end{tabular}
\begin{tabular}{|l|l|}
\hline \multicolumn{2}{|c|}{$X^2 \varphi^2 D^2$} \\
\hline$Q_{G^2 \varphi^2 D^2}^{(1)}$ & $\left(D^\mu \varphi^{\dagger} D^\nu \varphi\right) G_{\mu \rho}^A G_\nu^{A \rho}$ \\
$Q_{G^2\varphi^2 D^2}^{(2)}$ & $\left(D^\mu \varphi^{\dagger} D_\mu \varphi\right) G_{\nu \rho}^A G^{A \nu \rho}$ \\
$Q_{G^2 \varphi^2 D^2}^{(3)}$ & $\left(D^\mu \varphi^{\dagger} D_\mu \varphi\right) G_{\nu \rho}^A \widetilde{G}^{A \nu \rho}$ \\
\hline
\end{tabular}
\begin{tabular}{|c|c|c|c|c|c|}
\hline
\multicolumn{2}{|c|}{$\varphi^8$} &
\multicolumn{2}{c|}{$\varphi^6D^2$} &
\multicolumn{2}{c|}{$\varphi^4D^4$} \\
\hline
$Q_{\varphi^8}$  &  $(\varphi^\dagger \varphi)^4$ &
$Q_{\varphi^6 \square}$ &
$(\varphi^\dagger \varphi)^{2} \square (\varphi^\dagger \varphi)$ &
$Q_{\varphi^4 D^4}^{(1)}$  &
$(D_{\mu} \varphi^{\dagger} D_{\nu} \varphi) (D^{\nu} \varphi^{\dagger} D^{\mu} \varphi)$ \\
& & $Q_{\varphi^{6} D^{2}}$ & $(\varphi^{\dagger} \varphi)
(\varphi^\dagger D_\mu\varphi)^{\ast} (\varphi^\dagger D^\mu\varphi)$  &
$Q_{\varphi^4 D^4}^{(2)}$  &
$(D_{\mu} \varphi^{\dagger} D_{\nu} \varphi) (D^{\mu} \varphi^{\dagger} D^{\nu} \varphi)$ \\
& & & & $Q_{\varphi^4 D^4}^{(3)}$  &
$(D_{\mu} \varphi^{\dagger} D^{\mu} \varphi) (D_{\nu} \varphi^{\dagger} D^{\nu} \varphi)$ \\
\hline
\end{tabular}
    \caption{Subset of dimension-8 SMEFT operators used in this work, basis of~\cite{Murphy:2020rsh}, with slightly modified class $\varphi^6 D^2$ following~\cite{Dedes:2023zws}. }
    \label{tab:Dim8:SMEFT}
\end{table}

\begin{table}[h!]
    \centering
    \begin{tabular}{|c|c|}
    \hline \multicolumn{2}{|c|}{$X^2 \varphi^4 D^2$} \\
    \hline$Q_{G^2\varphi^4 D^2 }^{(1)}$ & $(\varphi^{\dagger} \varphi)\left(D^\mu \varphi^{\dagger} D^\nu \varphi\right) G_{\alpha \mu}^A G_\nu^{A \alpha}$ \\
    \hline
    \end{tabular}
        \begin{tabular}{|c|c|}
    \hline \multicolumn{2}{|c|}{$X^2 \varphi^4 D^4$} \\
    \hline$Q_{G^2\varphi^4 D^4 }^{(1)}$ & $
    (\varphi^{\dagger} \varphi)\left(D^\mu D^\nu \varphi^{\dagger} D^\alpha D^\beta \varphi\right) G_{\mu \alpha}^A G_{\nu \beta}^{A}$ \\
\hline
\end{tabular}
    \caption{Dimension-10 and -12 SMEFT operators used in this work.}
    \label{tab:Dim10:Dim12:SMEFT}
\end{table}

\subsection{HEFT}
\label{appendix:HEFT}

In this work, we adopt the conventions of Ref.~\cite{Brivio:2016fzo} for the LO and NLO HEFT operator bases. For NNLO, we are aware of the complete basis constructed in Ref.~\cite{Sun:2022snw}, which extends the NLO basis of Ref.~\cite{Sun:2022ssa}. However, we employ a slightly modified NNLO basis, closer in structure to that of Ref.~\cite{Brivio:2016fzo}, in which the operators are written in a form more directly comparable to their SMEFT counterparts. This choice significantly simplifies the comparison between the two EFT frameworks. Where necessary, we provide the correspondence between operators in our basis and those in Ref.~\cite{Sun:2022snw}. This approach is justified since our analysis focuses on a single process, sensitive only to a subset of the NNLO HEFT operators.

\subsubsection{HEFT at NLO}

We use the following set of NLO HEFT operators in the basis of Ref.~\cite{Brivio:2016fzo},
\begin{equation}
     \mathcal{L}_{\text{NLO}}^{\text{SMEFT}} = \mathcal{P}(h)_{DH} + \mathcal{P}(h)_{GH}
     + \mathcal{P}(h)_{\tilde{G}H} \, ,
\end{equation}
with,
\begin{equation}
    \begin{aligned}
        \mathcal{P}(h)_{DH}&=\left( \partial_\mu \mathcal{F}(h)_{DH} \, \partial^\mu \mathcal{F}'(h)_{DH} \right)^2,\quad 
        \mathcal{F}^{(\prime)}_{DH} = \sum_{n=1}^{\infty} C_{DH}^{(\prime)(n)}\left( \frac{h}{v} \right)^n, \\
        \mathcal{P}(h)_{GH}&=-\frac{1}{4} g_s^2  G_{\mu \nu}^A G^{A \mu \nu}\mathcal{F}_{GH},\quad 
        \mathcal{F}_{GH} =  \sum_{n=1}^{\infty} C_{GH}^{(n)}\left( \frac{h}{v} \right)^n,\\
        \mathcal{P}(h)_{\tilde{G}H}&=-\frac{1}{4} g_s^2  \tilde{G}_{\mu \nu}^A G^{A \mu \nu}\mathcal{F}_{\tilde{G}H},\quad 
        \mathcal{F}_{\tilde{G}H} =  \sum_{n=1}^{\infty} C_{\tilde{G}H}^{(n)}\left( \frac{h}{v} \right)^n.
    \end{aligned}
    \label{eq:HEFT:ops:NLO:Ilaria}
\end{equation}
The operator $\mathcal{P}(h)_{DH}$ corresponds to $\mathcal{O}_{15}^{UhD^4}$ in Ref.~\cite{Sun:2022ssa}, while operators
$\mathcal{P}(h)_{GH}$ and $\mathcal{P}(h)_{\tilde{G}H}$ match 
$\mathcal{O}_3^{X^2 Uh}$ and $\mathcal{O}_4^{X^2 Uh}$.

\subsubsection{HEFT at NNLO}

We use the following set of NNLO HEFT operators in our own basis, similar to NLO basis of Ref.~\cite{Brivio:2016fzo},

\begin{equation}
     \mathcal{L}_{\text{NNLO}}^{\text{HEFT}} = \mathcal{P}(h)_{GHD}^{(1)} + \mathcal{P}(h)_{GHD}^{(2)}
     + \mathcal{P}(h)_{\tilde{G}HD} \, .
\end{equation}
with,
\begin{equation}
    \begin{aligned}
        \mathcal{P}(h)_{GHD}^{(1)}&=-\frac{1}{4} g_s^2 (\partial^\alpha h \partial_\alpha h) G_{\mu \nu}^A G^{A \mu \nu}\mathcal{F}_{GHD},\quad 
        \mathcal{F}_{GHD}^{(1)} =\sum_{n=0}^{\infty} C^{(n)}_{GHD (1)}\left( \frac{h}{v} \right)^n,\\
        \mathcal{P}(h)_{GHD}^{(2)}&=-\frac{1}{4} g_s^2 (\partial^\alpha h \partial_\nu h) G_{\mu \alpha}^A G^{A \mu \nu}\mathcal{F}_{GHD},\quad 
        \mathcal{F}_{GHD}^{(2)} =\sum_{n=0}^{\infty} C^{(n)}_{GHD (2)}\left( \frac{h}{v} \right)^n,\\
        \mathcal{P}(h)_{\tilde{G}HD}&=-\frac{1}{4} g_s^2 (\partial^\alpha h \partial_\alpha h) \tilde{G}_{\mu \nu}^A G^{A \mu \nu}\mathcal{F}_{GHD},\quad 
        \mathcal{F}_{\tilde{G}HD} =\sum_{n=0}^{\infty} C^{(n)}_{\tilde{G}HD}\left( \frac{h}{v} \right)^n.
    \end{aligned}
    \label{eq:HEFT:ops:NNLO:Ilaria}
\end{equation}
Alternatively, one can use the NNLO basis of Ref.~\cite{Sun:2022snw}, with the following translation between the two bases,
\begin{equation}
    \begin{aligned}
        \mathcal{O}^{G_L^2 h^2 D^2}+\text{h.c.} & \rightarrow \mathcal{P}(h)_{GHD}^{(1)} + \mathcal{P}(h)_{\tilde{G}HD},\\
         \mathcal{O}^{G_L h^2 G_R D^2}+\text{h.c.} & \rightarrow \mathcal{P}(h)_{GHD}^{(2)}.\\
    \end{aligned}
    \label{eq:HEFT:ops:NNLO:Translation}
\end{equation}

\subsubsection{HEFT at N$^3$LO}
Additionally, we included the following N$^3$LO operator used for generating a non-zero $c_{gghhh}^{++,(2)}$ on-shell contribution,
\begin{equation}
    \mathcal{P}(h)_{GHD^4}^{(1)}=-\frac{1}{4} g_s^2 (\partial^\mu \partial^\nu h \partial_\alpha \partial_\beta h) G_{\mu \alpha}^A G^{A \nu \beta}\mathcal{F}^{(1)}_{GHD^4},\quad
    \mathcal{F}_{GHD^4}^{(1)} = \sum_{n=0}^{\infty} C_{GHD^4(1)}^{(n)}\left( \frac{h}{v} \right)^n.
        \label{eq:HEFT:op:N$^3$LO}
\end{equation}

\section{The importance of matching complete amplitudes}
\label{app:matching_subtleties}

In this appendix, we show
two phenomena that highlight the importance of matching complete amplitudes and lead to results that would otherwise be unexpected. Both phenomena have been discussed in the literature before, in particular in App. A of~\cite{Balkin:2021dko} and of~\cite{Liu:2023jbq}.
First, we show how the non-factorizable piece of the on-shell amplitude absorbs some contributions that, in the EFT, are generated by factorizable diagrams. Second, we show how the first phenomenon and non-trivial cancelations allow the on-shell amplitude to correctly account for momentum-dependent lower-point vertices in the EFT.

\subsection{Absorption into non-factorizable pieces}
\label{app:full_matching_importance}

The results of the four-point matching are enough to explain the importance of matching complete amplitudes. We take the $++00$~ $gg\to hh$ amplitude and consider the SMEFT matching results in Eq.~\eqref{eq:match_Result_gghh_++00_00_rational_Gf}, we will focus on the $C_{\varphi G}^2$ contribution. This contribution arises via the gluon-field kinetic normalization induced by $Q_{\varphi G}$, which then causes the shift $C_{\varphi G}\to C_{\varphi G}\left( 1+\frac{\sqrt{2}}{G_{F}}\frac{C_{\varphi G}}{\Lambda^2} \right)$. This correlation between the $C_{\varphi G}^2$ and $C_{\varphi G}$ contributions can be observed in the SMEFT matching result for $c_{ggh}$, Eq.~\eqref{eq:match_ons_smeft_hhh_Gf}, but not in the one for $c_{gghh}^{++,(00)}$, Eq.~\eqref{eq:match_Result_gghh_++00_00_rational_Gf}. This indicates that the latter receives contributions from diagrams with virtual particles. Here, we will show how the latter contributions are absorbed by the non-factorizable piece. 

The pole structure of the $gg\to hh$ amplitude tells us that the $++00$ helicity configuration has only a pole along the s-channel and hence our on-shell amplitude is composed of,
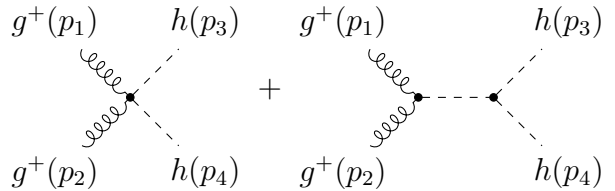
\begin{equation*}
\begin{tikzpicture}[baseline=(v1.base)]
        \begin{feynman}[inline=v1.base]
            \vertex (v1) at (0,0) [dot] {};
            \vertex (g1) at (-1,-1) {\( g^{+}(p_2) \)};
            \vertex (g2) at (-1,1) {\( g^{+}(p_1) \)};
            \vertex (h1) at (1,1) {\( h(p_3) \)};
            \vertex (h2) at (1,-1) {\( h(p_4) \)};
            \diagram* {
                (g1) -- [gluon] (v1),
                (g2) -- [gluon] (v1),
                (v1) -- [scalar] (h1),
                (v1) -- [scalar] (h2),
            };
        \end{feynman}
    \end{tikzpicture}
+
\begin{tikzpicture}[baseline=(v1.base)]
        \begin{feynman}[inline=v1.base]
            \vertex (v1) at (0,0) [dot] {};
            \vertex (v2) at (1,0) [dot] {};
            \vertex (g1) at (-1,-1) {\( g^{+}(p_2) \)};
            \vertex (g2) at (-1,1) {\( g^{+}(p_1) \)};
            \vertex (h1) at (2,1) {\( h(p_3) \)};
            \vertex (h2) at (2,-1) {\( h(p_4) \)};
            \diagram* {
                (g1) -- [gluon] (v1),
                (g2) -- [gluon] (v1),
                (v1) -- [scalar] (v2),
                (v2) -- [scalar] (h1),
                (v2) -- [scalar] (h2),
            };
        \end{feynman}
    \end{tikzpicture},
\end{equation*}
which can be written as,
\begin{equation}
\begin{aligned}
   \mcM_{++00}^{ONS} = & i\, c_{gghh}^{++} [1|2]^2 + c_{ggh}\,c_{3h} \frac{[1|2]^2}{s-m_{h}^2} \\
   = & i\, c_{gghh}^{++} [1|2]^2 + \left( - 6\sqrt{2}\, \frac{m_h^2}{G_F}\, \frac{C_{\varphi G}^2}{\Lambda^4}  \right) \frac{[1|2]^2}{s-m_{h}^2} \, ,
\end{aligned}
\label{eq:smeft_amp_semi_matched}
\end{equation}
where we have used Eq.~\eqref{eq:match_ons_smeft_ggh_Gf},~\eqref{eq:match_ons_smeft_hhh_Gf}, and kept the term proportional to $C_{\varphi G}^2$ in their product.
The spinor structure yields $[1|2]^2=-s$.
On the other hand, we have computed the same amplitude in the SMEFT up to order $1/\Lambda^4$. Focusing on the piece proportional to $C_{\varphi G}^2$ we get,
\begin{equation}
    \mcM_{++00}^{SMEFT} = 6\,\frac{\sqrt{2}}{G_F^2}\,\frac{C_{\varphi G}^2}{\Lambda^4} \frac{s^2}{s-m_{h}^2}+...,
\end{equation}
where $...$ stands for all the other SMEFT contributions.
The result of the matching procedure can easily be seen from a simple manipulation of $\mcM_{++00}^{SMEFT}$,
\begin{equation}
    \begin{aligned}
    \mcM_{++00}^{SMEFT} = & 6\,\frac{\sqrt{2}}{G_F^2}\,\frac{C_{\varphi G}^2}{\Lambda^4} \frac{s^2}{s-m_{h}^2} \\
    = & 6\,\frac{\sqrt{2}}{G_F^2}\,\frac{C_{\varphi G}^2}{\Lambda^4} \frac{s\,(s-m_h^2+ m_h^2)}{s-m_{h}^2} \\
    = & 6\,\frac{\sqrt{2}}{G_F^2}\,\frac{C_{\varphi G}^2}{\Lambda^4} s + 6 \frac{\sqrt{2}}{G_F^2}\,\frac{C_{\varphi G}^2}{\Lambda^4} m_{h}^2 \frac{s}{s-m_h^2}.
\end{aligned}
\label{eq:manipulation_smeft}
\end{equation}
The second term in the last line above matches exactly the last term in the last line in ~\eqref{eq:smeft_amp_semi_matched}, while the first term in the last line of \eqref{eq:manipulation_smeft} has the right Lorentz dependence to be absorbed by the first term in the expansion of $c_{gghh}^{++}$, i.e. $c_{gghh}^{++,(00)}$. Hence, we get,
\begin{equation}
    - i\, c_{gghh}^{++,(00)} = 6\,\frac{\sqrt{2}}{G_F^2}\,\frac{C_{\varphi G}^2}{\Lambda^4},
\end{equation}
which agrees with Eq.~\eqref{eq:match_Result_gghh_++00_00_rational_Gf}.

What we have shown here is commonplace for Higgs SMEFT operators that generate lower-point amplitudes upon setting the Higgs field to its VEV~\cite{Balkin:2021dko}. This phenomenon has further consequences in the case of massive external vector bosons, where the high-energy limit of an amplitude with a Higgs boson can be related to a low-energy amplitude with a longitudinal massive vector boson. In our case, the absence of massive external vector bosons allows us to not distinguish between high- and low-energy limits.

\subsection{Cancellation of momentum dependence from three- and four-Higgs vertices}
\label{app:cancellations_5_higgses}

In this appendix, we show that the momentum dependence in the three- and four-Higgs vertices induced by operators such as $\mcO_{\varphi D}$ and $\mcO_{\varphi \Box}$ does not represent an obstacle to match the EFT amplitudes with the bootstrapped on-shell ones. 

This is a generalization of the phenomenon shown in the previous appendix and in the literature~\cite{Balkin:2021dko,Liu:2023jbq}. To the best of our knowledge, we show it here for the first time for five-point EFT amplitudes.

To illustrate this, we perform a fully analytical
matching between SMEFT and the bootstrapped on-shell three-, four-, and five-Higgs amplitudes considering only the SM, $C_\varphi$ and $C_{\varphi D}$ contributions\footnote{We notice that $Q_{\varphi\Box}$ generates the same amplitudes than $Q_{\varphi D}$ up to a constant overall factor, hence this proof would be equally valid for it and we omit it for simplicity.}. 
As can be seen in App.~\ref{app:full_matching_res}, the dim-6 operators only generate contributions to the constant piece in the on-shell coefficient $c_{4h}$, $c_{4h}^{(00)}$. Thus, the factorizable piece of the on-shell five-point amplitude will depend on the momenta only via the propagators. The SMEFT five-Higgs amplitude does not present this feature a priori due to the momentum-dependence introduced by $Q_{\varphi D}$.

Before showing how the discrepancy is solved, we review some useful kinematic relations. For $i,\,j,\,k,\,\ell,\,m\in\lbrace1,2,3,4,5\rbrace$, all different, momentum conservation dictates that,
\begin{equation}
    s_{ij}=s_{k\ell}+s_{km}+s_{\ell m}-3\,m_{h}^2 \, .
\end{equation}
If we define the inverse propagator as, $   \Delta_{ij} =s_{ij} - m_{h}^2$, it can be shown that,
\begin{equation}
    \sum_{k\neq i} \Delta_{ik} = 2 m_{h}^2,\quad \forall\, i\in\lbrace1,2,3,4,5\rbrace \, .
\end{equation}
Finally, all the possible inverse propagators are related by the equation,
\begin{equation}
    \sum_{i=1}^{4}\sum_{j>i}^{5}\Delta_{ij} = 5\,m_{h}^2 \, .
    \label{eq:master_simplifying_relation}
\end{equation}

For simplicity and without loss of generality, we focus on terms of order $1/\Lambda^4$ that are quadratic on $C_{\varphi D}$ and absorb $\Lambda^2$ inside the WC. We start by looking at the diagrams,
\begin{equation}
       \mcM_{2}^{\mathrm{SMEFT}}(5h)=
       \begin{tikzpicture}[baseline=(b1.base)]
\begin{feynman}[inline=b1.base]
\vertex (i1) {$h_{1}$};
\vertex[below=1cm of i1] (i2) {$h_{2}$};
\vertex[below right=0.5cm and 0.75cm of i1, blob, minimum height=0.25cm,minimum width=0.25cm] (b1) {};
\vertex[right=0.75cm of b1, blob, minimum height=0.25cm,minimum width=0.25cm] (b2) {};
\vertex[below right=0.5cm and 0.75cm of b2] (h1) { $h_5$ };
\vertex[right=0.75cm of b2] (h2) { $h_4$ };
\vertex[above right=0.5cm and 0.75cm of b2] (h3) { $h_3$ };
\diagram* {
(i1) -- [scalar] (b1),
(i2) -- [scalar] (b1),
(b1) -- [scalar] (b2),
(b1) -- [scalar] (h1),
(b2) -- [scalar] (h2),
(b2) -- [scalar] (h3)
};
\end{feynman}
\end{tikzpicture}+\text{perm. of 3, 4, 5}
\end{equation}
which contain terms like,
\begin{equation}
    \mcM_{2}^{\mathrm{SMEFT}}(5h) \supset -i\,\frac{15}{8}C_{\varphi D}^2\, \lambda \frac{\Delta_{12}+\Delta_{13}+\Delta_{14}+\Delta_{15}+\Delta_{23}+\Delta_{25}-2\Delta_{34}+\Delta_{35}+\Delta_{45}}{\Delta_{34}},
\end{equation}
and there are no other contributions within this topology that goes like $\sim C_{\varphi D}^2/\Delta_{34}$. By using Eq.~\eqref{eq:master_simplifying_relation}, one can simplify the above expression to,
\begin{equation}
    \mcM_{1}^{\mathrm{SMEFT}}(5h) \supset i\,\frac{45}{8}C_{\varphi D}^2\, \lambda - i\,\frac{75}{8} \frac{C_{\varphi D}^2 \,v^2\lambda^2}{\Delta_{34}},
\end{equation}
and hence we have removed all momentum dependence from the numerator of terms with a propagator. This can be achieved in all diagrams with a single propagator to obtain amplitudes that are explicitly in the form of what is obtained from on-shell gluing.

Now, we move on to diagrams with two propagators. Consider the topology,
\begin{equation} 
\mcM_{4}^{\mathrm{SMEFT}}\left(5h\right)=
\begin{tikzpicture}[baseline=(b1.base)]
\begin{feynman}[inline=h3.base]
\vertex (i1) {$h_1$};
\vertex[below=1cm of i1] (i2) {$h_2$};
\vertex[below right=0.5cm and 0.5cm of i1, blob,minimum height=0.25cm,minimum width=0.25cm] (b1) {};
\vertex[right=1cm of b1, blob, minimum height=0.25cm,minimum width=0.25cm] (b2) {};
\vertex[below right=0.33cm and 1cm of b2] (h4) {$h_k$};
\vertex[above right=0.25cm and 0.75cm of b2, blob, minimum height=0.25cm,minimum width=0.25cm] (h1) {};
\vertex[above right=0.25cm and 0.5cm of h1] (h2) {$h_i$};
\vertex[below right=0.25cm and 0.5cm of h1] (h3) {$h_j$};
\diagram* {
(i1) -- [scalar] (b1),
(i2) -- [scalar] (b1),
(b1) -- [scalar] (b2),
(b2) -- [scalar] (h1),
(h1) -- [scalar] (h2),
(h1) -- [scalar] (h3),
(b2) -- [scalar] (h4)
};
\end{feynman}
\end{tikzpicture},
\end{equation}
with $i,\,j,\,k\in\lbrace 3,4,5\rbrace$. These diagrams in SMEFT contain the following terms,
\begin{equation}
    \begin{aligned}
\mcM_{4}^{\mathrm{SMEFT}}\left(5h\right) \supset&\, i\,\frac{2925}{16}\frac{C_{\varphi D}^2 \lambda^3 v^7}{\Delta_{12}\Delta_{34}}+i\, 15 C_{\varphi D}^2\, v^5\lambda^2 \frac{\Delta_{12}+\Delta_{34}}{\Delta_{12}\,\Delta_{34}}\\
&-i\,\frac{45}{4}C_{\varphi D}^2\, v^5\lambda^2 \frac{ \sum_{i=1}^{2}\sum_{j=3}^{5}\Delta_{ij}\,+\,\sum_{k=3}^{4}\Delta_{k5} }{\Delta_{12}\Delta_{34}} \\
& - i \frac{3}{4}C_{\varphi D}^{2}\, v^3\lambda\frac{ \sum_{i=1}^{2}\sum_{j=3}^{5}\Delta_{ij}\,+\,\sum_{k=3}^{4}\Delta_{k5} }{\Delta_{12} } \\
& - i \frac{3}{4}C_{\varphi D}^{2}\, v^3\lambda\frac{ \sum_{i=1}^{2}\sum_{j=3}^{5}\Delta_{ij}\,-\Delta_{34}+\,\sum_{k=3}^{4}\Delta_{k5}}{\Delta_{34}},
    \end{aligned}
\end{equation}
which can be simplified to,
\begin{equation}
\begin{aligned}
    \mcM_{4}^{\mathrm{SMEFT}}\left(5h\right) \supset & \, i\,\frac{9}{4}C_{\varphi D}^2 v^3\lambda+i\frac{45}{2}C_{\varphi D}^{2} v^5\lambda^2\left(\frac{1}{\Delta_{12}}+\frac{1}{\Delta_{34}}\right) \\    
    & +i\frac{2025}{16}\frac{C_{\varphi D}^2 \lambda^3 v^7}{\Delta_{12}\Delta_{34}}+i\frac{3}{4}C_{\varphi D}^{2}v^3\lambda\left(\frac{\Delta_{34}}{\Delta_{12}}+\frac{\Delta_{12}}{\Delta_{34}}\right),
\end{aligned}
\end{equation}
by using Eq.~\eqref{eq:master_simplifying_relation}. The first three terms pose no problem for the matching to the on-shell amplitudes. However, the last term cannot be reproduced by them due to being a rational function of the propagators. In fact, the first dangerous term, $\propto\frac{\Delta_{34}}{\Delta_{12}}$, cancels out within this same class of diagrams once one considers the terms proportional to $\frac{1}{\Delta_{12}\Delta_{35}}$ and $\frac{1}{\Delta_{12}\Delta_{45}}$, which in turn generate other dangerous pieces. In total, the relevant contributions from these diagrams are,
\begin{equation}
\begin{aligned}
    \mcM_{4}^{\mathrm{SMEFT}}\left(5h\right) \supset&\, i\,\frac{15}{2}C_{\varphi D}^2 v^3\lambda+i\frac{45}{2}C_{\varphi D}^{2} v^5\lambda^2\left(\frac{91}{30\,\Delta_{12}}+\frac{1}{\Delta_{34}}+\frac{1}{\Delta_{35}}+\frac{1}{\Delta_{45}}\right)\\
    & +i\frac{2025}{16}\frac{C_{\varphi D}^2 \lambda^3 v^7}{\Delta_{12}}\left( \frac{1}{\Delta_{34}}+\frac{1}{\Delta_{35}}+\frac{1}{\Delta_{45}} \right)\\
    & +i\frac{3}{4}C_{\varphi D}^{2}v^3\lambda\left(\frac{\Delta_{12}}{\Delta_{34}}+\frac{\Delta_{12}}{\Delta_{35}}+\frac{\Delta_{12}}{\Delta_{45}}\right),
    \end{aligned}
\end{equation}
where we collected at the end the terms that cannot be reproduced within the on-shell approach.

These dangerous terms cancel out within the SMEFT amplitude when considering all diagrams, hence allowing for a clean match to the bootstrapped on-shell amplitude. Consider the class of diagrams,
\begin{equation}
 \mcM_ {5}^{\mathrm{SMEFT}}\left(5h\right)=
\begin{tikzpicture}[baseline=(h1.base)]
\begin{feynman}[inline=h4.base]
\vertex (i1) {$h_1$};
\vertex[below=1cm of i1] (i2) {$h_2$};
\vertex[right=1cm of i1, blob, minimum height=0.25cm,minimum width=0.25cm] (b1) {};
\vertex[right=1cm of i2, blob, minimum height=0.25cm,minimum width=0.25cm] (b2) {};
\vertex[right=1cm of b2, blob, minimum height=0.25cm,minimum width=0.25cm] (b3) {};
\vertex[above right=0.5cm and 0.866cm of b3] (h1) { $h_j$ };
\vertex[right=1cm of b1] (h2) { $h_i$ };
\vertex[below right=0.5cm and 0.866cm of b3] (h3) { $h_k$ };
\diagram* {
(i1) -- [scalar] (b1),
(i2) -- [scalar] (b2),
(b1) -- [scalar] (b2),
(b2) -- [scalar] (b3),
(b1) -- [scalar] (h2),
(b3) -- [scalar] (h1),
(b3) -- [scalar] (h3)
};
\end{feynman}
\end{tikzpicture} +\text{perm.}
\end{equation}
After several simplifications, one can see that this class of diagrams generates terms like,
\begin{equation}
    \mcM_{5}^{\mathrm{SMEFT}}\left(5h\right) \supset\, -i\frac{3}{4}C_{\varphi D}^{2}v^3\lambda\left(\frac{\Delta_{12}}{\Delta_{34}}+\frac{\Delta_{12}}{\Delta_{35}}+\frac{\Delta_{12}}{\Delta_{45}}\right),
\end{equation}
which exactly cancels the contribution from $\mcM_{4}^{\mathrm{SMEFT}}\left(5h\right)$. This cancellation is not limited to an specific order in WCs. In addition, $\mcM_{5}^{\mathrm{SMEFT}}\left(5h\right)$ generates further dangerous terms that are cancelled against the following amplitude,
\begin{equation}
\mcM_{6}^{\mathrm{SMEFT}}(5h)=
\begin{tikzpicture}[baseline=(h1.base)]
\begin{feynman}[inline=h4.base]
\vertex (i1) {$h_1$};
\vertex[below=2cm of i1] (i2) {$h_2$};
\vertex[right=0.866cm of i1, blob, minimum height=0.25cm,minimum width=0.25cm] (b1) {};
\vertex[right=0.866cm of i2, blob, minimum height=0.25cm,minimum width=0.25cm] (b2) {};
\vertex[below=1cm of b1, blob, minimum height=0.25cm,minimum width=0.25cm] (b3) {};
\vertex[right=0.866cm of b3] (h1) { $h_4$ };
\vertex[right=0.866cm of b1] (h2) { $h_3$ };
\vertex[right=0.866cm of b2] (h3) { $h_5$ };
\diagram* {
(i1) -- [scalar] (b1),
(i2) -- [scalar] (b2),
(b1) -- [scalar] (b3),
(b2) -- [scalar] (b3),
(b3) -- [scalar] (h1),
(b1) -- [scalar] (h2),
(b2) -- [scalar] (h3)
};
\end{feynman}
\end{tikzpicture}
+\text{perm. }
\end{equation}
Finally, notice that these three classes of amplitudes are of the same topology and we have differentiated them only for the sake of clarity.

\section{On-shell amplitudes}
\label{app:all_os_amps}
In this section, we collect expression for all on-shell three-, four- and five-point on-shell amplitudes derived and used in this work.

\subsection{Three-point}
\label{app:3pt_os_amps}

\subsubsection{$gg\to h$}

\begin{align}
    \mcM\left( g_{1}^{a,+};g_{2}^{b,+};h \right) & = i \, \delta^{ab} \left[1|2\right]^{2} c_{ggh} \, .     \label{eq:ggh:++:app} \\
    \mcM\left( g_{1}^{a,+};g_{2}^{b,-};h \right) & = 0 \,.
    \label{eq:ggh:+-:app}
\end{align}

\subsubsection{$hh\to h$}
\begin{equation}
    \mcM\left( h;h;h \right) = i\,c_{3h} \, . 
    \label{eq:hhh:app}
\end{equation}

\subsection{Four-point}
\label{app:4pt_os_amps}

\subsubsection{$gg\to hh$}
\subsubsection*{Non-factorizable contributions}

\begin{align}
&\mcM\left(g_{1}^{a,+};g_{2}^{b,+};h_3;h_4\right)_{\text{NF}}= i\,\delta^{ab} \left[1|2\right]^2 c_{gghh}^{++},\quad 
    c_{gghh}^{++}=\frac{1}{\bar{\Lambda}^2}\sum_{i,j} \frac{{c_{gghh}^{++,(ij)}}}{\bar\Lambda^{2\left(i+j\right)}} s_{12}^{i}\ s_{13}^{j} \, . \label{eq:amp_gghh:NF:++:app} \\
&\mcM\left(g_{1}^{a,+};g_{2}^{b,-};h_3;h_4\right)_{\text{NF}}= \, i\, \delta^{ab} \left[ 1| \mathbf{3} - \mathbf{4} | 2 \rangle \right.^2 c_{gghh}^{+-},\quad 
    c_{gghh}^{+-}=\frac{1}{\bar{\Lambda}^4}\sum_{i,j} \frac{{c_{gghh}^{+-,(ij)}}}{\bar\Lambda^{2\left(i+j\right)}} s_{12}^{i}\ s_{13}^{j} \,.
    \label{eq:amp_gghh:NF:+-:app}
\end{align}

\subsubsection*{Factorizable contributions}
\begin{align}
    & \mcM\left( g_{1}^{a,+};g_{2}^{b,+};h_3;h_4 \right)_{s\text{--ch.}} = - \delta^{ab} \frac{c_{3h} \, c_{ggh} }{s_{12}-m_h^2}\squaket{1}{2}^2 \, . 
    \label{eq:amp_gghh:F:++:app} \\        
    &\mcM\left(g_{1}^{a,+}; g_{2}^{b,-}; h_{3}; h_{4}\right)_{t+u\text{--ch.}} = -\delta^{ab} \frac{|c_{ggh}|^2}{4} \left[ 1|\mathbf{3}-\mathbf{4}|2\rangle\right.^2 \left( \frac{1}{s_{13}} + \frac{1}{s_{23}} \right) \, .
\label{eq:amp_gghh:F:+-:app}
\end{align}

\subsubsection{$hh\to hh$}

\subsubsection*{Non-factorizable contribution}
\begin{equation}
\begin{aligned}
    \mcM\left(h;h;h;h\right)_{\text{NF}} = i \, c_{4h},\quad 
    c_{4h}=\sum_{i,j} \frac{c_{4h}^{(ij)}}{\bar\Lambda^{2\left(i+j\right)}} s_{12}^{i}\ s_{13}^{j}\, .
\end{aligned}
\label{eq:amp_hhhh_NF:app}
\end{equation}

\subsubsection*{Factorizable contributions}
\begin{equation}
\mcM\left( h;h;h;h \right)_{s+t+u\text{--ch.}}= c_{3h}^2\left(\frac{1}{s_{12}-m_h^2}+\frac{1}{s_{13}-m_h^2}+\frac{1}{s_{23}-m_h^2}\right) \, .
\label{eq:amp_hhhh_F:app}
\end{equation}

\subsection{Five-point}
\label{app:5pt_os_amps}

\subsubsection{$gg\to hhh$}
\subsubsection*{Non-factorizable contributions}
\label{app:5pt_os_amps_gghhh}
\begin{align}
   \mcM\left(g_{1}^{a,+}; g_{2}^{b,+}; h_{3}; h_{4}; h_{5}\right)_{\text{NF}}
   & = i \,\delta^{ab}\, c_{gghhh}^{++,\, (1)} [1|2]^2 \nonumber \\
   &+ i\,\delta^{ab}\, c_{gghhh}^{++,\, (2)}\left( [1|\mathbf{3}\mathbf{4}|2]  [1|\mathbf{4}\mathbf{3}|2] +
   [1|\mathbf{3}\mathbf{5}|2]  [1|\mathbf{5}\mathbf{3}|2] +
   [1|\mathbf{4}\mathbf{5}|2]  [1|\mathbf{5}\mathbf{4}|2]\right) , \nonumber \\
   c_{gghhh}^{++,\, (1)} & = \frac{1}{\bar\Lambda^3}\sum_{i,j,k,l,m=0} \frac{c_{gghhh}^{++,\, (1)(ijklm)}}{\bar\Lambda^{2\left(i+j+k+l+m\right)}} s_{12}^{i}\ s_{13}^{j}\ s_{14}^{k}\ s_{23}^{l}\ s_{24}^{m} \, , 
     \label{eq:gghhh:OS:NF:++:app}\\
   c_{gghhh}^{++,\, (2)} & = \frac{1}{\bar\Lambda^7}\sum_{i,j,k,l,m=0} \frac{c_{gghhh}^{++,\, (2)(ijklm)}}{\bar\Lambda^{2\left(i+j+k+l+m\right)}} s_{12}^{i}\ s_{13}^{j}\ s_{14}^{k}\ s_{23}^{l}\ s_{24}^{m} \, . \nonumber
\end{align}

\begin{align}
\mcM\left(g_{1}^{a,+}; g_{2}^{b,-}; h_{3}; h_{4}; h_{5}\right)_{\text{NF}}
         & = i \, 
         \delta^{ab}\, c_{gghhh}^{+-} \left([1|\mathbf{3}\tret{2}{}{}{}^2 + [1|\mathbf{4}\tret{2}{}{}{}^2 + [1|\mathbf{5}\tret{2}{}{}{}^2 \right) \, , \nonumber \\
  c_{gghhh}^{+-} & = \frac{1}{\bar\Lambda^5}\sum_{i,j,k,l,m=0} \frac{c_{gghhh}^{+-,\, (ijklm)}}{\bar\Lambda^{2\left(i+j+k+l+m\right)}} s_{12}^{i}\ s_{13}^{j}\ s_{14}^{k}\ s_{23}^{l}\ s_{24}^{m} \, .     \label{eq:gghhh:OS:NF:+-:app}
\end{align}

\subsubsection*{Factorizable contributions}

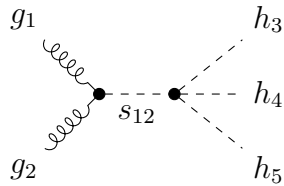
\begin{equation*}
\vcenter{\hbox{
\begin{tikzpicture}
\begin{feynman}
\vertex (i1) {$g_1$};
\vertex[below=2cm of i1] (i2) {$g_2$};
\vertex[below right=1cm and 1cm of i1, dot] (b1) {};
\vertex[right=1cm of b1, dot] (b2) {};
\vertex[below right=1cm and 1.25cm of b2] (h5) {$h_5$};
\vertex[right=1.25cm of b2] (h4) {$h_4$};
\vertex[above right=1cm and 1.25cm of b2] (h3) {$h_3$};
\diagram* {
  (i1) -- [gluon] (b1),
  (i2) -- [gluon] (b1),
  (b1) -- [scalar, edge label'=\(s_{12}\)] (b2),
  (b2) -- [scalar] (h5),
  (b2) -- [scalar] (h4),
  (b2) -- [scalar] (h3)
};
\end{feynman}
\end{tikzpicture}
}}
\end{equation*}
\begin{align}
& \mcM \left(g_{1}^{a,+}; g_{2}^{b,+}; h_{3}; h_{4}; h_{5}\right)_{\text{F}1} = \delta^{ab} {\squaket{1}{2}}^2\ \frac{ c_{ggh} \ c_{4h}\left(s_{34},s_{35}\right)  }{s_{12}-m_h^2} \, . 
\label{eq:gghhh:OS:F1:++:app}\\
& \mcM \left(g_{1}^{a,+}; g_{2}^{b,-}; h_{3}; h_{4}; h_{5}\right)_{\text{F}1} =  0 \, .
\label{eq:gghhh:OS:F1:+-:app}
\end{align}

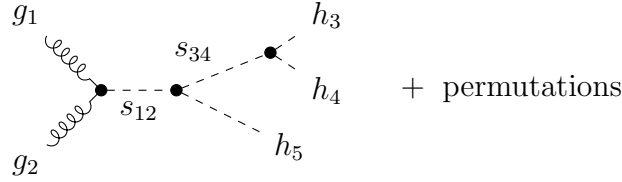
\begin{equation*}
\vcenter{\hbox{%
\begin{tikzpicture}
\begin{feynman}
\vertex (i1) {$g_1$};
\vertex[below=2cm of i1] (i2) {$g_2$};
\vertex[below right=1cm and 1cm of i1, dot] (b1) {};
\vertex[right=1cm of b1, dot] (b2) {};
\vertex[below right=0.75cm and 1.5cm of b2] (h4) {$h_5$};
\vertex[above right=0.5cm and 1.25cm of b2, dot] (h1) {};
\vertex[above right=0.5cm and 0.75cm of h1] (h2) {$h_3$};
\vertex[below right=0.5cm and 0.75cm of h1] (h3) {$h_4$};
\diagram* {
  (i1) -- [gluon] (b1),
  (i2) -- [gluon] (b1),
  (b1) -- [scalar, edge label'=\(s_{12}\)] (b2),
  (b2) -- [scalar, edge label=\(s_{34}\)] (h1),
  (h1) -- [scalar] (h2),
  (h1) -- [scalar] (h3),
  (b2) -- [scalar] (h4)
};
\end{feynman}
\end{tikzpicture}
}}
\quad  +\ \vcenter{\hbox{\text{permutations}}}
\end{equation*}
\begin{align}
& \mcM \left(g_{1}^{a,+}; g_{2}^{b,+}; h_{3}; h_{4}; h_{5}\right)_{\text{F}2} = -i \, \delta^{ab} {\squaket{1}{2}}^2\, c_{ggh} \, c_{3h}^2 \, 
\sum_{ij\in\{34,35,45\}}
\left(
  \frac{1}{s_{12}-m_h^2}\,
  \frac{1}{s_{ij} - m_h^2}
\right). \label{eq:gghhh:OS:F2:++:app} \\
& \mcM \left(g_{1}^{a,+}; g_{2}^{b,-}; h_{3}; h_{4}; h_{5}\right)_{\text{F}2} =  0\, .
\label{eq:gghhh:OS:F2:+-:app}
\end{align}

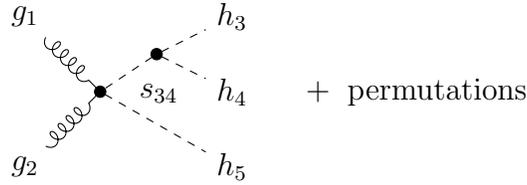
\begin{equation*}
\vcenter{\hbox{
\begin{tikzpicture}
\begin{feynman}
\vertex (i1) {$g_1$};
\vertex[below=2cm of i1] (i2) {$g_{2}$};
\vertex[below right=1cm and 1cm of i1, dot] (b1) {};
\vertex[above right=0.5cm and 0.75cm of b1, dot] (b2) {};
\vertex[right=2.75cm of i2] (h3) {$h_5$};
\vertex[above=1cm of h3] (h4) {$h_4$};
\vertex[above=1cm of h4] (h5) {$h_3$};
\diagram* {
  (i1) -- [gluon] (b1),
  (i2) -- [gluon] (b1),
  (b2) -- [scalar, edge label=\(s_{34}\)] (b1),
  (b1) -- [scalar] (h3),
  (b2) -- [scalar] (h4),
  (b2) -- [scalar] (h5)
};
\end{feynman}
\end{tikzpicture}
}}
\quad +\ \vcenter{\hbox{\text{permutations}}}
\end{equation*}
\begin{align}
\mcM \left(g_{1}^{a,+}; g_{2}^{b,+}; h_{3}; h_{4}; h_{5}\right)_{\text{F}3} & = \delta^{ab} c_{3h} [1|2]^2 \,
\left(\frac{c_{gghh}^{++}\left(s_{12},s_{25}\right)}{s_{34}-m_h^2}+
\frac{c_{gghh}^{++}\left(s_{12},s_{24}\right)}{s_{35}-m_h^2}+
\frac{c_{gghh}^{++}\left(s_{12},s_{23}\right)}{s_{45}-m_h^2}\right). 
\label{eq:gghhh:OS:F3:++:app} \\
\mcM \left(g_{1}^{a,+}; g_{2}^{b,-}; h_{3}; h_{4}; h_{5}\right)_{\text{F}3} & = \delta^{ab} c_{3h} \, \Bigg(\frac{c_{gghh}^{+-}\left(s_{12},s_{25}\right)}{s_{34}-m_h^2}[1|\mathbf{3+4-5}\tret{2}{}{}{}^2 
\label{eq:gghhh:OS:F3:+-:app} \\ 
&+ 
\frac{c_{gghh}^{+-}\left(s_{12},s_{24}\right)}{s_{35}-m_h^2}[1|\mathbf{5+3-4}\tret{2}{}{}{}^2 +
\frac{c_{gghh}^{+-}\left(s_{12},s_{23}\right)}{s_{45}-m_h^2}[1|\mathbf{5+4-3}\tret{2}{}{}{}^2 \Bigg). \nonumber
\end{align}

\newpage

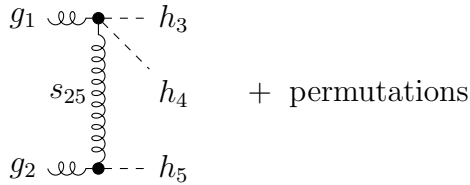
\begin{equation*}
\vcenter{\hbox{
\begin{tikzpicture}
\begin{feynman}
\vertex (i1) {$g_1$};
\vertex[below=2cm of i1] (i2) {$g_2$};
\vertex[right=1cm of i1, dot] (b1) {};
\vertex[right=1cm of i2, dot] (b2) {};
\vertex[right=1cm of b2] (h3) {$h_5$};
\vertex[below right=1cm and 1cm of b1] (h4) {$h_4$};
\vertex[right=1cm of b1] (h5) {$h_3$};
\diagram* {
  (i1) -- [gluon] (b1),
  (i2) -- [gluon] (b2),
  (b2) -- [gluon, edge label=\(s_{25}\)] (b1),
  (b2) -- [scalar] (h3),
  (b1) -- [scalar] (h4),
  (b1) -- [scalar] (h5)
};
\end{feynman}
\end{tikzpicture}
}}
\quad +\ \vcenter{\hbox{\text{permutations}}}
\end{equation*}

\begin{align}
\mcM \left(g_{1}^{a,+}; g_{2}^{b,+}; h_{3}; h_{4}; h_{5}\right)_{\text{F}4} & = \delta^{ab} c_{ggh} \times 
\sum_{\substack{i,j\in\{1,2\}\\ i\neq j}} \Bigg(c_{gghh}^{++}\left(s_{34}, s_{i3}\right) \frac{{\squa{j}{}{}(j+\mathbf{5})(\mathbf{3-4})\squet{i}{}{}}^2
}{s_{j5}} \label{eq:gghhh:OS:F4:++:app} \\
 & + c_{gghh}^{++}\left(s_{35}, s_{i5}\right) \frac{
{\squa{j}{}{}(j+\mathbf{4})(\mathbf{3-5})\squet{i}{}{}}^2
}{s_{j4}} \nonumber \\
& +
c_{gghh}^{++}\left(s_{45}, s_{i4}\right) \frac{
{\squa{j}{}{}(j+\mathbf{3})(\mathbf{4-5})\squet{i}{}{}}^2
}{s_{j3}} \Bigg) . \nonumber
\end{align}

\begin{align}
\mcM \left(g_{1}^{a,+}; g_{2}^{b,-}; h_{3}; h_{4}; h_{5}\right)_{\text{F}4} & = -\delta^{ab} \times \label{eq:gghhh:OS:F4:+-:app} \\ & \sum_{i=3}^{5}\left(\frac{ c_{ggh}\,\left(c_{gghh}^{++}\left(s_{12},s_{1i}\right)\right)^*}{s_{1i}}+
\frac{c_{ggh}^{*}\,c_{gghh}^{++}\left(s_{12},s_{2i}\right)}{s_{2i}}\right){\squa{1}{}{}\mathbf{i}\tret{2}{}{}}^2. \nonumber
\end{align}

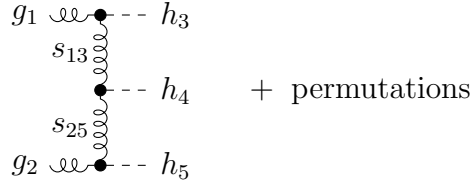
\begin{equation*}
\vcenter{\hbox{
\begin{tikzpicture}
\begin{feynman}
\vertex (i1) {$g_1$};
\vertex[below=2cm of i1] (i2) {$g_2$};
\vertex[right=1cm of i1, dot] (b1) {};
\vertex[right=1cm of i2, dot] (b2) {};
\vertex[above=1cm of b2, dot] (b3) {};
\vertex[right=1cm of b2] (h3) {$h_5$};
\vertex[right=1cm of b3] (h4) {$h_4$};
\vertex[right=1cm of b1] (h5) {$h_3$};
\diagram* {
  (i1) -- [gluon] (b1),
  (i2) -- [gluon] (b2),
  (b2) -- [gluon, edge label=\(s_{25}\)] (b3),
  (b3) -- [gluon, edge label=\(s_{13}\)] (b1),
  (b2) -- [scalar] (h3),
  (b3) -- [scalar] (h4),
  (b1) -- [scalar] (h5)
};
\end{feynman}
\end{tikzpicture}
}}
\quad +\ \vcenter{\hbox{\text{permutations}}}
\end{equation*}
\begin{align}
 \mcM \left(g_{1}^{a,+}; g_{2}^{b,+}; h_{3}; h_{4}; h_{5}\right)_{\text{F}5} & = i \, \delta^{ab} \, c_{ggh}|c_{ggh}|^2 \times  \label{eq:gghhh:OS:F5:++:app} \\
& \sum_{\substack{i,j\in\{1,2\}\\ i\neq j}} \left(\frac{ {\squa{i}{}{}(i+\mathbf{3})(j+\mathbf{5})\squet{j}{}{}}^2
}{s_{i3}\, s_{j5}} +
\frac{ 
{\squa{i}{}{}(i+\mathbf{5})(j+\mathbf{4})\squet{j}{}{}}^2
}{s_{i5}\, s_{j4}} \right. \nonumber \\ 
 & \left. +
\frac{ {\squa{i}{}{}(i+\mathbf{4})(j+\mathbf{3})\squet{j}{}{}}^2}{s_{i 4}\, s_{j3}} \right) . \nonumber\\
 \mcM \left(g_{1}^{a,+}; g_{2}^{b,-}; h_{3}; h_{4}; h_{5}\right)_{\text{F}5} & =  0 \, . \label{eq:gghhh:OS:F5:+-:app}
\end{align}

\newpage

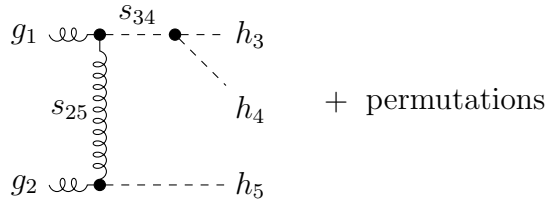
\begin{equation*}
\vcenter{\hbox{
\begin{tikzpicture}
\begin{feynman}
\vertex (i1) {$g_1$};
\vertex[below=2cm of i1] (i2) {$g_2$};
\vertex[right=1cm of i1, dot] (b1) {};
\vertex[right=1cm of i2, dot] (b2) {};
\vertex[right=1cm of b1, dot] (b3) {};
\vertex[right=2cm of b2] (h3) {$h_5$};
\vertex[below right=1cm and 2cm of b1] (h4) {$h_4$};
\vertex[right=1cm of b3] (h5) {$h_3$};
\diagram*{
  (i1) -- [gluon] (b1),
  (i2) -- [gluon] (b2),
  (b2) -- [gluon, edge label=\(s_{25}\)] (b1),
  (b1) -- [scalar, edge label=\(s_{34}\)] (b3),
  (b2) -- [scalar] (h3),
  (b3) -- [scalar] (h4),
  (b3) -- [scalar] (h5)
};
\end{feynman}
\end{tikzpicture}
}}
\quad +\ \vcenter{\hbox{\text{permutations}}}
\end{equation*}
\begin{align}
\mcM \left(g_{1}^{a,+}; g_{2}^{b,+}; h_{3}; h_{4}; h_{5}\right)_{\text{F}6} & = 0 \, .\label{eq:gghhh:OS:F6:++:app} \\
\mcM \left(g_{1}^{a,+}; g_{2}^{b,-}; h_{3}; h_{4}; h_{5}\right)_{\text{F}6} & = i \, \delta^{ab} \,  c_{3h} \, |c_{ggh}|^2 \times \label{eq:gghhh:OS:F6:+-:app}\\
&\sum_{i\in\{1,2\}} \left( 
\frac{{\squa{1}{}{}\mathbf{3}\tret{2}{}{}}^2}{s_{45}-m_h^2}\frac{1}{s_{i3}} + \frac{{\squa{1}{}{}\mathbf{4}\tret{2}{}{}}^2}{s_{35}-m_h^2}\frac{1}{s_{i4}} + \frac{{\squa{1}{}{}\mathbf{5}\tret{2}{}{}}^2}{s_{34}-m_h^2}\frac{1}{s_{i5}}
\right)\, . \nonumber
\end{align}

\subsubsection{$hh\to hhh$}

\subsubsection*{Non-factorizable contributions}
\begin{equation}
\begin{aligned}
\mcM\left(h;h;h;h;h\right)_{\text{NF}}=i\, c_{5h} \, ,\quad 
    c_{5h}=\frac{1}{\bar\Lambda}\sum_{i,j,k,l,m=0} \frac{c_{5h}^{(ijklm)}}{\bar\Lambda^{2\left(i+j+k+l+m\right)}} s_{12}^{i}\ s_{13}^{j}\ s_{14}^{k}\ s_{23}^{l}\ s_{24}^{m} \, .
\end{aligned}
\label{eq:hhhhh:OS:NF:app}
\end{equation}

\subsubsection*{Factorizable contributions}

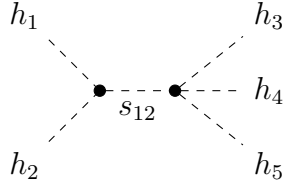
\begin{equation*}
\vcenter{\hbox{
\begin{tikzpicture}
\begin{feynman}
\vertex (i1) {$h_1$};
\vertex[below=2cm of i1] (i2) {$h_2$};
\vertex[below right=1cm and 1cm of i1, dot] (b1) {};
\vertex[right=1cm of b1, dot] (b2) {};
\vertex[below right=1cm and 1.25cm of b2] (h5) {$h_5$};
\vertex[right=1.25cm of b2] (h4) {$h_4$};
\vertex[above right=1cm and 1.25cm of b2] (h3) {$h_3$};
\diagram* {
  (i1) -- [scalar] (b1),
  (i2) -- [scalar] (b1),
  (b1) -- [scalar, edge label'=\(s_{12}\)] (b2),
  (b2) -- [scalar] (h5),
  (b2) -- [scalar] (h4),
  (b2) -- [scalar] (h3)
};
\end{feynman}
\end{tikzpicture}
}}
\end{equation*}

\begin{equation}
\mcM\!\left(h;h;h;h;h\right)_{\mathrm{F}1}
= \frac{c_{3h}\,c_{4h}\!\left(s_{34},\,s_{35}\right)}{s_{12}-m_h^2}.
\label{eq:hhhhh:OS:F1:app}
\end{equation}

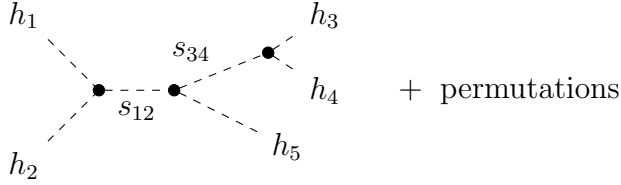
\begin{equation*}
\vcenter{\hbox{
\begin{tikzpicture}
\begin{feynman}
\vertex (i1) {$h_1$};
\vertex[below=2cm of i1] (i2) {$h_2$};
\vertex[below right=1cm and 1cm of i1, dot] (b1) {};
\vertex[right=1cm of b1, dot] (b2) {};
\vertex[below right=0.75cm and 1.5cm of b2] (h4) {$h_5$};
\vertex[above right=0.5cm and 1.25cm of b2, dot] (h1) {};
\vertex[above right=0.5cm and 0.75cm of h1] (h2) {$h_3$};
\vertex[below right=0.5cm and 0.75cm of h1] (h3) {$h_4$};
\diagram* {
  (i1) -- [scalar] (b1),
  (i2) -- [scalar] (b1),
  (b1) -- [scalar, edge label'=\(s_{12}\)] (b2),
  (b2) -- [scalar, edge label=\(s_{34}\)] (h1),
  (h1) -- [scalar] (h2),
  (h1) -- [scalar] (h3),
  (b2) -- [scalar] (h4)
};
\end{feynman}
\end{tikzpicture}
}}
\quad +\ \vcenter{\hbox{\text{permutations}}}
\end{equation*}
\begin{equation}
\mcM\!\left(h;h;h;h;h\right)_{\mathrm{F}2}
= -\,i\, c_{3h}^3 \,
\sum_{ij\in\{34,35,45\}}
\left(
  \frac{1}{s_{12}-m_h^2}\,
  \frac{1}{s_{ij} - m_h^2}
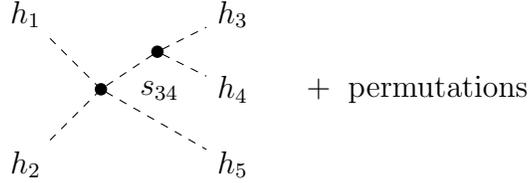
\right).
\label{eq:hhhhh:OS:F2:app}
\end{equation}

\begin{equation*}
\vcenter{\hbox{
\begin{tikzpicture}
\begin{feynman}
\vertex (i1) {$h_1$};
\vertex[below=2cm of i1] (i2) {$h_{2}$};
\vertex[below right=1cm and 1cm of i1, dot] (b1) {};
\vertex[above right=0.5cm and 0.75cm of b1, dot] (b2) {};
\vertex[right=2.75cm of i2] (h3) {$h_5$};
\vertex[above=1cm of h3] (h4) {$h_4$};
\vertex[above=1cm of h4] (h5) {$h_3$};
\diagram* {
  (i1) -- [scalar] (b1),
  (i2) -- [scalar] (b1),
  (b2) -- [scalar, edge label=\(s_{34}\)] (b1),
  (b1) -- [scalar] (h3),
  (b2) -- [scalar] (h4),
  (b2) -- [scalar] (h5)
};
\end{feynman}
\end{tikzpicture}
}}
\quad +\ \vcenter{\hbox{\text{permutations}}}
\end{equation*}

\begin{equation}
\mcM\!\left(h;h;h;h;h\right)_{\mathrm{F}3}
=  c_{3h} \left(\frac{\,c_{4h}(s_{12},\,s_{25})}{s_{34}-m_h^2} + 
\frac{\,c_{4h}(s_{12},\,s_{24})}{s_{35}-m_h^2} + 
\frac{\,c_{4h}(s_{12},\,s_{23})}{s_{45}-m_h^2}\right).
\label{eq:hhhhh:OS:F3:app}
\end{equation}

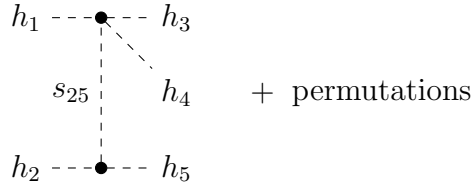
\begin{equation*}
\vcenter{\hbox{%
\begin{tikzpicture}
\begin{feynman}
\vertex (i1) {$h_1$};
\vertex[below=2cm of i1] (i2) {$h_2$};
\vertex[right=1cm of i1, dot] (b1) {};
\vertex[right=1cm of i2, dot] (b2) {};
\vertex[right=1cm of b2] (h3) {$h_5$};
\vertex[below right=1cm and 1cm of b1] (h4) {$h_4$};
\vertex[right=1cm of b1] (h5) {$h_3$};
\diagram* {
  (i1) -- [scalar] (b1),
  (i2) -- [scalar] (b2),
  (b2) -- [scalar, edge label=\(s_{25}\)] (b1),
  (b2) -- [scalar] (h3),
  (b1) -- [scalar] (h4),
  (b1) -- [scalar] (h5)
};
\end{feynman}
\end{tikzpicture}
}}
\quad +\ \vcenter{\hbox{\text{permutations}}}
\end{equation*}

\begin{equation}
\mcM\!\left(h;h;h;h;h\right)_{\mathrm{F}4}
= c_{3h}\,
\sum_{\substack{i,j\in\{1,2\}\\ i\neq j}}
\left(
   \frac{c_{4h}(s_{34},s_{i3})}{s_{j5}-m_h^2}
 + \frac{c_{4h}(s_{45},s_{i4})}{s_{j3}-m_h^2}
 + \frac{c_{4h}(s_{35},s_{i5})}{s_{j4}-m_h^2}
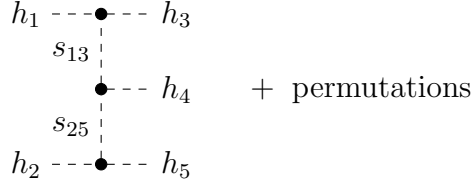
\right) \, .
\label{eq:hhhhh:OS:F4:app}
\end{equation}

\begin{equation*}
\vcenter{\hbox{%
\begin{tikzpicture}
\begin{feynman}
\vertex (i1) {$h_1$};
\vertex[below=2cm of i1] (i2) {$h_2$};
\vertex[right=1cm of i1, dot] (b1) {};
\vertex[right=1cm of i2, dot] (b2) {};
\vertex[above=1cm of b2, dot] (b3) {};
\vertex[right=1cm of b2] (h3) {$h_5$};
\vertex[right=1cm of b3] (h4) {$h_4$};
\vertex[right=1cm of b1] (h5) {$h_3$};
\diagram* {
  (i1) -- [scalar] (b1),
  (i2) -- [scalar] (b2),
  (b2) -- [scalar, edge label=\(s_{25}\)] (b3),
  (b3) -- [scalar, edge label=\(s_{13}\)] (b1),
  (b2) -- [scalar] (h3),
  (b3) -- [scalar] (h4),
  (b1) -- [scalar] (h5)
};
\end{feynman}
\end{tikzpicture}
}}
\quad +\ \vcenter{\hbox{\text{permutations}}}
\end{equation*}

\begin{equation}
\mcM\!\left(h;h;h;h;h\right)_{\mathrm{F}5}
= -\,i\,c_{3h}^{\,3}\,
\sum_{\substack{i,j \in \{3,4,5\}\\ i\neq j}}
\frac{1}{s_{1i}-m_h^2}\,
\frac{1}{s_{2j}-m_h^2} \, .
\label{eq:hhhhh:OS:F5:app}
\end{equation}

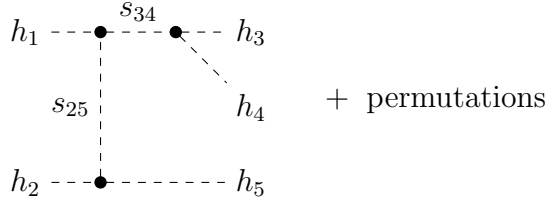
\begin{equation*}
\vcenter{\hbox{%
\begin{tikzpicture}
\begin{feynman}
\vertex (i1) {$h_1$};
\vertex[below=2cm of i1] (i2) {$h_2$};
\vertex[right=1cm of i1, dot] (b1) {};
\vertex[right=1cm of i2, dot] (b2) {};
\vertex[right=1cm of b1, dot] (b3) {};
\vertex[right=2cm of b2] (h3) {$h_5$};
\vertex[below right=1cm and 2cm of b1] (h4) {$h_4$};
\vertex[right=1cm of b3] (h5) {$h_3$};
\diagram*{
  (i1) -- [scalar] (b1),
  (i2) -- [scalar] (b2),
  (b2) -- [scalar, edge label=\(s_{25}\)] (b1),
  (b1) -- [scalar, edge label=\(s_{34}\)] (b3),
  (b2) -- [scalar] (h3),
  (b3) -- [scalar] (h4),
  (b3) -- [scalar] (h5)
};
\end{feynman}
\end{tikzpicture}
}}
\quad +\ \vcenter{\hbox{\text{permutations}}}
\end{equation*}

\begin{align}
\mcM\!\left(h;h;h;h;h\right)_{\mathrm{F}6}
  & =  -\,i\,c_{3h}^{\,3} \times \label{eq:hhhhh:OS:F6:app}
\\
&\sum_{i\in\{1,2\}}
\left(
  \frac{1}{s_{i5}-m_h^2}\,\frac{1}{s_{34}-m_h^2}
 +\frac{1}{s_{i4}-m_h^2}\,\frac{1}{s_{35}-m_h^2}
 +\frac{1}{s_{i3}-m_h^2}\,\frac{1}{s_{45}-m_h^2}
\right) \, .
\nonumber
\end{align}

\section{Complete matching results}
\label{app:full_matching_res}

In this appendix, we collect the complete results of matching the on-shell scattering amplitudes to dim-8 SMEFT or NNLO HEFT. For the matching to SMEFT, we identify the on-shell scale $\bar\Lambda$ with the SMEFT cutoff $\Lambda$. The identification of $\bar\Lambda$ with a HEFT parameter is discussed in the main text and here we present the HEFT results without equating $\bar\Lambda$ to anything. 

In presenting the results, we neglect coefficients that vanish trivially due to the dimensional analysis or symmetry arguments (see, e.g., section~\ref{sec:HH_nonfact}).

\subsection{Three-point amplitudes}
\label{appendix:ggh}

\subsubsection{$gg\to h$}
SMEFT:
\begin{equation}
    \begin{aligned}
-i\, c_{ggh}=& \frac{\,2^{\frac{3}{4}}}{\sqrt{G_F}\,\Lambda^2}\left(C_{\varphi G} + i C_{\varphi \tilde{G}} \right)\\
        & +\frac{2^{\frac{1}{4}}}{G_{F}^{\frac{3}{2}}\,\Lambda^4}\left(C_{\varphi G} + i C_{\varphi \tilde{G}} \right)\left( C_{\varphi\Box} - \frac{1}{4} C_{\varphi D} + 2 C_{\varphi G} \right)\\
        & + \frac{2^{\frac{1}{4}}}{G_{F}^{\frac{3}{2}}\,\Lambda^4} \left( C_{G^2\varphi^4}^{(1)} + i C_{G^2\varphi^4}^{(2)} \right),
    \end{aligned}
    \label{eq:match_ons_smeft_ggh_Gf}
\end{equation}

HEFT:
\begin{equation}
    \begin{aligned}
        -i\, c_{ggh}=& -\frac{g_S^2 }{2^{3/4}} \sqrt{G_F} \left( C_{GH}^{(1)}+ i\, C_{\tilde{G}H}^{(1)} \right).
    \end{aligned}
    \label{eq:match_ons_smeft_ggh_HEFT}
\end{equation}

\subsubsection{$hh\to h$}
\label{appendix:hhh}

For the coefficient of the 3-Higgs boson on-shell amplitude, the matching with SMEFT yields,
\begin{equation}
    \begin{aligned}
    -i\,c_{3h} = \ & 3 \sqrt[\leftroot{-3}\uproot{3} 4]{2} \sqrt{G_F} m_{h}^2 + \frac{3}{8\,G_F^{3/2}\,\Lambda^2}\left(-8 \sqrt[4]{2} C^\varphi + 12\ 2^{3/4} \,C_{\varphi\Box}\,
   G_F\ m_h^2 - 3\ 2^{3/4} C_{\varphi D} G_F\ m_h^2 \right)\\
   &-\frac{3}{64\,G_{F}^{5/2} \Lambda^4 }\left( 96\ 2^{3/4} C_{\varphi} C_{\varphi\Box}-24\ 2^{3/4} C_{\varphi}
   C_{\varphi D} -192 \sqrt[4]{2} C_{\varphi^6 \Box}\, G_{F}
   \, m_{h}^2 + 64 \ 2^{3/4} C_{\varphi^8} \right. \\ & \left. +20 \sqrt[4]{2} C_{\varphi^6 D^2} G_{F} m_{h}^2 - 240 \sqrt[4]{2} (C_{\varphi\Box})^2 G_{F}
   m_{h}^2 \right. \\ & \left.+120 \sqrt[4]{2} C_{\varphi\Box} C_{\varphi D} G_{F}
   m_{h}^2 - 15 \sqrt[4]{2} (C_{\varphi D})^2 G_{F} m_{h}^2 \right) .
    \end{aligned}
    \label{eq:match_ons_smeft_hhh_Gf}
\end{equation}

In HEFT, we have,
\begin{equation}
    \begin{aligned}
    -i\,c_{3h} = \ & 3\,m_h^2\sqrt[4]{2}\sqrt{G_F}\left(1+\Delta\lambda_{3} \right) .
    \end{aligned}
    \label{eq:match_ons_smeft_hhh_HEFT}
\end{equation}
    
\subsection{Four-point amplitudes}

\subsubsection{$hh\to hh$}
\label{appendix:hhhh}

Here, we expand the on-shell coefficient as,
\begin{equation}
    c_{4h} = c_{4h}^{(00)}+\frac{c_{4h}^{(10)}}{\Lambda^2}\, s_{12} + \frac{c_{4h}^{(01)}}{\Lambda^2}\, s_{13} + \frac{c_{4h}^{(20)}}{\Lambda^4}\, s_{12}^2 + \frac{c_{4h}^{(02)}}{\Lambda^4}\, s_{13}^2 + \frac{c_{4h}^{(11)}}{\Lambda^4}\, s_{12}\,s_{13}+... \, ,
\label{eq:coeff_onshell_hhhh_expansion}
\end{equation}
which can be matched to SMEFT up to dimension-8 in the $G_F$ input scheme to yield,
\begin{equation}
    \begin{aligned}
        -i\,c_{4h}^{(00)} = & 3\ m_h^2 \sqrt{2}\ G_F + \frac{2}{\Lambda^2} \left(-\frac{9 \sqrt{2} C_{\varphi}}{G_{F}}+25
   C_{\varphi \Box} m_{h}^2-\frac{25 C_{\varphi D}
   m_{h}^2}{4}\right)\\
   &-\frac{3}{G_F^2\ \Lambda^4}\left(36 C_{\varphi} C_{\varphi \Box}-9 C_{\varphi} C_{\varphi D}+16 C_{\varphi^8}\right) 
   -6\frac{m_h^4}{\Lambda^4}\left(C_{\varphi^4 D^4}^{(1)}+C_{\varphi^4 D^4}^{(2)}+C_{\varphi^4 D^4}^{(3)}\right)
   \\
   &+\frac{3}{4\sqrt{2}}\frac{m_h^2}{G_F\ \Lambda^4} \left(172 C_{\varphi^6\Box}-21 C_{\varphi^6D^2}+22 (C_{\varphi D}-4
   C_{\varphi \Box})^2\right) , \\
        -i\,c_{4h}^{(10)} = & \frac{4 m_{h}^2}{\Lambda
   ^2}
   \left(C_{\varphi^4 D^4}^{(1)}+C_{\varphi^4 D^4}^{(2)}+C_{\varphi^4 D^4}^{(3)}\right) , \\
        c_{4h}^{(01)} = & c_{4h}^{(10)}\, ,\\
        -i\,c_{4h}^{(20)} = & - \left(C_{\varphi^4 D^4}^{(1)}+C_{\varphi^4 D^4}^{(2)}+C_{\varphi^4 D^4}^{(3)}\right) , \\
        c_{4h}^{(02)} = & c_{4h}^{(20)}\, ,\\
        c_{4h}^{(11)} = & c_{4h}^{(20)}\, .
    \end{aligned}
\label{eq:match_Result_hhhh_rational_SMEFT_Gf}
\end{equation}

In HEFT up to NNLO, we find
\begin{equation}
    \begin{aligned}
        -i\,c_{4h}^{(00)} = & 3\ m_h^2 \sqrt{2}\ G_F \left(1+\Delta\lambda_4\right)- 16\ m_h^2 G_F^2 ( C_{DH}^{(1)}\,{C'}_{DH}^{(1)} )^2 , \\
        -i\,c_{4h}^{(10)} = & 32\,  m_{h}^2 G_F^2 \bar\Lambda^2 (C_{DH}^{(1)}\,{C'}_{DH}^{(1)})^2 , \\
        -i\ c_{4h}^{(01)} = & 32\,  m_{h}^2 G_F^2 \bar\Lambda^2 (C_{DH}^{(1)}\,{C'}_{DH}^{(1)})^2 , \\
        -i\,c_{4h}^{(20)} = & -8\, G_F^2 \bar\Lambda^4 (C_{DH}^{(1)}\,{C'}_{DH}^{(1)})^2 , \\
        c_{4h}^{(02)} = & c_{4h}^{(20)}\, , \\
        c_{4h}^{(11)} = & c_{4h}^{(20)}\, .
    \end{aligned}
\label{eq:match_Result_hhhh_rational_HEFT}
\end{equation}

\subsubsection{$gg\to hh$}
\label{appendix:gghh}

The result of matching onto SMEFT at dimension-8 in the $G_F$ input scheme is,
\begin{align}
    &\begin{aligned}
     -i\, c_{gghh}^{++,(00)} =& 2 \left(
        C_{\varphi G} + i C_{\varphi \tilde{G}} \right) + 3 \frac{\sqrt{2}}{G_F\, \Lambda^2} \left( C_{G^2\varphi^4}^{(1)} + i C_{G^2\varphi^4}^{(2)}  \right) 
   \\
& +   \frac{\sqrt{2}}{G_F \Lambda^2} \left( C_{\varphi G} + i C_{\varphi \tilde{G}}\right) \left( 4 C_{\varphi \Box} - C_{\varphi D} + 6 C_{\varphi G} + 4 i C_{\varphi \tilde{G}} \right) \\
& +  2 \frac{m_h^2}{\Lambda^2} \left(\frac{1}{4} C_{G^2\varphi^2D^2}^{(1)} + C_{G^2\varphi^2D^2}^{(2)} + i C_{G^2\varphi^2D^2}^{(3)} \right),
\label{eq:match_Result_gghh_++00_00_rational_Gf}
\end{aligned}\\
&\begin{aligned}
     -i\, c_{gghh}^{++,(10)} = & -
   \left(\frac{1}{4} C_{G^2\varphi^2D^2}^{(1)}+
  C_{G^2\varphi^2D^2}^{(2)} + i C_{G^2\varphi^2D^2}^{(3)}\right).
\end{aligned} \nonumber 
\end{align}

\begin{align}
    &\begin{aligned}
     -i\, c_{gghh}^{+-,(00)} =&  \frac{1}{8}\,C^{(1)}_{G^2\varphi^2 D^2}\, .
\label{eq:match_Result_gghh_+-00_00_rational_Gf}
\end{aligned}
\end{align}

In the case of HEFT, up to NNLO order, we have,
\begin{align}
&
\begin{aligned}
     -i\ c_{gghh}^{++,(00)} =& - \sqrt{2}\, g_S^2 \,\bar\Lambda^2 \, G_F\left( C_{GH}^{(2)} +i\,C_{\widetilde{G}H}^{(2)} \right) \\
     & +\frac{g_S^4}{\sqrt{2}}\,G_F \bar\Lambda^2\, (C_{GH}^{(1)}+ i\, C_{\widetilde{G}H}^{(1)} )^2 \\
     &- m_h^2\, \frac{g_S^2\ G_F^2}{2} \,\bar\Lambda^2 \, \left( C_{GHD(1)}^{(0)} + 4 C_{GHD(2)}^{(0)} - 4 i\, C_{\widetilde{G}HD}^{(0)} \right) ,
     \label{eq:match_Result_gghh_++00_00_HEFTNNLO}
\end{aligned}\\
&
\begin{aligned}
     -i\ c_{gghh}^{++,(10)} = \frac{g_S^2\ G_F^2}{4} \bar\Lambda^4 \left( C_{GHD(1)}^{(0)} +4\,C_{GHD(2)}^{(0)} + i\,4\, C_{\tilde G HD}^{(0)}  \right) .
\end{aligned} \nonumber 
\end{align}
\begin{align}
&
\begin{aligned}
     -i\ c_{gghh}^{+-,(00)} =& -\frac{g_S^2\ G_F^2}{8} \bar\Lambda^4 C_{GHD(1)}^{(0)} \, .
\label{eq:match_Result_gghh_+-00_00_HEFTNNLO}
\end{aligned}
\end{align}

\subsection{Five-point amplitudes}
\subsubsection{$hh\to hhh$}
\label{appendix:hhhhh}
Matching on to dim-8 SMEFT in the $G_F$ scheme yields,
\begin{equation}
    \begin{aligned}
        -i\ c_{5h}^{(00000)} = & - \frac{45\ 2^{3/4}}{\sqrt{G_F}}\frac{C_{\varphi}}{\Lambda}+30\sqrt[4]{2}\frac{\sqrt{G_F}m_h^2}{\Lambda} \left( 4\, C_{\varphi\Box}-C_{\varphi D} \right) \\ 
        & - \frac{ 15 }{ 2^{7/4} G_{F}^{3/2} \Lambda ^3}\left(43\ C_{\varphi} (4 C_{\varphi\Box}-C_{\varphi D})+56 i
   C_{\varphi^8}\right)\\
        & + 18 \sqrt[4]{2} \frac{\sqrt{G_F}\ m_h^4}{\Lambda^3} \left( C_{\varphi^4D^4}^{(1)}+C_{\varphi^4D^4}^{(2)}+C_{\varphi^4D^4}^{(3)} \right) \\
        & + \frac{15}{2^{5/4}} \frac{m_h^2}{\sqrt{G_F}\Lambda^3} \left(64 C_{\varphi^6\Box}-(8+2 i)
   C_{\varphi^6D^2}+208 C_{\varphi\Box}^2\right. \\ 
    & \left. -104 C_{\varphi\Box}
   C_{\varphi D}+16 i C_{\varphi\Box}+13 C_{\varphi D}^2\right), \\
   -i\ c_{5h}^{(10000)} = & - 6 \sqrt[4]{2} \frac{m_h^2}{\Lambda} \,\sqrt{G_F}\,\left( C_{\varphi^4D^4}^{(1)}+C_{\varphi^4D^4}^{(2)}+C_{\varphi^4D^4}^{(3)} \right) ,  \\
   -i\ c_{5h}^{(01000)} = & - 3 \sqrt[4]{2} \frac{m_h^2}{\Lambda} \,\sqrt{G_F}\,\left( C_{\varphi^4D^4}^{(1)}+C_{\varphi^4D^4}^{(2)}+C_{\varphi^4D^4}^{(3)} \right) ,  \\
   -i\ c_{5h}^{(00100)} = & 0\, ,\\
   -i\ c_{5h}^{(00010)} = & - 3 \sqrt[4]{2} \frac{m_h^2}{\Lambda} \,\sqrt{G_F}\,\left( C_{\varphi^4D^4}^{(1)}+C_{\varphi^4D^4}^{(2)}+C_{\varphi^4D^4}^{(3)} \right) ,  \\
   -i\ c_{5h}^{(00001)} = & 0\, .
    \label{eq:match_Result_hhhhh_rational_SMEFT_Gf}
    \end{aligned} 
\end{equation}
If the matching is performed to HEFT to NNLO order, the result is instead,
\begin{equation}
    \begin{aligned}
   -i\ c_{5h}^{(00000)} = & 120\ \kappa_{\lambda_5}\ \bar\Lambda\ \sqrt[4]{2} \sqrt{G_F} \\
   & + 24 \ m_{h}^4 \ \bar\Lambda \sqrt[4]{2} G_F^{5/2} \left( 6\ (C_{DH}^{(1)} {C' }_{DH}^{(1)})^2 (1+\Delta\lambda_3) - 52\ C_{DH}^{(1)} {C' }_{DH}^{(1)} \left( C_{DH}^{(1)} {C' }_{DH}^{(2)} + C_{DH}^{(2)} {C' }_{DH}^{(1)} \right) \right) , \\
   -i\ c_{5h}^{(10000)} = & - 12\ m_h^2 \ \bar\Lambda^3 \sqrt[4]{2}\ G_{F}^{5/2} \left( 4 (C_{DH}^{(1)} {C' }_{DH}^{(1)})^2 (1+\Delta\lambda_3) - 36\  C_{DH}^{(1)} {C' }_{DH}^{(1)} \left( C_{DH}^{(1)} {C' }_{DH}^{(2)} + C_{DH}^{(2)} {C' }_{DH}^{(1)} \right) \right) ,  \\
   -i\ c_{5h}^{(01000)} = & - 12\ m_h^2 \ \bar\Lambda^3 \sqrt[4]{2}\ G_{F}^{5/2} \left( 2 (C_{DH}^{(1)} {C' }_{DH}^{(1)})^2 (1+\Delta\lambda_3) -24\  C_{DH}^{(1)} {C' }_{DH}^{(1)} \left( C_{DH}^{(1)} {C' }_{DH}^{(2)} + C_{DH}^{(2)} {C' }_{DH}^{(1)} \right) \right)  , \\
   -i\ c_{5h}^{(00100)} = & 288\ m_h^2 \ \bar\Lambda^3 \sqrt[4]{2}\ G_{F}^{5/2}  C_{DH}^{(1)} {C'}_{DH}^{(1)} \left( C_{DH}^{(1)} {C'}_{DH}^{(2)} + C_{DH}^{(2)} {C' }_{DH}^{(1)} \right) , \\
   -i\ c_{5h}^{(00010)} = & - 12\ m_h^2 \ \bar\Lambda^3 \sqrt[4]{2}\ G_{F}^{5/2} \left( 2 (C_{DH}^{(1)} {C' }_{DH}^{(1)})^2 (1+\Delta\lambda_3) -24\  C_{DH}^{(1)} {C' }_{DH}^{(1)} \left( C_{DH}^{(1)} {C' }_{DH}^{(2)} + C_{DH}^{(2)} {C' }_{DH}^{(1)} \right) \right)  , \\
   -i\ c_{5h}^{(00001)} = & 288\ m_{h}^{2} \ \bar\Lambda^3 \sqrt[4]{2}\ G_{F}^{5/2} C_{DH}^{(1)} {C' }_{DH}^{(1)} \left( C_{DH}^{(1)} {C' }_{DH}^{(2)} + C_{DH}^{(2)} {C' }_{DH}^{(1)} \right), \\
   -i\ c_{5h}^{(20000)} = & - 48\ \bar\Lambda^5 \sqrt[4]{2}\ G_{F}^{5/2} C_{DH}^{(1)} {C' }_{DH}^{(1)} \left( C_{DH}^{(1)} {C' }_{DH}^{(2)} + C_{DH}^{(2)} {C' }_{DH}^{(1)} \right), \\
   -i\ c_{5h}^{(11000)} = & - 48\ \bar\Lambda^5 \sqrt[4]{2}\ G_{F}^{5/2} C_{DH}^{(1)} {C' }_{DH}^{(1)} \left( C_{DH}^{(1)} {C' }_{DH}^{(2)} + C_{DH}^{(2)} {C' }_{DH}^{(1)} \right), \\
   -i\ c_{5h}^{(10100)} = & - 48\ \bar\Lambda^5 \sqrt[4]{2}\ G_{F}^{5/2} C_{DH}^{(1)} {C' }_{DH}^{(1)} \left( C_{DH}^{(1)} {C' }_{DH}^{(2)} + C_{DH}^{(2)} {C' }_{DH}^{(1)} \right), \\
   -i\ c_{5h}^{(10010)} = & - 48\ \bar\Lambda^5 \sqrt[4]{2}\ G_{F}^{5/2} C_{DH}^{(1)} {C' }_{DH}^{(1)} \left( C_{DH}^{(1)} {C' }_{DH}^{(2)} + C_{DH}^{(2)} {C' }_{DH}^{(1)} \right), \\
   -i\ c_{5h}^{(10001)} = & - 48\ \bar\Lambda^5 \sqrt[4]{2}\ G_{F}^{5/2} C_{DH}^{(1)} {C' }_{DH}^{(1)} \left( C_{DH}^{(1)} {C' }_{DH}^{(2)} + C_{DH}^{(2)} {C' }_{DH}^{(1)} \right), \\
   -i\ c_{5h}^{(02000)} = & - 32\ \bar\Lambda^5 \sqrt[4]{2}\ G_{F}^{5/2} C_{DH}^{(1)} {C' }_{DH}^{(1)} \left( C_{DH}^{(1)} {C' }_{DH}^{(2)} + C_{DH}^{(2)} {C' }_{DH}^{(1)} \right), \\
   -i\ c_{5h}^{(01100)} = & - 32\ \bar\Lambda^5 \sqrt[4]{2}\ G_{F}^{5/2} C_{DH}^{(1)} {C' }_{DH}^{(1)} \left( C_{DH}^{(1)} {C' }_{DH}^{(2)} + C_{DH}^{(2)} {C' }_{DH}^{(1)} \right), \\
   -i\ c_{5h}^{(01010)} = & - 32\ \bar\Lambda^5 \sqrt[4]{2}\ G_{F}^{5/2} C_{DH}^{(1)} {C' }_{DH}^{(1)} \left( C_{DH}^{(1)} {C' }_{DH}^{(2)} + C_{DH}^{(2)} {C' }_{DH}^{(1)} \right), \\
   -i\ c_{5h}^{(01001)} = & - 16\ \bar\Lambda^5 \sqrt[4]{2}\ G_{F}^{5/2} C_{DH}^{(1)} {C' }_{DH}^{(1)} \left( C_{DH}^{(1)} {C' }_{DH}^{(2)} + C_{DH}^{(2)} {C' }_{DH}^{(1)} \right), \\
   -i\ c_{5h}^{(00200)} = & - 32\ \bar\Lambda^5 \sqrt[4]{2}\ G_{F}^{5/2} C_{DH}^{(1)} {C' }_{DH}^{(1)} \left( C_{DH}^{(1)} {C' }_{DH}^{(2)} + C_{DH}^{(2)} {C' }_{DH}^{(1)} \right), \\
   -i\ c_{5h}^{(00110)} = & - 16\ \bar\Lambda^5 \sqrt[4]{2}\ G_{F}^{5/2} C_{DH}^{(1)} {C' }_{DH}^{(1)} \left( C_{DH}^{(1)} {C' }_{DH}^{(2)} + C_{DH}^{(2)} {C' }_{DH}^{(1)} \right), \\
   -i\ c_{5h}^{(00101)} = & - 32\ \bar\Lambda^5 \sqrt[4]{2}\ G_{F}^{5/2} C_{DH}^{(1)} {C' }_{DH}^{(1)} \left( C_{DH}^{(1)} {C' }_{DH}^{(2)} + C_{DH}^{(2)} {C' }_{DH}^{(1)} \right), \\
   -i\ c_{5h}^{(00020)} = & - 32\ \bar\Lambda^5 \sqrt[4]{2}\ G_{F}^{5/2} C_{DH}^{(1)} {C' }_{DH}^{(1)} \left( C_{DH}^{(1)} {C' }_{DH}^{(2)} + C_{DH}^{(2)} {C' }_{DH}^{(1)} \right), \\
   -i\ c_{5h}^{(00011)} = & - 32\ \bar\Lambda^5 \sqrt[4]{2}\ G_{F}^{5/2} C_{DH}^{(1)} {C' }_{DH}^{(1)} \left( C_{DH}^{(1)} {C' }_{DH}^{(2)} + C_{DH}^{(2)} {C' }_{DH}^{(1)} \right), \\
   -i\ c_{5h}^{(00002)} = & - 32\ \bar\Lambda^5 \sqrt[4]{2}\ G_{F}^{5/2} C_{DH}^{(1)} {C' }_{DH}^{(1)} \left( C_{DH}^{(1)} {C' }_{DH}^{(2)} + C_{DH}^{(2)} {C' }_{DH}^{(1)} \right) .\\
    \end{aligned}
    \label{eq:match_Result_hhhhh_rational_HEFT}
\end{equation}

\subsubsection{$gg\to hhh$}
\label{appendix:gghhh}
For the $++000$ helicity configuration, the result of matching onto dim-8 SMEFT in the $G_F$ scheme is,
\begin{equation}
    \begin{aligned}
        -i\ c_{gghhh}^{++,(1)(00000)}= & \frac{9\,m_h^2 \sqrt{G_F}}{2^{7/4}\Lambda}\left( C_{G^2\varphi^2D^2}^{(1)} + 4\ C_{G^2\varphi^2D^2}^{(2)} + i 4\ C_{G^2\varphi^2D^2}^{(3)} \right)+\frac{6\ 2^{3/4}}{\Lambda\ \sqrt{G_F}}\left( C_{G^2\varphi^4}^{(1)}+i\ C_{G^2\varphi^4}^{(2)} \right) \\
        &+\frac{8\ 2^{3/4}}{ \Lambda\ \sqrt{G_F} }\left(C_{\varphi\Box}-\frac{1}{4}C_{\varphi D}\right)\left( C_{\varphi G} + { i} C_{\varphi\widetilde{G}} \right)+\frac{12\ 2^{3/4}}{\Lambda\ \sqrt{G_F}}\left( C_{\varphi G} + { i} C_{\varphi\widetilde{G}} \right)^2,
    \end{aligned}
    \label{eq:match_Result_gghhh_pp_rational_SMEFT_Gf}
\end{equation}
and all other on-shell coefficients vanish.
HEFT at NNLO already provides a non-vanishing contribution to the coefficients that are linear in the Mandelstam variables,
\begin{equation}
    \begin{aligned}
    -i\ c_{gghhh}^{++,(1)(00000)}= & -4\ m_h^2 \sqrt[4]{2} G_F^{5/2} g_S^2 \ \bar\Lambda^3 \left(C_{DH}^{(1)}\,\,{C'}_{DH}^{(1)}\right)^2 \left( C_{GH}^{(1)} +i\ C_{\widetilde{G}H}^{(1)} \right) \\
    & + 3\ 2^{3/4}\ G_{F}^{3/2}\ g_{S}^{4}\ \bar\Lambda^3 \left( C_{GH}^{(1)} +i\ C_{\widetilde{G}H}^{(1)} \right) \left( C_{GH}^{(2)} +i\ C_{\widetilde{G}H}^{(2)} \right) \\
    & - 3\ 2^{3/4}\ G_{F}^{3/2} g_{S}^{2} \bar\Lambda^3 \left( C_{GH}^{(3)} +i\ C_{\widetilde{G}H}^{(3)} \right)\\ & -\frac{3\ m_h^2}{ 2^{7/4} }\ G_F^{5/2}\ g_S^2\ \bar\Lambda^3 \left[ 3\ C_{GHD(1)}^{(0)}(\Delta\lambda_3+1) + C_{GHD(1)}^{(1)} \right. \\  
    & \left. + \ 4\ \left( 3\ C_{GHD(2)}^{(0)} (\Delta\lambda_3+1) + C_{GHD(2)}^{(1)} + i\ (3\ C_{\widetilde{G}HD}^{(0)}(\Delta\lambda_3+1) + C_{\widetilde{G}HD}^{(1)}) \right) \right], \\
    -i\ c_{gghhh}^{++,(1)(10000)} = &\ \frac{G_F^{5/2}\ g_S^2\ \bar\Lambda^5 }{2^{7/4}}\left(  C_{GHD(1)}^{(1)}+ { 4} C_{GHD(2)}^{(1)} + { 4}i\ C_{\widetilde{G}HD}^{(1)} \right),\\
    -i\ c_{gghhh}^{++,(1)(01000)} = &\ 4\sqrt[4]{2} G_{F}^{5/2} g_{S}^2\ \bar\Lambda^5 {C_{DH}^{(1)}}^2 {{C'}_{DH}^{(1)}}^2 \left( C_{GH}^{(1)} + i\ C_{\widetilde{G} H}^{(1)}  \right),\\
    -i\ c_{gghhh}^{++,(1)(00100)} = &\ 0 \, ,\\
    -i\ c_{gghhh}^{++,(1)(00010)} = &\ 4\sqrt[4]{2} G_{F}^{5/2} g_{S}^2\ \bar\Lambda^5 {C_{DH}^{(1)}}^2 {{C'}_{DH}^{(1)}}^2 \left( C_{GH}^{(1)} + i\ C_{\widetilde{G} H}^{(1)}  \right),\\
    -i\ c_{gghhh}^{++,(1)(00001)} = &\ 0 \, .
    \end{aligned}
    \label{eq:match_Result_gghhh_pp_rational_HEFT_NNLO}
\end{equation}

The $+-000$ helicity configuration can not be generated at tree-level by dim-8 SMEFT, so there are no corresponding matching results. HEFT, instead, yields the following at NNLO order,

\begin{equation}
-i\ c_{gghhh}^{+-,(00000)}=\bar{\Lambda} \, \dfrac{g_s^2 \sqrt{G_F} }{2^{9/4}} \; C^{(1)}_{GHD(2)}\,.
    \label{eq:match_Result_gghhh_pm_rational_HEFT_NNLO}
\end{equation}

\bibliographystyle{JHEP.bst}
\bibliography{bibliography}

\end{document}